\documentclass[review=false]{jfp-epi}

\usepackage{lipsum} 

\usepackage{stmaryrd}
\usepackage[all]{xy}
\usepackage{array}
\usepackage{booktabs} 
\newtheorem*{theorem*}{Theorem}  
\theoremstyle{acmplain}
\newtheorem{theorem}{Theorem}

\theoremstyle{acmdefinition}

\usepackage{listings}
\jfpVolume{99}
\jfpArticle{13}
\jfpDOI{10.46298/jfp.01234}
\jfpYear{2089}

\received[Submitted]{October 2087}
\received[accepted]{April 2089}

\begin{document}

\title{Functional design of efficient and parallelizable combinatorial generators
	using convolution}

\author{Xi He}
\affiliation{%
  \institution{Peking University}
  \city{Beijing}
  \country{China}
  \authoremail{xihe@pku.edu.cn}
}
\author{Zhenjiang Hu}
\affiliation{%
	\institution{Peking University}
	\city{Beijing}
	\country{China}
	\authoremail{huzj@pku.edu.cn}
}
\author{Max A. Little}
\affiliation{%
	\institution{University of Birmingham}
	\city{Birmingham}
	\country{UK}
	\authoremail{maxl@mit.edu}
}

\begin{abstract}
The application of program transformation and algebraic methods to
the development of efficient combinatorial optimization (CO) algorithms
relies on an exhaustive combinatorial generator for the problem specification,
followed by the fusion of thinning or filtering processes into this
specification. However, the effectiveness of such fusion transformations
critically depends on the \emph{structural compatibility} between
the objective function and the generator, which is highly \emph{problem
	dependent}. In practice, when the majority of candidate solutions
remain unfiltered or are not eliminated---as is the case for most
intractable CO problems---the overall efficiency of the resulting
fused program is largely determined by the intrinsic efficiency of
the combinatorial generator. Consequently, if the specification itself
exhibits suboptimal performance, the fused program will inherit a
correspondingly inferior level of efficiency.

We argue that a genuine designed process should also account for \emph{hardware
	compatibility} and \emph{parallelizability}---particularly the ability
to support efficient parallel execution on modern hardware architectures,
including multi-level cache hierarchies and GPUs. 

However, does achieving formal correctness necessarily conflict with designing algebraically
elegant algorithms that support fusion? Can we achieve both at the same time?
More importantly, can techniques from functional programming help
us systematically and formally construct hardware-compatible generators?
If so, it may be possible to derive efficient optimization algorithm from a exhaustive specification
that guarantees both formal correctness and efficient hardware-oriented
execution.

In this paper, we show that techniques from functional programming---most
notably Bird’s algebra of programs---together with semiring algebras,
provide powerful formal tools for the systematic construction of such
hardware-compatible and parallelizable combinatorial generators.

This paper investigates generators for two of the most fundamental
combinatorial structures---\emph{combinations} and \emph{permutations}---together
with their natural extension to \emph{nested generators} (e.g., combinations/permutations
of combinations/permutations). To the best of our knowledge, nested
generators have received virtually no systematic treatment beyond
trivial sequential composition. We present concise recursive formulations
of nested generators based on the \emph{join-list homomorphism}.

Finally, we demonstrate how \emph{Gray code ordering} can be fruitfully
incorporated into the construction of the combinatorial generators
developed in this work, yielding further improvements in both effectiveness
and efficiency.

\end{abstract}

\maketitle

\section{Introduction}

The significance of efficient combinatorial generators is twofold.
First, certain applications demand the exhaustive enumeration of combinations.
For example, in automated proof search tactics (such as the Crush
tactics in \citet{chlipala2013cpdt}) relies on the exhaustive enumeration
of combinations of possible proof steps to automatically discover
proofs. In combinatorial geometry, the cell enumeration problem for
hyperplane arrangements \citep{avis1996reverse}, requires the complete
enumeration of all cells. Notably, \citet{gerstner2006algorithms}
demonstrated that this problem can be reduced to a combination generation
tasks.

Second, in the study of program calculation \citep{bird2020algorithm,bird1996algebra},
the development of combinatorial optimization algorithms relies on
the availability of a well-defined combinatorial generator that guarantees
the exhaustive enumeration of all candidate combinations. Such generators
provide the clearest specification for deriving correct-by-construction
algorithms. However, as noted by \citet{DEMOOR1993Bridging}, these
formal development of the algorithm often lead to less efficient algorithm
compared ad-hoc designed algorithms.

We argue that a principal source of this inefficiency lies in the
widespread neglect of hardware compatibility during the early stages
of algorithm design. When fusion is applied to an inherently inefficient
initial specification, and the vast majority of candidate combinations
remain neither filtered nor eliminated---as is the case for most
intractable combinatorial optimization problems---the resulting fused
program inevitably inherits, and often retains, this inefficiency.
This situation is typical of most nontrivial combinatorial problems,
for which exhaustive enumeration of all candidates constitutes the
default baseline.

Algorithms that are more efficient and exhibit stronger hardware compatibility
are therefore more likely to succeed in practice, a phenomenon termed
the \emph{hardware lottery} \citep{hooker2021hardware}. This term
captures the observation that certain research ideas prevail \textbf{not}
because they are intrinsically superior to competing approaches, but
because they \emph{align well with the prevailing software and hardware
	ecosystem}. The most prominent example is deep neural networks which
benefited substantially from optimized hardware and software. A similar
``lottery'' was observed in our previous investigation of the decision
tree problem \citep{he2025FoODTII}, where a hardware-compatible brute-force
algorithm---readily parallelizable on GPUs---was compared against
a fused dynamic programming (DP) recursive algorithm, our empirical
results showed that a carefully optimized DP algorithm can be less
efficient than a brute-force approach with superior hardware compatibility.

Nevertheless, we argue that techniques from algebra and functional
programming can also be employed to derive efficient combinatorial
generators that substantially \textbf{enhance cache performance} and
are inherently based on \emph{array operations}, thereby \textbf{facilitating
	efficient GPU implementation}. This paper serves as an exposition
on the design of such \emph{GPU-compatible generators} with \emph{streamlined
	memory management} to enhance cache performance for some of the most
fundamental combinatorial structures---namely, combinations and permutations.
The significance of the combination generator is emphasized by \citet{knuth2005art}
in his influential book \emph{The Art of Computer Programming}, where
he wrote
\begin{quote}
	``Even the apparently lowly topic of combination generation turns
	out to be surprisingly rich,.... I strongly believe in building up
	a firm foundation.''
\end{quote}
In this paper, we demonstrate how techniques from Bird’s algebra of
programming \citep{bird1996algebra} and semiring algebra \citet{little2024polymor}
can be employed to develop efficient, hardware-compatible generators.
The contributions of this paper are organized as follows:
\begin{enumerate}
	\item We present two efficient \textbf{$K$-combination }and\textbf{ permutation
		generators} based on a \emph{generic convolution operator}. The generators
	are defined using join-list homomorphisms, thereby naturally enabling
	\emph{embarrassingly parallelizable execution}. Since our construction
	of these generators is grounded in semiring algebra, the semiring
	fusion techniques discussed in \citet{little2024polymor,emoto2012filter}
	are directly applicable.
	\item We extend classical combination generators to their natural extension
	as \textbf{nested generators} (combinations/permutations of combinations/permutations),
	wherein a combination/permutation generator takes the output of another
	combination/permutation generator as input.
	\item We incorporate \textbf{Gray code ordering} (\textbf{revolving door
		ordering}) into the nested combination generators, which not only
	substantially reduces memory usage but also establishes a perfect
	indexing scheme without the need for additional hashing functions.
\end{enumerate}
The paper is organized as follows. In Section \ref{sec: section 2},
we discuss related work. Section \ref{sec: section3} introduces the
notation used throughout the paper. Section \ref{sec: section4} presents
the detailed construction of the D\&C $K$-combination/permutation
generators and examines their connection to revolving door algorithms.
Section \ref{sec: section 5} provides a detailed treatment of the
construction of the nested generator and demonstrates how revolving
door ordering can be incorporated to facilitate the development of
nested generators.

A more detailed exposition of how various combinatorial and optimization
problems can be related to the combinatorial generators proposed in
this paper is provided in Section \ref{sec:Applications}. In the
experimental section \ref{sec:Experimental-results}, we evaluate
the performance of the generators on a representative problem closely
related to combination generation. The results show that, thanks to
the parallelism enabled by our approach, we achieve nearly two orders
of magnitude speedup compared with a highly optimized C-based Python
library, and also outperform state-of-the-art algorithms based on
mixed-integer programming solvers

and experiments on the generation performance, and we choose one example
problem that is closely related to the combination generation to illustrate,
the experiments shows that due to the parallability that we provide,
not only we can achieve two order magnitude speed up compared with
a C-based python library, but also outperforms the state-of-art algorithm
based on mixed-integer programming solvers \citep{ren2022global}.

Finally we present our conclusion in Section \ref{sec: Conclusion}.

\section{Related studies\label{sec: section 2}}

Combinations and permutations are among the simplest, yet most fundamental
and important, combinatorial objects. Numerous algorithms have been
developed for generating combinations. Although all combinatorial
generators exhaustively enumerate the corresponding combinatorial
objects, they often differ in two key aspects: the \emph{ordering}
(\emph{ranking}) imposed on the generated configurations and \emph{generation
	efficiency}.

Most studies in combinatorial generation focus on the first aspects.
A prominent example is lexicographical ordering, which compares elements
within a configuration from left to right; for instance, $\left[1,2,3\right]$
is ranked lower than $\left[1,3,2\right]$. Another, more widely used
ordering relation is \emph{Gray code ordering}, which consists of
lists of combinatorial objects in which successive objects are ``close''
to one another in some prescribed sense. For example, the \emph{revolving
	door ordering} is an example of Gray code ordering, which requires
that every pair of adjacent combinations in $\mathit{cs}_{k}$, including
the first and the last, differ by exactly two elements. The revolving
door algorithm is perhaps the most widely known algorithm for generation
combination, the term ``revolving door'' is coined by \citet{NijWilf1976}
though the method was originally invented by \citet{tang1973distance}.
\citet{bitner1976efficient} later proposed a \emph{loopless}\footnote{A loopless algorithm is one in which, in the worst case, only a small
	number of operations (or bounded by a constant) are required between
	successive confgurations, this concept was first introduced by \citet{ehrlich1973loopless}. } version of the algorithm. 

A more restrictive class of Gray code algorithms is known as \emph{adjacent-change}
\emph{algorithms}, in which successive bitstrings (representing combinations)
differ by a \textbf{single} interchange of adjacent bits. \citet{knuth2011art}
refers to such algorithms as \emph{perfect} \emph{generators}. It
has been proven by \citet{ruskey1984adjacent} and \citet{eades1984some}
that adjacent-change algorithms do \textbf{not} exist in general for
combination generation.

However, due to their intrinsic ordering requirements, Gray code orderings
inherently exhibit a sequential nature and are therefore unable to
exploit associativity to construct definitions based on divide-and-conquer
(D\&C) algorithms (i.e., join-list homomorphisms). It has long been
recognized that this style of recursion is crucial for constructing
MapReduce-style algorithms \citep{emoto2012filter}. 

For the second aspect, it may appear counterintuitive that different
generators will exhibit different levels of efficiency, since all
of them must exhaustively enumerate all combinations. If the number
of configurations grows exponentially with the input size, any generator
for such objects inherently has exponential (or worse) complexity.
However, as \citet{ruskey2003combinatorial} noted, an ideal combinatorial
generator should have \emph{constant amortized time}---that is, generating
each configuration should take a constant amount of time on average,
regardless of the input size. 

On the other hand, exploring D\&C algorithms are crucial for constructing
efficient (parallelizable) generators. As \citet{gupta2000automatic}
noted that ``The \emph{divide-and-conquer style of writing array based
	computations} is likely to gain popularity in the future, as it allows
better exploitation of the memory hierarchy in modern machines, particularly
when used together with a \emph{recursively blocked array data layout}.''

Therefore, to achieve superior efficiency for combination generators---whose
worst-case complexity cannot be improved---it is essential to exploit
parallelism by designing D\&C-style generator with a recursively blocked
array data layout.

D\&C-style recursion for combination and permutation generation has
been studied previously. For instance, \citet{emoto2012filter} and
\citet{jeuring1993theories} provide definitions of combination and
permutation generators based on join-list homomorphisms, respectively.
However, although their definitions are intuitively straightforward,
they are not ideal, as neither incorporates a recursively blocked
array data layout. Consequently, both definitions are inherently list-based,
which complicates memory management. To address this gap, in this
paper we propose D\&C-style combination and permutation generators
with a recursively blocked array layout, thereby further enhancing
their parallelizability and streamlining the process of memory management.

Regarding nested generators, to the best of our knowledge, a systematic
study of nested combinatorial generators has not been reported previously.
We hope that our exploration demonstrates how techniques from functional
programming can facilitate the discovery of new algorithmic structures
in a sound and concise manner.

Although this previously unexplored problem may initially appear to
be motivated primarily by novelty rather than practical importance,
we find that such generators support a wide range of significant applications.
In statistical analysis, procedures such as analysis of variance \citep{anderson2003permutation}
inherently involve combinations of permutations over selected factors.
In optimization, models such as two-layer neural networks \citep{he2025deepice}
and optimal hyperplane decision trees involve combinations or permutations
of hyperplanes \citep{he2025FoODTI}, which can in turn be formulated
as nested combination/permutation problems.

In combinatorial optimization, the nested combination--combination
generator is closely related to the 2-layer ReLU/maxout neural network
\citep{he2025deepice}, while the nested permutation--combination
generator arises in the solution of the hypersurface decision tree
problem considered by \citet{he2025FoODTI}. Notably, both applications
involving nested permutation/combination generators have been solved
only very recently, which may help explain the historical lack of
systematic study of nested generators. 

\section{Notations\label{sec: section3}}

The types of real and natural numbers are denoted as $\mathbb{R}$
and $\mathbb{N}$, respectively. We use square brackets $\left[\mathcal{X}\right]$
to denote the set of all finite lists of elements $x:\mathcal{X}$,
where $\mathcal{X}$ (or letters $\mathcal{Y}$ and $\mathcal{Z}$
at the end of the alphabet) represent \emph{type variables}. Variables
of these types are written in lowercase, e.g. $x:\mathcal{X}$. A
list of such variables is denoted as $\mathit{xs}:\left[\mathcal{X}\right]$,
and a list of lists is written as $\mathit{xss}=\left[\left[\mathcal{X}\right]\right]$.
We use $\mathit{xs}{}_{D}$ or $!\left(D,xs\right)$ to refer to the
$D$th element of the list $\mathit{xs}$ (start from zero), and use
the symbol $\cup$ to denote either set union or list concatenation,
depending on the context. Function composition is denoted using the
infix symbol ``$\cdot$'', such that $\left(f\cdot g\right)\left(a\right)=f\left(g\left(a\right)\right)$.
Functions can be partially applied; for example, $!\left(D\right):\left[\mathcal{X}\right]\to\mathcal{X}$
is a partially applied function that takes a list and returns its
$D$th element. The list comprehension is defined as $\mathit{xs}=\left[x\mid x\leftarrow\mathit{xs}\right]$.

Let $C_{K}^{N}$ denote the number of possible combinations for choosing
$K$ elements from a set of size $N$ set, i.e., the binomial coefficient
$\left(\begin{array}{c}
	N\\
	K
\end{array}\right)$, $\mathit{cs}_{K}^{N}$ denote the set of all $K$-combinations from
a size $N$ set, and $\mathinner{css}_{K}^{N}=\left[\mathit{cs}_{1}^{N},\ldots,\mathit{cs}_{K}^{N}\right]$denotes
the list of all combination sets of sizes up to $K$. When the context
is clear, we omit the superscript $N$ or subscript $K$ for brevity.
Note that since the subscript in $\mathit{cs}_{K}^{N}$ and $\mathinner{css}_{K}^{N}$
conveys different meanings, we use notation ($!$) to index their
elements. In other words, $!\left(k,\mathinner{css}_{K}^{N}\right)=\mathit{cs}_{k}^{N}$.

\section{Novel array-based combination generators using convolution\label{sec: section4}}

\subsection{Divide-and-conquer version}

\emph{Vandermonde's identity} is an identity for binomial coefficients,
which states that
\begin{equation}
	C_{K}^{N+M}=\sum_{k=0}^{K}\left(C_{K-k}^{N}C_{k}^{M}\right)\label{eq: Vandermonde=002019s identity - arithmetic semiring}
\end{equation}
Equation (\ref{eq: Vandermonde=002019s identity - arithmetic semiring})
can be proved by a simple combinatorial counting argument: Consider
two sets $A$ and $B$ with distinct elements, where $\left|A\right|=N$
and $\left|B\right|=M$. To select $K$ elements from the union $A\cup B$,
we can choose $K-k$ elements from $A$ and $k$ elements from $B$.
There are $C_{K-k}^{N}C_{k}^{M}$ ways to do this, and $k$ ranging
from $0$ to $K$.

The combinatorial proof of Vandermonde's identity is constructive
and follows a pattern similar to D\&C algorithms. Specifically, the
original problem $C_{K}^{N+M}$ can be decomposed into two \emph{smaller},
\emph{independent} subproblems $C_{K-k}^{N}$ and $C_{k}^{M}$ for
arbitrary $0\leq M,N\leq M+N$ which can be solved separately. This
observation inspires us to design an efficient array-based D\&C algorithm
for generating combinations.

If we replace the sum and product operations in (\ref{eq: Vandermonde=002019s identity - arithmetic semiring})---originally
defined over the arithmetic semiring $\left(\mathbb{N},+,\times,0,1\right)$---with
the operations of\emph{ }the\emph{ generator semiring} $\left(\left[\left[\mathcal{X}\right]\right],\cup,\circ,\left[\;\right],\left[\left[\;\right]\right]\right)$
\citep{little2024polymor}, we obtain 
\begin{equation}
	\mathit{cs}_{K}^{N+M}=\bigcup_{k=0}^{K}\left(\mathit{cs}_{K-k}^{N}\circ\mathit{cs}_{k}^{M}\right)\label{eq: Vandermonde's identity-generator semiring}
\end{equation}
where $\circ$ is the cross-join operator defined by $\mathit{xs}\circ\mathit{ys}=\left[x\cup y\mid x\leftarrow\mathit{xs},y\leftarrow\mathit{ys}\right]$.
For example $\left[\left[1\right],\left[2\right]\right]\circ\left[\left[3\right],\left[4\right]\right]=\left[\left[1,3\right],\left[1,4\right],\left[2,3\right],\left[2,4\right]\right]$.
This yields the combinatorial proof of Vandermonde's identity under
the generator semiring: all possible $K$-combinations of a set of
size $M+N$ ($\mathit{cs}_{K}^{N+M}$) can be constructed by cross-joining
all $K-k$-combinations of the size-$N$ subset ($\mathit{cs}_{K-k}^{N}$)
with all $k$-combinations of the size $M$ subsets $\mathit{cs}_{k}^{M}$).

To apply program transformation, we need to manipulate the actual
program rather than just the mathematical formula. The right-hand
side of (\ref{eq: Vandermonde's identity-generator semiring}) can
be implemented as 
\begin{equation}
	\mathit{cs}_{K}^{N+M}=\mathit{concat}\left(\mathit{zipWith}\left(\circ,\mathit{reverse}\left(\mathinner{css}_{K}^{N}\right),\mathinner{css}_{K}^{M}\right)\right)\label{eq: singe-combination}
\end{equation}
where $\mathit{zipWith}:\left(\mathcal{X}\times\mathcal{Y}\to\mathcal{Z}\right)\times\left[\mathcal{X}\right]\times\left[\mathcal{Y}\right]\to\left[\mathcal{Z}\right]$
applies a binary function $f:\mathcal{X}\times\mathcal{Y}\to\mathcal{Z}$
pairwise to elements of two lists $\left[\mathcal{X}\right]$ and
$\left[\mathcal{Y}\right]$ (assumed to have the same length), $\mathit{reverse}:\left[\mathcal{X}\right]\to\left[\mathcal{X}\right]$
reverses the order of a list, and $\mathit{concat}:\left[\left[\mathcal{X}\right]\right]\to\left[\mathcal{X}\right]$
flattens a list of lists into a single list. See Appendix \ref{subsec:Standard-functions-in Haskell}
for definitions and example executions. 

Equation (\ref{eq: singe-combination}) \emph{nearly} defines a D\&C
algorithm, where the larger problem $\mathit{cs}_{K}^{N+M}$ is constructed
from the smaller problems $\mathinner{css}_{K}^{N}$ and $\mathinner{css}_{K}^{M}$.
However, it does \emph{not} yet fully exhibit a D\&C structure, as
(\ref{eq: singe-combination}) is not recursive: \emph{the type of
	the variable on the left-hand side ($\mathinner{cs}_{K}$) does not
	match the type of the variable on the right-hand side $\mathinner{css}_{K}$}.
Specifically, $\mathinner{cs}_{K}:\left[\left[\mathcal{X}\right]\right]$
is a list of lists, whereas $\mathinner{css}_{K}:\left[\left[\left[\mathcal{X}\right]\right]\right]$
is a list of lists of lists (a list of $\mathinner{cs}_{K}$). Despite
this mismatch, $\mathinner{css}_{K}^{N}$ and $\mathinner{css}_{K}^{M}$
can be used not only to construct $\mathit{cs}_{K}^{N+M}$, but also
all smaller combinations $\mathit{cs}_{k}^{N+M}$ such that $0\le k\le K$.
In other words, the following equality holds

\begin{equation}
	\mathinner{css}_{K}^{N+M}=\left[\mathit{cs}_{1}^{N+M},\ldots,\mathit{cs}_{K}^{N+M}\right]=\left[\bigcup_{k=0}^{0}\left(\mathit{cs}_{K-k}^{N}\circ\mathit{cs}_{k}^{M}\right),\ldots,\bigcup_{k=0}^{K}\left(\mathit{cs}_{K-k}^{N}\circ\mathit{cs}_{k}^{M}\right)\right]
\end{equation}
where $\bigcup_{k=0}^{0}x_{k}=\left[\left[\:\right]\right]$. This
construction is essentially the \emph{discrete convolution} of two
disjoint sets of combinations of size $K$. If we define the convolution
operator\\
$\mathit{convol}\left(\oplus,\otimes,\mathit{xss},\mathit{yss}\right)=\left[\bigoplus_{k=0}^{0}\left(\mathit{xss}_{K-k}\otimes\mathit{yss}_{K}\right),\ldots,\bigoplus_{k=0}^{K}\left(\mathit{xss}_{K-k}\otimes\mathit{yss}_{K}\right)\right]$.
Using this notation, we have
\begin{equation}
	\mathinner{css}_{K}^{N+M}=\mathit{convol}\left(\cup,\circ,\mathinner{css}_{K}^{N},\mathinner{css}_{K}^{M}\right).
\end{equation}
This convolution operator can be implemented as 
\begin{align}
	\mathit{convol} & :\left(\mathcal{X}\to\mathcal{Y}\to\left[\mathcal{Z}\right]\right)\to\left[\mathcal{X}\right]\to\left[\mathcal{Y}\right]\to\left[\left[\mathcal{Z}\right]\right]\\
	\mathit{convol} & \left(\circ,\mathit{xss},\mathit{yss}\right)=\left[\mathit{concat}\left(\mathit{zipWith}\left(\circ,\mathit{reverse}\left(\mathinner{css}\right),\mathit{yss}\right)\right)\mid\mathinner{css}\leftarrow\mathit{inits}\left(\mathit{xss}\right)\right],
\end{align}
where the $\mathit{inits}:\left[\mathcal{X}\right]\to\left[\left[\mathcal{X}\right]\right]$
returns all (non-empty) initial segments of a list. See Appendix \ref{subsec:Standard-functions-in Haskell}
for definitions and examples. Note that if $\mathit{xss}$ has length
$N$, then $\mathit{map}\left(\mathit{reverse},\mathit{inits}\left(\mathit{xss}\right)\right)$
(where $\mathit{map}\left(f,\mathit{xs}\right)=\left[f\left(x\right)\mid x\leftarrow\mathit{xs}\right]$
which applies $f$ to each element $x$ in $\mathit{xs}$) takes $O\left(N^{3}\right)$
operations because $\mathit{inits}$ produces a list of length $O\left(N^{2}\right)$
and $\mathit{reverse}$ takes $O\left(N\right)$ time. Using the list
accumulation lemma \citep{bird1987introduction}, we can implement
a more efficient function $\mathit{revinits}\left(\mathit{xss}\right)=\mathit{map}\left(\mathit{reverse},\mathit{inits}\left(\mathit{xss}\right)\right)$
(Appendix \ref{subsec:Standard-functions-in Haskell} for definition)
in $O\left(N^{2}\right)$ time by accumulating results. Thus, the
convolution operator can be efficiently implemented as $\mathit{convol}\left(\circ,\mathit{xss},\mathit{yss}\right)=\left[\mathit{concat}\left(\mathit{zipWith}\left(\circ,\mathinner{css},\mathit{yss}\right)\right)\mid\mathinner{css}\leftarrow\mathit{revinits}\left(\mathit{xss}\right)\right]$.

We can now construct the D\&C algorithm $\mathit{kcombs}$ for generating
all $K$-combinations ($0\leq k\le K$). Given an input list $\mathit{xs}\cup\mathit{ys}$,
$\mathit{kcombs}$ is defined as

\begin{equation}
	\begin{aligned}\mathit{kcombs} & \left(K,\left[\;\right]\right)=\left[\left[\left[\:\right]\right],\left[\:\right]^{K}\right]\\
		\mathit{kcombs} & \left(K,\left[x\right]\right)=\left[\left[\left[\:\right]\right],\left[\left[x\right]\right],\left[\:\right]^{K-1}\right]\\
		\mathit{kcombs} & \left(K,\mathit{xs}\cup\mathit{ys}\right)=\mathit{convol}\left(\circ,\mathit{kcombs}\left(K,xs\right),\mathit{kcombs}\left(K,ys\right)\right).
	\end{aligned}
	\label{eq:kcombs-generator-D=000026C}
\end{equation}
where $\left[\:\right]^{K}=\left[\left[\:\right]\mid k\leftarrow\left[1,\ldots,K\right]\right]$
denotes the empty list replicated $\left[\:\right]$ $K$ times. In
the base cases, there is exactly one way to form the 0-combination,
one way to form the 1-combination for a singleton list, and no way
to form larger combinations.

Because the discrete convolution\\
 $\mathit{convol}\left(\circ,\mathit{xss},\mathit{yss}\right)=\left[\bigcup_{k=0}^{0}\left(\mathit{xss}_{K-k}\otimes\mathit{yss}_{K}\right),\ldots,\bigcup_{k=0}^{K}\left(\mathit{xss}_{K-k}\otimes\mathit{yss}_{K}\right)\right]$
organizes combinations of the same size in the same list, all combinations
of size $k$ ($0\leq k\le K$) have the same length. Therefore, we
can organize combinations by size using $K$ different two-dimensional
arrays (matrices). The size of the matrix storing $\bigcup_{k=0}^{K}\left(\mathit{xss}_{K-k}\otimes\mathit{yss}_{K}\right)$
can be precomputed using $\sum_{k=0}^{K}\left(C_{K-k}^{N}C_{k}^{M}\right)$.
Hence, the recursive definition in (\ref{eq:kcombs-generator-D=000026C})
naturally induces a blocked array data layout, which enables full
vectorization and efficient implementation.

\subsection{Sequential version and relation to the revolving door algorithm}

\begin{figure}
	\centering{}\includegraphics[scale=0.15]{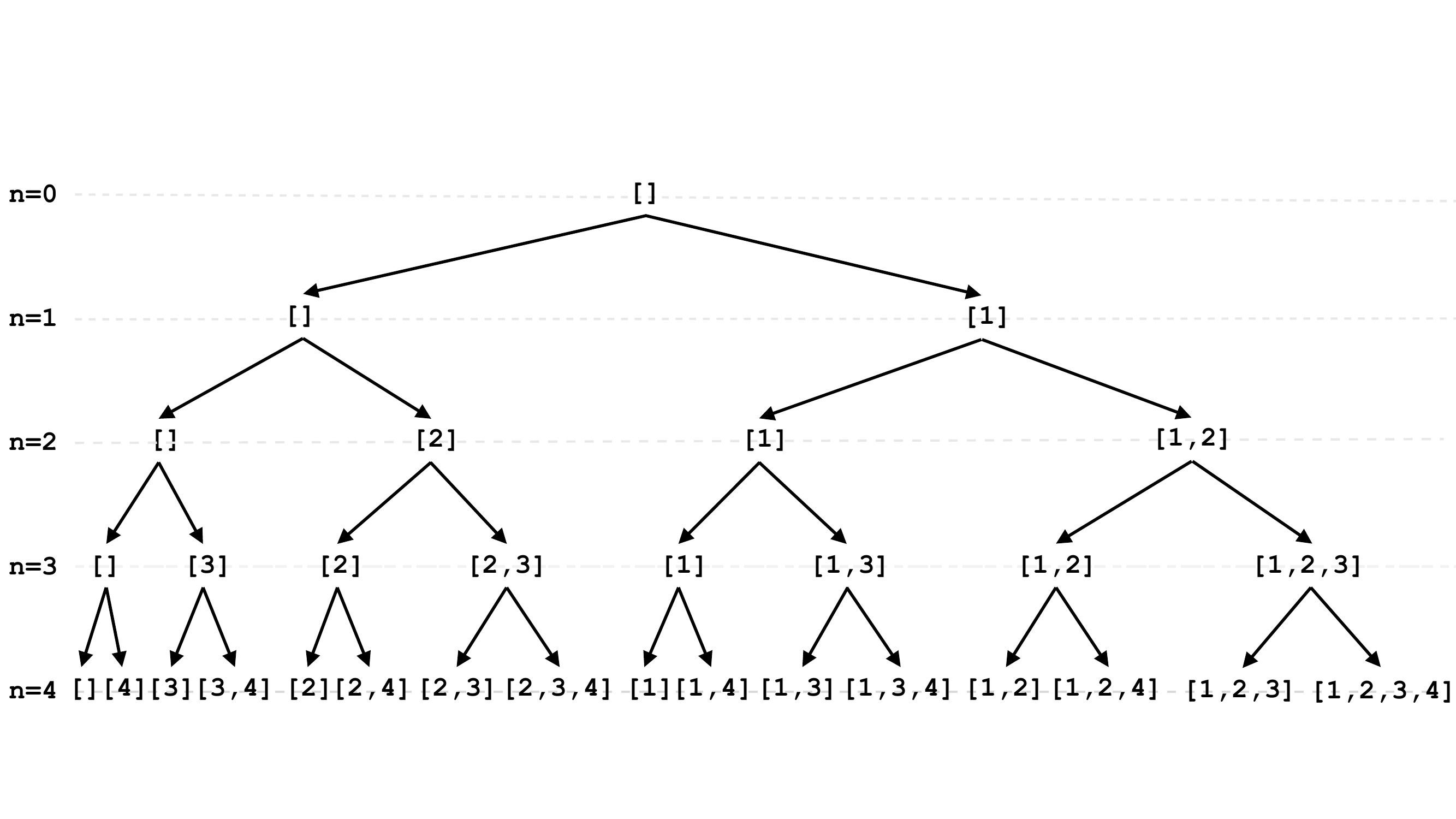}\includegraphics[scale=0.15]{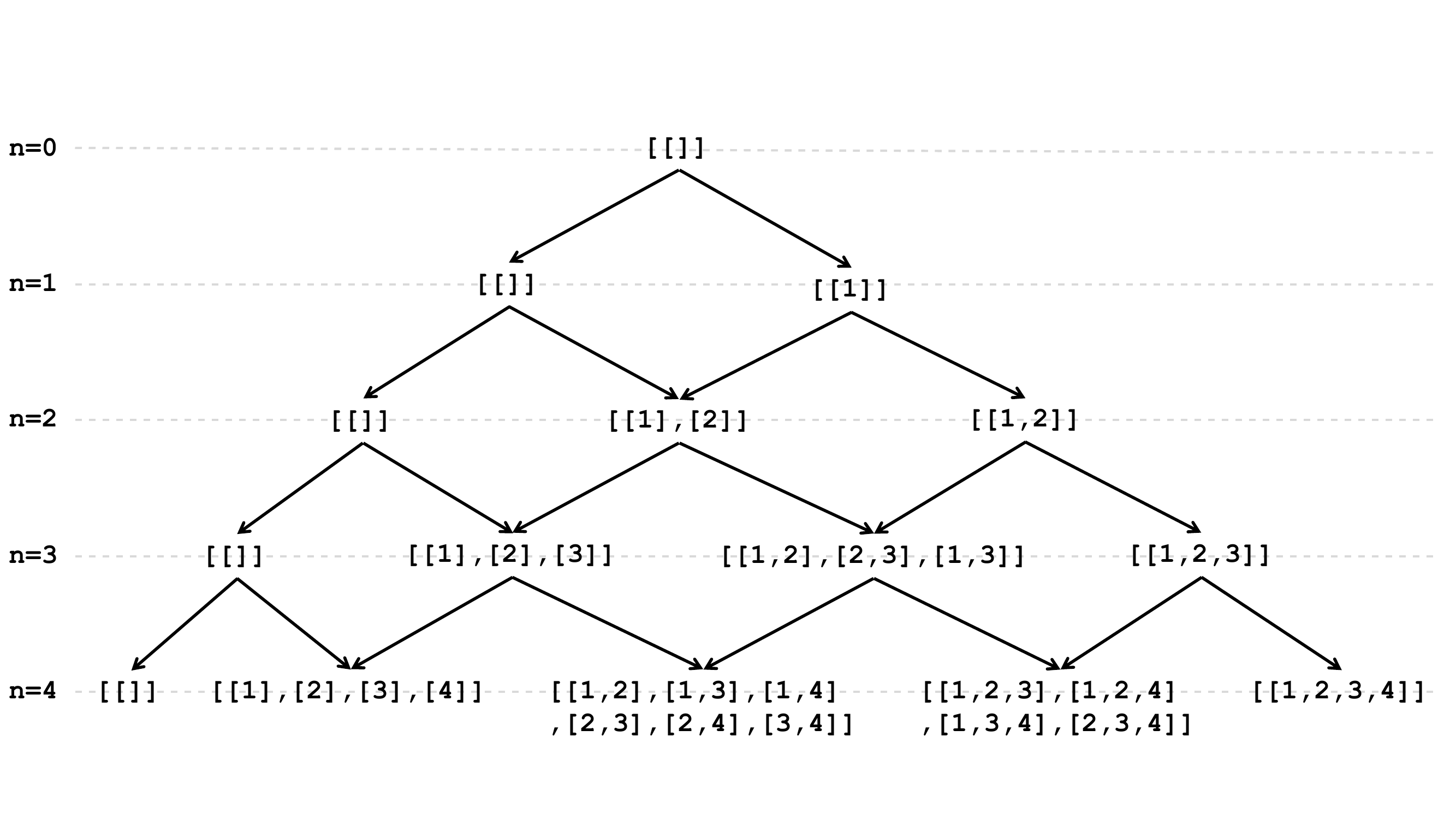}\caption{The sequential generation process of the classical combination generator
		(left) and the combination generator $\mathit{kcombs}$ proposed in
		this work (right) are illustrated for comparison. It is straightforward
		to observe that the generation process of $\mathit{kcombs}$ is structured
		more compactly by grouping combinations of the \textbf{\emph{same
				size}}\emph{ into a single list}, which can be viewed as a \textbf{matrix};
		consequently, the generator can be naturally implemented as\emph{
			a list of matrices}.This seemingly minor structural change yields
		substantial practical advantages in implementation. In particular,
		it simplifies memory management---since the size of each matrix can
		be determined via Vandermonde’s identity (\ref{eq: Vandermonde=002019s identity - arithmetic semiring})---and
		enables efficient parallelization on GPUs, as $\mathit{kcombs}$ relies
		solely on matrix multiplication.\label{fig: seq_gen classical kcombs}}
\end{figure}

\paragraph{Sequential version}

The \emph{sequential} (consume one data at a time) version of the
algorithm defined in (\ref{eq:kcombs-generator-D=000026C}) can be
derived straightforwardly by decomposing the input list sequentially.
Specifically, we have
\begin{equation}
	\begin{aligned}\mathit{kcombs}_{\text{seq}} & \left(K,\left[\right]\right)=\left[\left[\left[\;\right]\right],\left[\:\right]^{K}\right]\\
		\mathit{kcombs}_{\text{seq}} & \left(K,x:\mathit{xs}\right)=\mathit{convol}\left(\circ,\mathit{single}\left(x,K\right),\mathit{kcombs}_{\text{seq}}\left(K,\mathit{xs}\right)\right),
	\end{aligned}
\end{equation}
where $x:xs=\left[x\right]\cup xs$ and $\mathit{single}\left(x,K\right)=\left[\left[\left[\;\right]\right],\left[\left[x\right]\right],\left[\:\right]^{K-1}\right]$
for $K\geq1$ and $\mathit{single}\left(x,0\right)=\left[\left[\:\right]\right]$.
The convolution operator $\mathit{convol}\left(\cup,\circ,\mathit{xss},\mathit{yss}\right)$
though suitable for the D\&C version, is unnecessarily complex for
the sequential setting. Therefore, we can simplify the sequential
version of $\mathit{kcombs}$ by applying the following equational
reasoning
\begin{equation}
	\begin{aligned}\mathit{kcombs}_{\text{seq}}\left(K,x:xs\right)= & \:\mathit{convol}\left(\circ,\mathit{single}\left(x,K\right),\mathit{kcombs}_{\text{seq}}\left(K,xs\right)\right)\\
		=  & \text{  \{  definition of \ensuremath{convol} \} }\\
		& \left[\mathit{concat}\left(\mathit{zipWith}\left(\circ,\mathit{reverse}\left(\mathinner{css}\right),\mathit{yss}\right)\right)\mid\mathinner{css}\leftarrow\mathit{inits}\left(\mathit{single}\left(x,K\right)\right)\right]\\
		= & \text{ \{ definition of \ensuremath{inits} and let \ensuremath{\mathit{single}} \} }\\
		& \left[\mathit{concat}\left(\mathit{\mathit{zipWith}}\left(\circ,\mathit{reverse}\left(\mathit{single}\left(x,k\right)\right),\mathit{yss}\right)\right)\mid k\leftarrow\left[0,\ldots,K\right]\right]\\
		= & \text{ \{ definition of \ensuremath{\mathit{zipWith}} and \ensuremath{\mathit{concat}} \} }\\
		& \left[\left[\left[\:\right]\right]\right]\cup\left[\left[\left[x\right]\right]\circ\mathit{cs}_{k-1}\cup\left[\left[\:\right]\right]\circ\mathit{cs}_{k}\mid k\leftarrow\left[1,\ldots,K\right]\right]\\
		= & \text{ \text{ \{  let }\ensuremath{map\left(f,xs\right)=\left[f\left(x\right)\mid x\leftarrow xs\right]} \} }\\
		& \left[\left[\left[\:\right]\right]\right]\cup\left[\mathit{map}\left(x:,\mathit{cs}_{k-1}\right)\cup\mathit{cs}_{k}\mid k\leftarrow\left[1,\ldots,K\right]\right]
	\end{aligned}
\end{equation}
Using these, a much simpler recursive definition of the sequential
$\mathit{kcombs}$ function is given by
\begin{equation}
	\begin{aligned}\mathit{kcombs}_{\text{seq}} & \left(K,\left[\right]\right)=\left[\left[\left[\;\right]\right],\left[\right]^{K}\right]\\
		\mathit{kcombs}_{\text{seq}} & \left(K,x:xs\right)=\mathit{for}\left(x,\mathit{kcombs}_{\text{seq}}\left(K,xs\right)\right),
	\end{aligned}
	\label{eq: seq-kcombs}
\end{equation}
where $\mathit{for}\left(x,\mathit{css}_{K}^{n}\right)=\left[\left[\left[\:\right]\right]\right]\cup\left[\mathit{map}\left(x:,\mathit{cs}_{k-1}^{n}\right)\cup\mathit{cs}_{k}^{n}\mid k\leftarrow\left[1,\ldots,K\right]\right]$.
Note that when implementing the above recursion (\ref{eq: seq-kcombs})
in an imperative language, traversing the indices in decreasing order
$\left[K,K-1,\ldots,1\right]$ is more efficient than ascending order
$\left[1,\ldots,K-1,K\right]$, as it avoids unnecessary copying of
existing lists. Specifically, executing $\mathit{for}$ from $\left[1,\ldots,K\right]$
does not allow directly updating $\mathit{cs}_{i}^{n+1}$ to $\mathit{map}\left(x:,\mathit{cs}_{i-1}^{n}\right)\cup\mathit{cs}_{i}^{n}$,
because $\mathit{cs}_{i}$ will also be used when $k=i+1$. Consequently,
each recursive step requires copying an additional list. By contrast,
traversing in reverse order $\left[K,\ldots,1\right]$ avoids this
extra copying.

The sequential generation process of $\mathit{kcombs}$ is illustrated
in Figure \ref{fig: seq_gen classical kcombs}, the left panel illustrate
the generation process of the classical combination generator used
in \citet{bird1996algebra,jeuring1993theories,emoto2012filter}.

\paragraph{The revolving door ordering}

\begin{figure}
	\begin{centering}
		\includegraphics[scale=0.2]{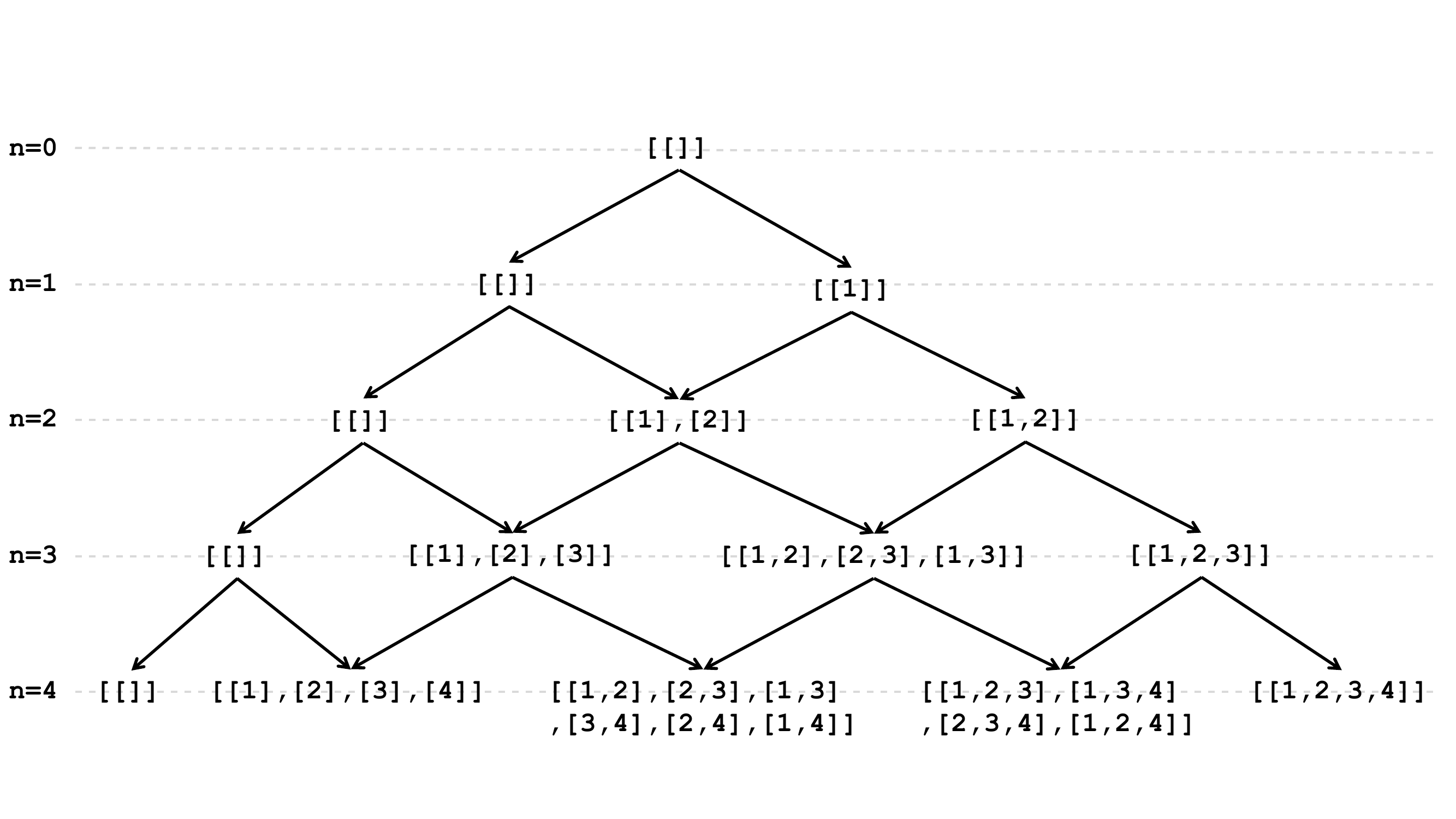}
		\par\end{centering}
	\caption{The sequential generation process of the combination generator $\mathit{kcombs}_{\text{revol}}$
		with combinations in revolving door ordering. The combinations generated
		by $\mathit{kcombs}_{\text{revol}}$ is different from $\mathit{kcombs}_{\text{seq}}$.
		Although both generate the same set of combinations, the ordering
		differs, as can be seen by comparing the combinations in this figure
		with those in the right panel of \ref{fig: seq_gen classical kcombs}.
		For example, $\mathit{kcombs}_{\text{seq}}\left(1,2,3,4\right)$ generate
		2-combinations in the order $\left[\left[1,2\right],\left[1,3\right],\left[1,4\right],\left[2,3\right],\left[2,4\right],\left[3,4\right]\right]$,
		while $\mathit{kcombs}_{\text{revol}}\left(4,3,2,1\right)$ generates
		them in the order $\left[\left[1,2\right],\left[2,3\right],\left[1,3\right],\left[3,4\right],\left[2,4\right],\left[1,4\right]\right]$.
		In the latter ordering, each pair of consecutive combinations differs
		by exactly two elements, including the first and last (i.e., their
		symmetric difference has size two), but $\left[1,2\right]$ and $\left[3,4\right]$
		has a symmetric difference 4. \label{fig: seq kcombs rev}}
\end{figure}

Assuming the input list is $\left[N,N-1,\ldots,1\right]$, by reorganizing
$\mathit{for}$, we can generate all $k$-combinations for $1\le k\leq K$
in revolving door order using following program

\begin{equation}
	\begin{aligned}\mathit{kcombs}_{\text{revol}} & \left(K,\left[\right]\right)=\left[\left[\left[\;\right]\right],\left[\right]^{K}\right]\\
		\mathit{kcombs}_{\text{revol}} & \left(K,\left[N,N-1,\ldots,1\right]\right)=\mathit{for}_{\text{revol}}\left(N,\mathit{kcombs}_{\text{revol}}\left(K,\left[N-1,N-2,\ldots,1\right]\right)\right)
	\end{aligned}
	\label{eq: seq-kcombs revol}
\end{equation}
where $\mathit{for}_{\text{revol}}\left(x,\mathit{css}_{K}^{n}\right)=\left[\left[\left[\:\right]\right]\right]\cup\left[\mathit{cs}_{k}^{n}\cup\mathit{reverse}\left(\mathit{map}\left(\cup\left[x\right],\mathit{cs}_{k-1}^{n}\right)\right)\mid k\leftarrow\left[1,\ldots,K\right]\right]$
and \\
$\cup\left[x\right]\left(xs\right)=xs\cup\left[x\right]$.
\begin{theorem}
	\emph{Equation (\ref{eq: seq-kcombs revol}) generates $\mathit{cs}_{k}^{N}$
		in revolving door order for all $1\le k\leq K$.}
\end{theorem}
\begin{proof}
	It is straightforward to verify that (\ref{eq: seq-kcombs revol})
	generates all valid combinations, as it merely reorders the list $\mathit{cs}_{k}^{n}$
	and $\mathit{map}\left(\cup\left[x\right]:,\mathit{cs}_{k-1}^{n}\right)$,
	the $\mathit{reverse}$ function only affects the ordering of elements,
	not the combinations themselves. To prove that (\ref{eq: seq-kcombs revol})
	satisfies the revolving door property, we proceed by induction. This
	property requires that every pair of adjacent combinations in $\mathit{cs}_{k}$,
	including the first and last, differ by exactly two elements (two
	lists with symmetric difference two). See Appendix \ref{Proof-of-Theorem}
	for detailed proof.
\end{proof}
The sequential generation process of $\mathit{kcombs}_{\text{revol}}$
is illustrated in (\ref{fig: seq kcombs rev}). Although this reorganization
of configurations may appear trivial or uninformative at first glance,
it is essential for certain applications that require the preservation
of revolving-door properties, such as position detection on a rotating
device; see \citet{ruskey2003combinatorial}, Chapter 5, for more
details. Moreover, as shown in Subsection (\ref{subsec:Incorporating-Gray-code}),
encoding such orderings is crucial for the construction of efficient
nested generators. In this context, the use of revolving-door algorithms
not only reduces memory consumption but also yields a perfect hashing
scheme for inner combinations without requiring any additional hashing
function.

\subsection{$K$-permutation generator}

\begin{figure}
	\begin{centering}
		\includegraphics[scale=0.2]{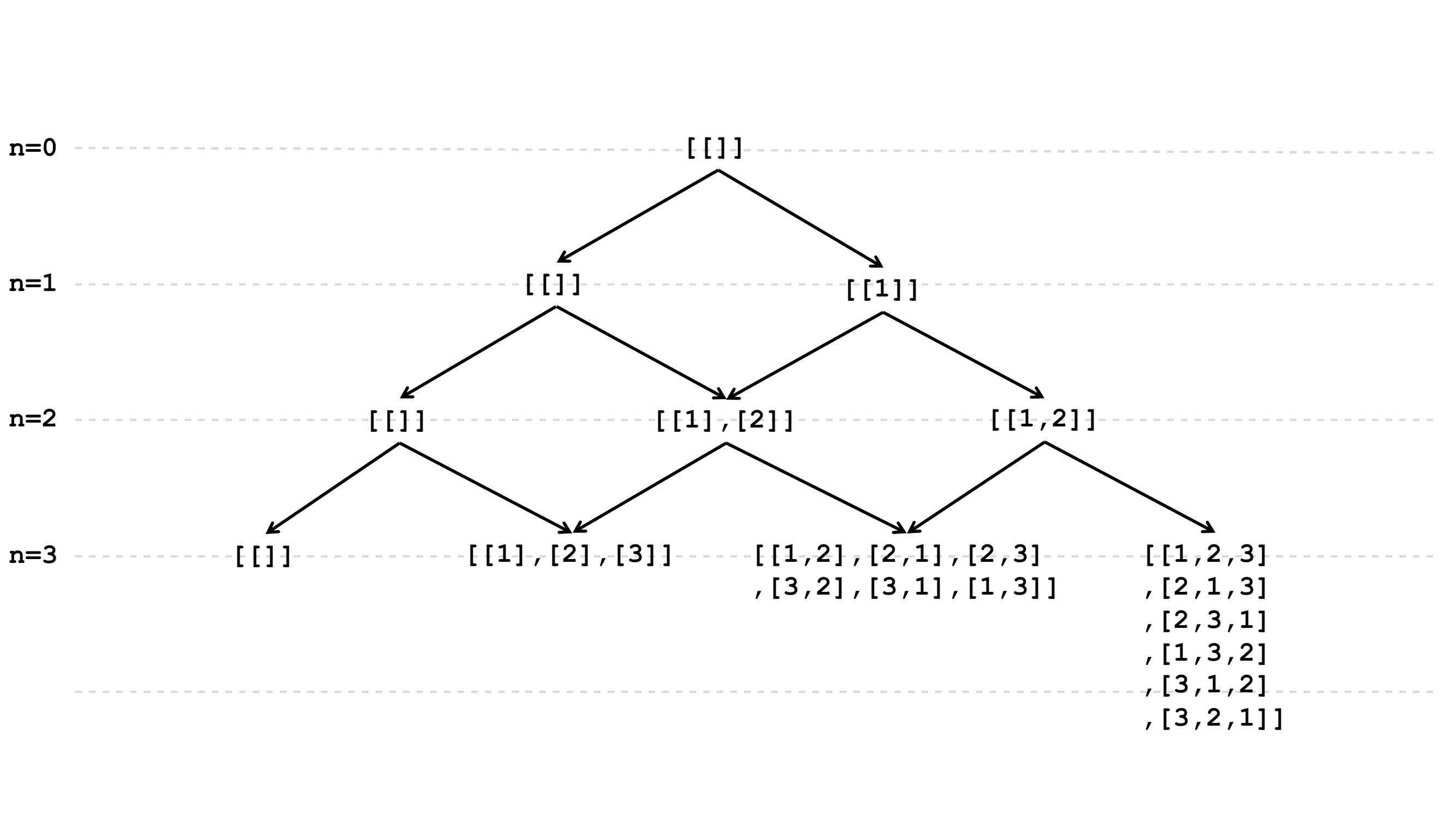}
		\par\end{centering}
	\caption{The sequential generation process of the $K$-permutation generator
		$\mathit{kperms}$.\label{fig: seq kperms}}
\end{figure}

$K$-permutations are essentially permutations of $K$-combinations.
A $K$-combination $x$ can be constructed by joining a $\left(K-k\right)$-combination
$y$ and a $k$-combination $z$, i.e., $x=y\cup z$. The $K$-permutations
$z$ can be defined by recursively inserting a $\left(K-k\right)$-combination
$x$ into all possible permutations of an $k$-combination $y$, in
other words, $\mathit{z}=\mathit{perms}\left(\mathit{x}\cup\mathit{y}\right)=\mathit{insert}\left(\mathit{x},\mathit{perms}\left(\mathit{y}\right)\right)$,
where $\mathit{perms}$ denote all possible permutations of a list.
The definition of $\mathit{insert}$ is defined as follows
\[
\begin{aligned}insert & :\left[a\right]\to\left[\left[a\right]\right]\to\left[\left[a\right]\right]\\
	insert & \left(\left[\:\right],\mathit{perms}\left(\mathit{y}\right)\right)=\mathit{perms}\left(\mathit{y}\right)\\
	insert & \left(a:\mathit{x},\mathit{perms}\left(\mathit{y}\right)\right)=insert\left(x,\mathit{concat}\left(\mathit{map}\left(f\left(a\right),\mathit{perms}\left(\mathit{y}\right)\right)\right)\right)
\end{aligned}
\]
where $f$ is defined as 
\[
\begin{aligned}f & :a\to\left[a\right]\to\left[a\right]\\
	f & \left(a,\left[\:\right]\right)=\left[\left[a\right]\right]\\
	f & \left(a,b:x\right)=\left(a:b:x\right):\mathit{map}\left(b:,\left(f\left(a,x\right)\right)\right)
\end{aligned}
\]
The $\mathit{insert}$ function is well established in the literature
\citep{bird1987introduction,bird1996algebra,bird2020algorithm}, which
sequentially inserting each element of $x$ into every possible position
within each permutation of $\mathit{perms}\left(y\right)$. However,
since the definition of $\mathit{insert}$ inherently sequential (insert
elements one-by-one), it is not suitable for designing D\&C algorithms
and therefore does not fit our construction.

Instead, \citet{jeuring1993theories} and \citet{bird2020algorithm}
previously proposed an elegant function for constructing D\&C \emph{permutation}
generators (not $K$-permutations). Rather than inserting elements
sequentially into partial permutations, one can \emph{interleave}
two partial combinations $y$ and $z$ to obtain all possible permutations
of $x$ \citep{bird2020algorithm}. Formally, we have 

\begin{align*}
	\mathit{perms}(x \cup y) &= \mathit{perms}(x) \merge \mathit{perms}(y) \\
	&= \mathit{concat}\Bigl(\bigl[\mathit{interleave}(a,b) \mid a \leftarrow \mathit{perms}(x),\ b \leftarrow \mathit{perms}(y)\bigr]\Bigr)
\end{align*}

where $\mathit{xs}\merge\mathit{ys}=\mathit{concat}\left[\mathit{interleave}\left(x,y\right)\mid x\leftarrow\mathit{xs},y\leftarrow\mathit{ys}\right]$.
The function $\mathit{interleave}:\left[\mathcal{X}\right]\times\left[\mathcal{X}\right]\to\left[\left[\mathcal{X}\right]\right]$
produces all lists that maintain the relative ordering of elements
in both input lists. It is defined as

\[
\begin{aligned}\mathit{interleave} & \left(\mathit{xs},\left[\:\right]\right)=\left[\mathit{xs}\right]\\
	\mathit{interleave} & \left(\left[\:\right],\mathit{ys}\right)=\left[\mathit{ys}\right]\\
	\mathit{interleave} & \left(\mathit{xs},\mathit{ys}\right)=\mathit{map}\left(x:,\mathit{interleave}\left(\mathit{xs},y:\mathit{ys}\right)\right)\cup\mathit{map}\left(y:,\mathit{interleave}\left(x:\mathit{xs},\mathit{ys}\right)\right).
\end{aligned}
\]
For instance,\\
 $\mathit{interleave}\left(\left[1,2\right],\left[3,4\right]\right)=\left[\left[1,2,3,4\right],\left[1,3,2,4\right],\left[1,3,4,2\right],\left[3,1,2,4\right],\left[3,1,4,2\right],\left[3,4,1,2\right]\right]$.
This structure defines a semiring $\left(\left[\left[\mathcal{X}\right]\right],\cup,\merge,\left[\;\right],\left[\left[\;\right]\right]\right)$
that satisfies the distributive law

\begin{align*}
	\left(\mathit{xs}\merge\mathit{ys}\right)\cup\left(\mathit{xs}\merge\mathit{zs}\right) 
	&= \mathit{concat}\Bigl[\mathit{interleave}\left(x,y\right)\mid x\leftarrow\mathit{xs},y\leftarrow\mathit{ys}\Bigr] \\
	&\quad \cup\; \mathit{concat}\Bigl[\mathit{interleave}\left(x,z\right)\mid x\leftarrow\mathit{xs},y\leftarrow\mathit{zs}\Bigr] \\
	&= \mathit{concat}\Bigl[\mathit{interleave}\left(x,y\right)\mid x\leftarrow\mathit{xs},y\leftarrow\mathit{ys}\cup\mathit{zs}\Bigr] \\
	&= \mathit{xs}\merge\left(\mathit{ys}\cup\mathit{zs}\right).
\end{align*}
Thus, we obtain the following recursion for generating $K$-permutations:

\begin{equation}
	\begin{aligned}\mathit{kperms} & \left(K,\left[\;\right]\right)=\left[\left[\left[\:\right]\right],\left[\:\right]^{K}\right]\\
		\mathit{kperms} & \left(K,\left[x\right]\right)=\left[\left[\left[\:\right]\right],\left[\left[x\right]\right],\left[\:\right]^{K-1}\right]\\
		\mathit{kperms} & \left(K,\mathit{xs}\cup\mathit{ys}\right)=\mathit{convol}\left(\merge,\mathit{kperms}\left(K,xs\right),\mathit{kperms}\left(K,ys\right)\right).
	\end{aligned}
	\label{eq:kperms-generator-D=000026C-}
\end{equation}
which generalizes the $\mathit{kcombs}$ generator by switching operator
$\circ$ as $\merge$.

The sequential generation process of $\mathit{kperms}$ is illustrated
in Figure (\ref{fig: seq kperms}).

\section{Nested generators\label{sec: section 5}}

In real-world applications, combinatorial objects with nested structures
are ubiquitous. For example, file systems and JSON documents can naturally
be understood as nested combinatorial objects.

Also, classical labeled binary trees can be interpreted as a nesting
of permutations with unlabeled trees \citet{flajolet2009analytic}.
Similarly, \citet{gibbons1991algebras} formalized a nested tree structure
in which each node itself represents a labeled subtree, and used this
formulation to derive accumulation functions over trees. 

However, exhaustive enumeration procedures for nested structures have
received comparatively little attention, despite their importance
across several domains. For instance, statistical analyses such as
analysis of variance \citet{anderson2003permutation} involve nested
combination-{}-permutation structures (i.e., combinations of permutations).
In optimization, models such as two-layer neural networks \citep{he2025deepice}
and optimal hyperplane decision trees involve combinations or permutations
of hyperplanes \citep{he2025FoODTI}, which can in turn be formulated
as nested combination/permutation problems.

Despite their importance, to the best of our knowledge, no prior work
has proposed algorithms for exhaustively enumerating nested combinations
or permutations beyond the trivial composition of two independent
generators. In this section, we investigate how techniques from functional
programming can be used to fuse such trivially composed combination/permutation
generators into a single recursive function that avoids materializing
large intermediate results. Furthermore, by incorporating Gray-code
ordering into the generation process, the fused generator naturally
supports efficient caching without extra overhead.

\subsection{Nested combination-combination generator}

\subsubsection{Divide-and-conquer version}

In this section, we introduce a new class of generator, called the
\emph{nested combination-combination generator} (NCCG). Denote $\text{\ensuremath{\left(K,D\right)}}$-NCCG
as a generator that produces all possible $K$-combinations of all
possible $D$-combinations with respect to an input list $\mathit{xs}$.
The $K$-combinations are referred to as outer combinations, denoted
$\mathit{ncss}:\mathit{NCss}$, while the $D$-combinations are referred
to as inner combinations, denoted $\mathit{css}:\mathit{Css}$.

A straightforward approach to constructing an NCCG involves first
generating all $d$-combination. Once this is complete, these intermediate
combinations are no longer needed and can be discarded by applying
the function $\mathit{setEmpty}\left(D,\mathit{xs}\right)$, which
sets the $D$th element of the list (implemented using array) $\mathit{xs}$
to an empty list (see Appendix \ref{subsec:Standard-functions-in Haskell}
for its definition). The result is then used as the input to the generator
$\mathit{kcombs}\left(K\right)$ to construct the $K$-combinations.
This approach can be specified compositionally as:

\begin{equation}
	\mathit{nestedCombs}\left(K,D\right)=\left\langle \mathit{setEmpty}\left(D\right),\mathit{kcombs}\left(K\right)\cdot!\left(D\right)\right\rangle \cdot\mathit{kcombs}\left(D\right),\label{nested combs specification}
\end{equation}
where $\left\langle f,g\right\rangle \left(a\right)=\left(f\left(a\right),g\left(a\right)\right)$.
This specification (\ref{nested combs specification}) has two key
limitations:
\begin{enumerate}
	\item \textbf{Intermediate storage overhead}: The approach requires storing
	all intermediate results produced by $\mathit{kcombs}\left(D\right)$
	which yields $O\left(N^{D}\right)$ combinations for an input of size
	$N$. This is both memory-intensive and computationally inefficient.
	\item \textbf{Incompatibility with recursive optimization methods}: In combinatorial
	optimization, the thinning method (also known as the dominance relation)
	is commonly used to prune infeasible or suboptimal partial configurations
	in branch-and-bound or dynamic programming algorithms. These methods
	crucially rely on recursive structures. However, (\ref{nested combs specification})
	is non-recursive, making it difficult to integrate thinning strategies.
\end{enumerate}
Instead, we propose fusing the function $\left\langle \mathit{setEmpty}\left(D\right),\mathit{kcombs}\left(K\right)\cdot\left(!\left(D\right)\right)\right\rangle $
with the recursive generator $\mathit{\mathit{kcombs}}\left(D\right)$,
thereby redefining the nested combination generator as a single recursive
process, then we can generate outer combinations without storing $O\left(N^{D}\right)$
combinations first. By fuse, we mean that given a function $f$ and
a recursive function $h$, if certain conditions are met, we can combine
them into a single recursive function $f\cdot h=g$. Formally, we
have the following theorem.
\begin{theorem}
	Fusion\emph{ }Theorem\emph{. Let $f$ be a function and let $h$ is
		a recursive function defined by the algebras $\mathit{alg}$, in the
		form
		\begin{equation}
			\begin{aligned}h & \left(\left[\;\right]\right)=alg_{1}\left(\left[\;\right]\right)\\
				h & \left(\left[a\right]\right)=alg_{2}\left(\left[a\right]\right)\\
				h & \left(x\cup y\right)=alg_{3}\left(h\left(x\right),h\left(y\right)\right).
			\end{aligned}
			\label{eq: join-list homomorphism}
		\end{equation}
		Let $\mathit{hom}\left(alg_{1},alg_{2},alg_{3}\right)$ be the unique
		solution in $h$ to the equation (\ref{eq: join-list homomorphism}),
		and $\mathit{hom}$ and $\mathit{hom}^{\prime}$ are short for $\mathit{hom}\left(alg_{1},alg_{2},alg_{3}\right)$
		and $\mathit{hom}\left(alg_{1}^{\prime},alg_{2}^{\prime},alg_{3}^{\prime}\right)$
		The fusion theorem states that $f\cdot\mathit{hom}=\mathit{hom}^{\prime}$
		if }$f\left(alg_{1}\left(\left[\;\right]\right)\right)=alg_{1}^{\prime}\left(\left[\;\right]\right)$
	\emph{and} $f\left(alg_{2}\left(\left[a\right]\right)\right)=alg_{2}^{\prime}\left(\left[a\right]\right)$\emph{
		and the recursive pattern satisfies}
	
	\emph{
		\begin{equation}
			f\left(alg_{3}\left(\mathit{hom}\left(x\right),\mathit{hom}\left(y\right)\right)\right)=alg_{3}^{\prime}\left(f\left(\mathit{hom}\left(x\right)\right),f\left(\mathit{hom}\left(y\right)\right)\right),\label{eq: join-list fusion condition}
		\end{equation}
		(\ref{eq: join-list fusion condition}) is known as the }fusion condition\emph{,
		in point-free style}\footnote{\emph{Point-free is a style of defining functions without explicitly
			mentioning their arguments.}}\emph{, (\ref{eq: join-list fusion condition}) can be expressed more
		succinctly as $f\cdot alg=alg^{\prime}\cdot\left(f\times f\right)$,
		where $f\times g\left(x,y\right)=\left(f\left(x\right),g\left(y\right)\right)$.\label{thm: join-list fusion law}}
\end{theorem}
\begin{proof}
	See appendix \ref{subsec:Proof-of-fusion-condition}.
\end{proof}
In our context, we want to construct a generator $\mathit{nestedCombs}$
defined by an algebra $\mathit{nestedCombsAlg}$ that satisfies the
following fusion condition
\begin{equation}
	f\cdot\mathit{\mathit{kcombsAlg}\left(d\right)}=\mathit{nestedCombsAlg}\left(k,d\right)\cdot f\times f\label{eq: nested-combs-fusion-condition}
\end{equation}
where $f=\left\langle \mathit{setEmpty}\left(D\right),\mathit{kcombs}\left(k\right)\cdot!\left(D\right)\right\rangle $,
and $\mathit{kcombsAlg}$ is defined as 
\begin{equation}
	\begin{aligned}\mathit{kcombsAlg}_{1} & \left(K,\left[\:\right]\right)=\left[\left[\left[\:\right]\right],\left[\:\right]^{K}\right]\\
		\mathit{kcombsAlg}_{2} & \left(K,\left[a\right]\right)=\mathit{single}\left(x,k\right)\\
		\mathit{kcombsAlg}_{3} & \left(K,x\cup y\right)=\mathit{convol}\left(\circ,\mathit{kcombs}\left(K,xs\right),\mathit{kcombs}\left(K,ys\right)\right).
	\end{aligned}
\end{equation}
In other words, we want the following diagram to commute:
\[
\xymatrix{\mathit{Css}\ar[d]_{f} &  &  &  & \left(\mathit{Css},\mathit{Css}\right)\ar[d]^{f\times f}\ar[llll]_{\mathit{\mathit{kcombsAlg}\left(D\right)}}\\
	\left(\mathit{Css},\mathit{NCss}\right) &  &  &  & \left(\left(\mathit{Css},\mathit{NCss}\right),\left(\mathit{Css},\mathit{NCss}\right)\right)\ar[llll]^{\mathit{nestedCombsAlg}\left(K,D\right)}
}
\]
The derivation of $\mathit{\mathit{nestedCombsAlg}}\left(K,D\right)$
for the empty and singleton cases is relatively straightforward. Since
we assume $D\geq2$, no outer combinations can be constructed in these
cases. We show that the fusion condition holds when this third pattern
of $\mathit{\mathit{nestedCombsAlg}}\left(K,D\right)$ is defined
as

\begin{equation}
	\begin{aligned} & \bigg<\mathit{setEmpty}\left(D_{l}\right)\cdot\mathit{KcombsAlg}\left(K\right)\cdot\mathit{Ffst},\\
		& \quad\quad\mathit{\mathit{\mathit{KcombsAlg}}\left(K\right)}\cdot\left\langle \mathit{Kcombs}\left(K\right)\cdot!\left(D\right)\cdot\mathit{KcombsAlg}\left(D\right)\cdot\mathit{Ffst},\mathit{KcombsAlg}\left(K\right)\cdot\mathit{Fsnd}\right\rangle \bigg>
	\end{aligned}
	\label{nested combs gen-abstract}
\end{equation}
where $\mathit{Ffst}\left(\left(a,b\right),\left(c,d\right)\right)=\left(a,c\right)$,
$\mathit{Fsnd}\left(\left(a,b\right),\left(c,d\right)\right)=\left(b,d\right)$.

The proof of equation (\ref{nested combs gen-abstract}) requires
expanding the commutative diagram into a more detailed diagram that
exposes the intermediate steps as follows:

\[
\xymatrix@C=1.2em@R=2.2em{\mathit{Css}\ar[d]^{\left\langle \mathit{SE}\left(D\right),!\left(D\right)\right\rangle } &  &  &  &  &  &  & \left(\mathit{Css},\mathit{Css}\right)\ar[d]_{\left\langle \mathit{SE}\left(D\right),!\left(D\right)\right\rangle \times\left\langle \mathit{SE}\left(D\right),!\left(D\right)\right\rangle }\ar[lllllll]^{\mathit{KCsA}\left(D\right)}\\
	\left(\mathit{Css},\mathit{Cs}\right)\ar[d]^{\mathit{id}\times\mathit{KCs}\left(K\right)} &  & \left(\mathit{Css},\left(\mathit{Cs},\mathit{Cs}\right)\right)\ar[ll]^{\mathit{SE}\left(D\right)\times\cup} &  &  &  &  & \left(\left(\mathit{Css},\mathit{Cs}\right),\left(\mathit{Css},\mathit{Cs}\right)\right)\ar[d]_{\left(\mathit{id}\times\mathit{KCs}\left(K\right)\right)\times\left(\mathit{id}\times\mathit{KCs}\left(K\right)\right)}\ar[lllll]^{\texttt{\texttt{\ensuremath{\left\langle \mathit{KCsA}\left(D\right)\cdot\mathit{Ffst},\left\langle !\left(D\right)\cdot\mathit{KCsA}\left(D\right)\cdot\mathit{Ffst},\cup\cdot\mathit{Fsnd}\right\rangle \right\rangle }}}}\\
	\left(\mathit{Css},\mathit{NCss}\right) &  & \left(\mathit{Css},\left(\mathit{NCss},\mathit{NCss}\right)\right)\ar[ll]^{\mathit{SE}\left(D\right)\times\mathit{KCsA}\left(K\right)} &  &  &  &  & \left(\left(\mathit{Css},\mathit{NCss}\right),\left(\mathit{Css},\mathit{NCss}\right)\right)\ar[lllll]^{\texttt{\ensuremath{\left\langle \mathit{KCsA}\left(D\right)\cdot\mathit{Ffst},\left\langle \mathit{KCs}\left(K\right)\cdot!\left(D\right)\cdot\mathit{KCsA}\left(D\right)\cdot\mathit{Ffst},\mathit{KCsA}\left(K\right)\cdot\mathit{Fsnd}\right\rangle \right\rangle }}}
}
\]

where $\cup\left(a,b\right)=a\cup b$, and $\mathit{\mathit{SE}}$,
$\mathit{KCs}$ and $\mathit{KCsA}$ are short for $\mathit{setEmpty}$,
$\mathit{Kcombs}$ and $\mathit{KcombsAlg}$. We then show that the
two paths between $\left(\mathit{Css},\mathit{Css}\right)$ and $\left(\mathit{Css},\mathit{Cs}\right)$,
as well as the two paths between $\left(\left(\mathit{Css},\mathit{Cs}\right),\left(\mathit{Css},\mathit{Cs}\right)\right)$
and $\left(\mathit{Css},\mathit{NCss}\right)$ are equivalent. We
leave the proof of fusion condition (\ref{nested combs gen-abstract})
in Appendix \ref{sec:Proofs}.

Given specification (\ref{nested combs gen-abstract}), an efficient
recursive program for $nestedCombs$ is defined as the following
\begin{equation}
	\begin{aligned}\mathit{nestedCombs} & \left(K,D,\left[\;\right]\right)=\left(\mathit{empty}\left(D\right),\left[\left[\left[\;\right]\right],\left[\:\right]^{K}\right]\right)\\
		\mathit{nestedCombs} & \left(K,D,\left[x\right]\right)=\left(\mathit{single}\left(x,D\right),\left[\left[\left[\;\right]\right],\left[\:\right]^{K}\right]\right)\\
		\mathit{nestedCombs} & \left(K,D,\mathit{xs}\cup\mathit{ys}\right)=\left(\mathit{setEmpty}\left(D,\mathit{css}\right),\mathit{ncss}\right),
	\end{aligned}
	\label{eq: nested combs--D=000026C}
\end{equation}
where $css=\mathit{kcombsAlg}\left(D,css_{1},css_{2}\right)$, such
that $\mathit{nestedCombs}\left(K,D,\mathit{xs}\right)=\left(\mathit{css}_{1},\mathit{ncss}_{1}\right)$
and $\mathit{nestedCombs}\left(K,D,\mathit{ys}\right)=\left(\mathit{css}_{2},\mathit{ncss}_{2}\right)$,
and $ncss$ is defined as 

\begin{equation}
	\mathit{ncss}=
	\begin{cases}
		\left[\left[\left[\;\right]\right],\left[\:\right]^{K}\right]
		& !\left(D,\mathit{css}\right)=\left[\;\right]
		\\[4pt]
		\begin{aligned}
			&\mathit{kcombsAlg}_{3}\Bigl(
			K,
			\mathit{kcombsAlg}_{3}(K,ncss_{1},ncss_{2}), \\
			&\qquad \qquad \qquad
			\mathit{kcombs}\bigl(K,!\left(D,\mathit{css}\right)\bigr)
			\Bigr)
		\end{aligned}
		& \text{otherwise}.
	\end{cases}
\end{equation}

See appendix \ref{subsec:Implementations-of-generators} for detailed
implementation in Haskell.

\subsubsection{Sequential version}

\begin{figure}
	\begin{centering}
		\includegraphics[scale=0.3]{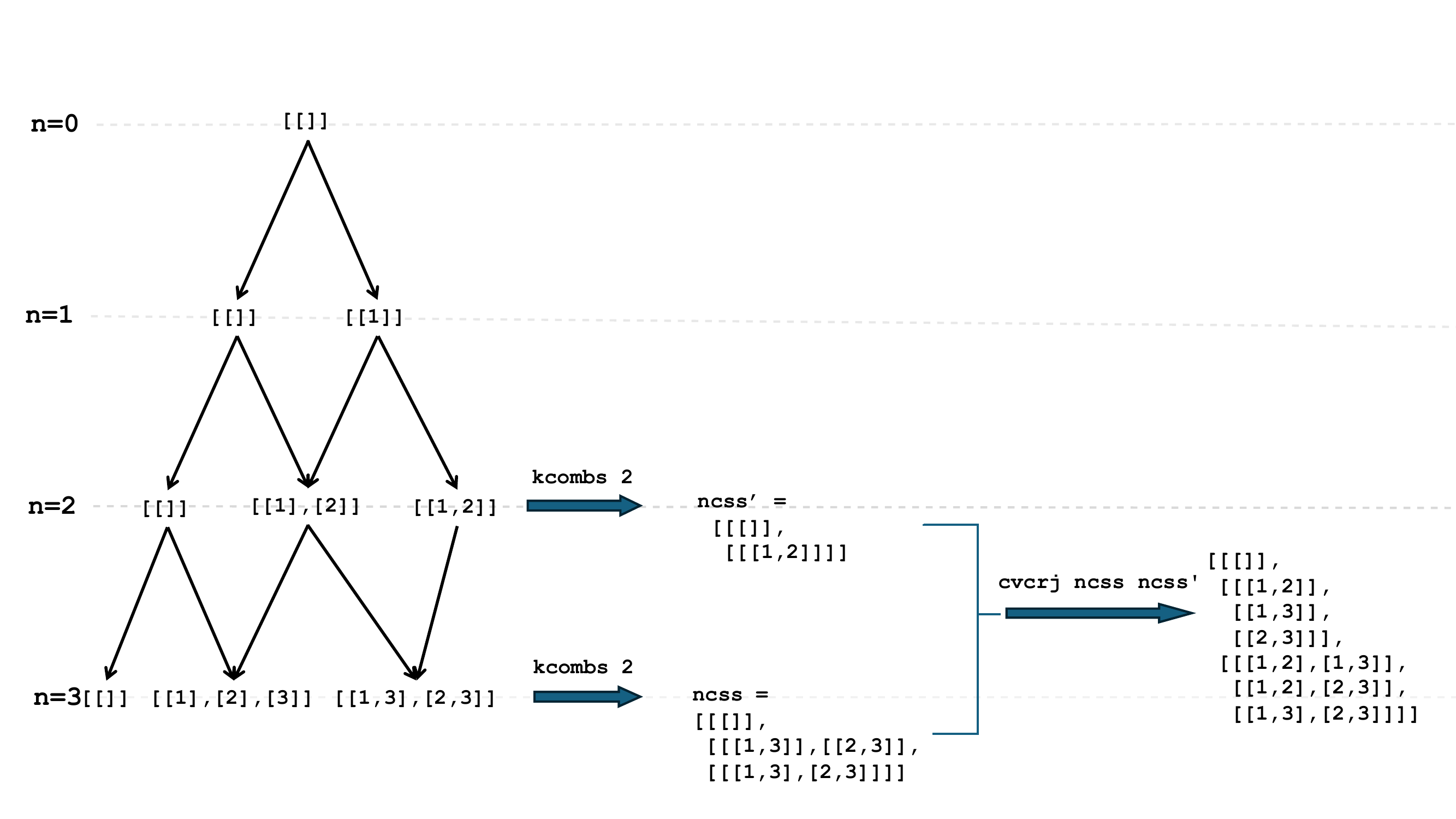}
		\par\end{centering}
	\caption{The sequential generation process of the ordinary nested combination-combination
		generator (NCCG). The previously generated nested combinations $\texttt{ncss'}$
		at stage 2 ($\texttt{n=2}$ level) and the newly generated nested
		combinations $\texttt{ncss}$ at stage 3 ($\texttt{n=2}$ level) are
		combined to form the overall nested combinations $\texttt{cvcrj ncss ncss'}$
		at stage 3. \label{fig: NCCGs}}
\end{figure}

Transforming the D\&C nested combination generator into a sequential
one is relatively straightforward. In sequential processing, a singleton
data element cannot construct any outer combinations; thus, $\mathit{nestedCombs}\left(K,D,\left[x\right]\right)=\left(\mathit{single}\left(x,D\right),\left[\left[\left[\;\right]\right],\left[\:\right]^{K}\right]\right)$,
which implies that $\mathit{ncss}_{1}=\left[\left[\left[\;\right]\right],\left[\:\right]^{K}\right]$.
Here, $\left[\left[\:\right]\right]$ and $\left[\:\right]$ are the
identity elements of the binary operations $\circ$ and $\cup$. It
follows that:\\
 $\mathit{kcombsAlg}\left(K,\left[\left[\left[\;\right]\right],\left[\:\right]^{K}\right],\mathit{ncss}^{\prime}\right)=\mathit{convol}\left(\cup,\circ,\left[\left[\left[\;\right]\right],\left[\:\right]^{K}\right],\mathit{ncss}^{\prime}\right)=\mathit{ncss}^{\prime}$.
Hence, the sequential version of $\mathit{nestedCombs}$ is defined
recursively as follows

\begin{equation}
	\begin{aligned}\mathit{nestedCombs} & \left(K,D,\left[\;\right]\right)=\left(\mathit{empty}\left(D\right),\left[\left[\left[\;\right]\right],\left[\:\right]^{K}\right]\right)\\
		\mathit{nestedCombs} & \left(K,D,x:\mathit{xs}\right)=\left(\mathit{setEmpty}\left(D,\mathit{css}\right),\mathit{ncss}\right),
	\end{aligned}
	\label{eq: nested combs --seq}
\end{equation}
where $css=\mathit{for}\left(x,\mathit{css}^{\prime}\right)$, such
that $\mathit{nestedCombs}\left(K,D,\mathit{xs}\right)=\left(\mathit{css}^{\prime},\mathit{ncss}^{\prime}\right)$
and $\mathit{ncss}$ is defined as

\begin{align*}
	\mathit{ncss} & =\begin{cases}
		\left[\left[\left[\;\right]\right],\left[\:\right]^{K}\right] & !\left(D,\mathit{css}\right)=\left[\;\right]\\
		\mathit{kcombsAlg}_{3}\left(K,\mathit{ncss}^{\prime},\mathit{kcombs}\left(K,!\left(D,\mathit{css}\right)\right)\right) & \text{otherwise}.
	\end{cases}
\end{align*}

The sequential generation process of $\mathit{nestedCombs}$ for input
$\left[1,2,3\right]$ are given in Figure (\ref{fig: NCCGs}).

\subsubsection{Incorporating Gray code ordering: encoding inner combinations as
	integers\label{subsec:Incorporating-Gray-code}}

\begin{figure}
	\begin{centering}
		\includegraphics[scale=0.15]{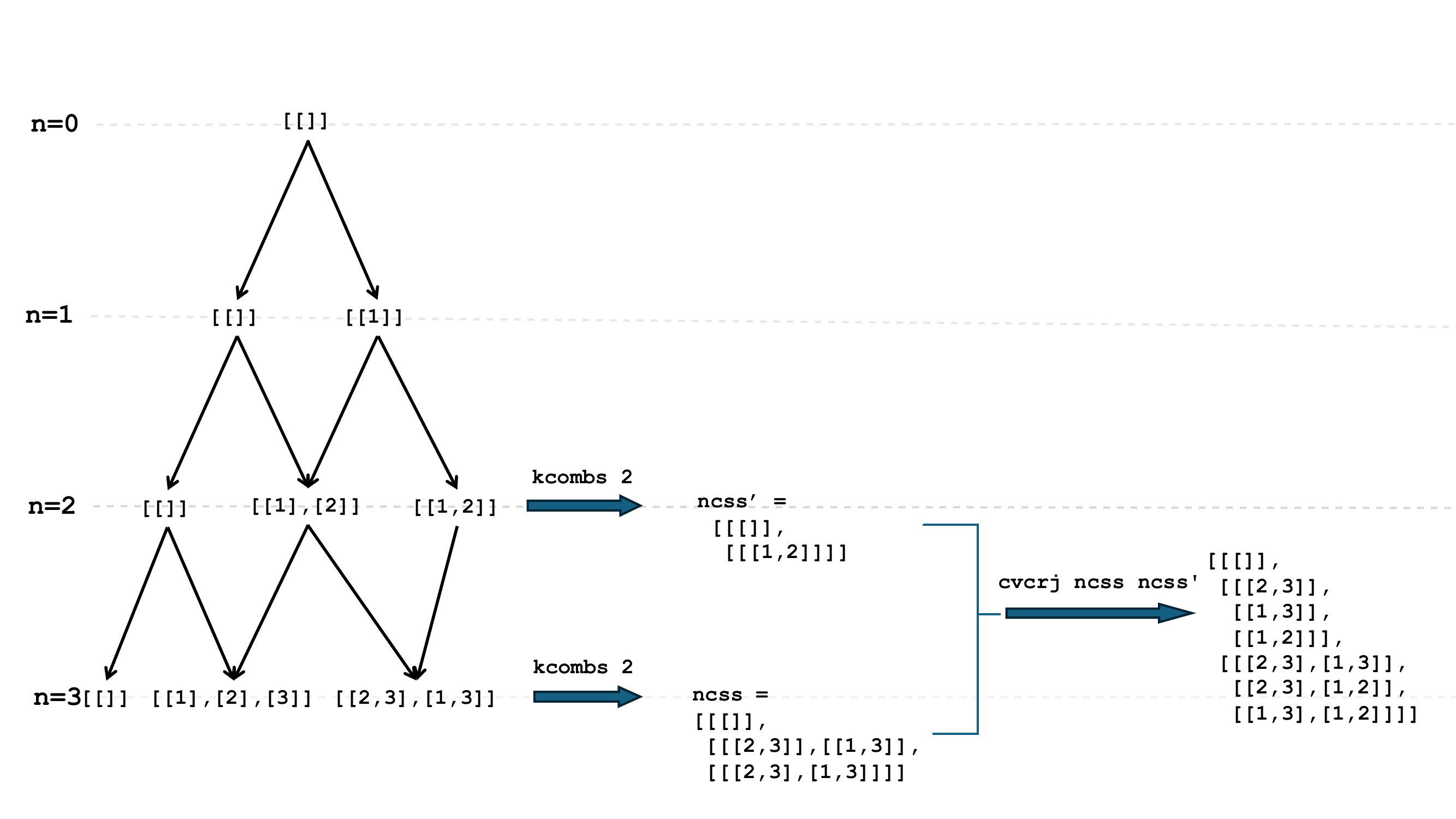}\includegraphics[scale=0.15]{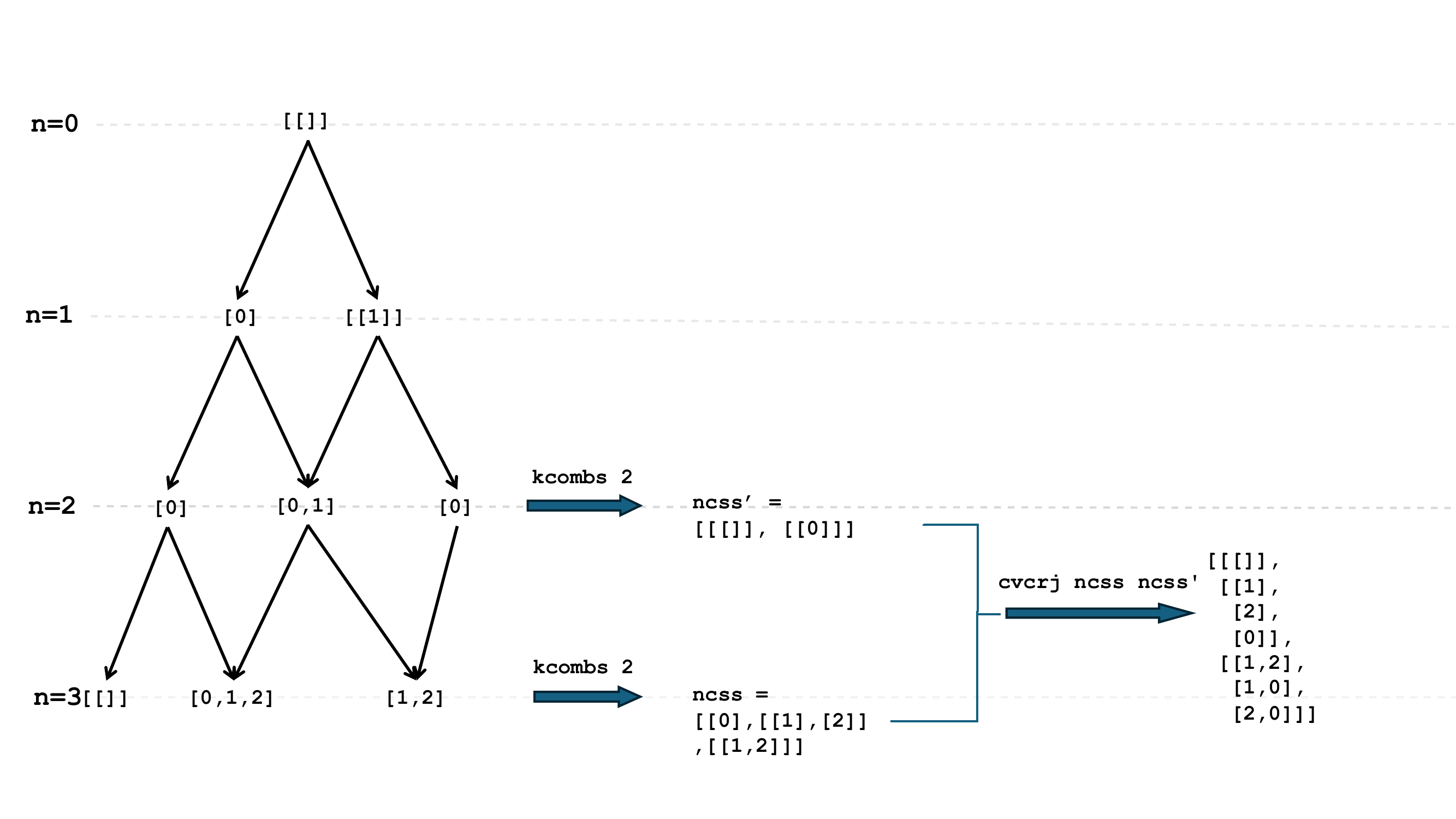}
		\par\end{centering}
	\caption{The sequential generation process of the ordinary nested combination-combination
		generator (NCCG) in revolving door ordering (left), and the NCGG by
		encoding the inner combination as integers (right). \label{fig: NCCGs-revol}}
\end{figure}

The sequential version of the nested combination generator can be
integrated with the revolving door algorithm introduced in (\ref{eq: seq-kcombs revol}).
By generating all inner combinations in revolving door order which
allows us to encode them as integers. This generalization is highly
practical due to two reasons:
\begin{enumerate}
	\item \textbf{Memory efficiency}: Encoding inner combinations as integers
	significantly reduces memory usage. Each combination can be represented
	by its rank, which requires at most $\log\left(C_{D}^{N}\right)$
	bits, compared to the $D\times64$ bits typically required when storing
	$D$-combination of 64-bit elements (e.g., floating-point values)
	or 64-bit pointer. This change can lead to substantial savings, especially
	when the number of combinations is large.
	\item \textbf{Efficient information indexing}: In many combinatorial optimization
	problems applications, the configurations represented by inner combinations
	are associated with auxiliary information. For instance, \citet{he2025deepice}
	encode hyperplanes as combinations, thus each combination is associated
	with the information about the prediction of each hyperplane. By encoding
	combinations using their ranks, this information can be precomputed
	and stored in a preallocated array indexed by the rank. As a result,
	lookups during the algorithm's execution can be performed in constant
	time, which not only improves efficiency but also reduces memory usage.
	This indeed can be seen as a perfect caching method for inner combinations
\end{enumerate}
Assume $C,I,M : \mathbb{N}$, and $\mathit{as} : \left[\mathbb{N}\right]$
is a list of integer. Define $\overrightarrow{I}=\left[0,\ldots,I-1\right]$,
$\overleftarrow{I}=\left[I-1,\ldots,0\right]$ and $M\pm\mathit{as}=\left[M\pm a\mid a\leftarrow\mathit{as}\right]$
(vectorized addition and subtraction). We have a property that $M-1-\overleftarrow{I}=C+\overrightarrow{I}$
if $C+I=M$ (consider $\overleftarrow{I}$ denote the distance to
$M-1$ for each element in $C+\overrightarrow{I}$). To transform
(\ref{eq: seq-kcombs revol}) to a integer version, we need to derive
a recursive pattern of exactly the same form in order to make sure
the combination in the previous stage is updated to the correct position
at the next stage, we derive this by following reasoning 
\begin{align*}
	\mathit{kcombs}_{\text{revolInt}}\left(K,N+1\right)= & \left[\overrightarrow{\mathit{C}_{k}^{N+1}}\mid k\leftarrow\left[0,\ldots,K\right]\right]\\
	= & \text{ $ \{ \mathit{C}_{k}^{N+1}=\mathit{C}_{k}^{N}+\mathit{C}_{k-1}^{N} \} $ }\\
	& \left[\left[0\right]\right]\cup\left[\overrightarrow{\mathit{C}_{k}^{N}+\mathit{C}_{k-1}^{N}}\mid k\leftarrow\left[1,\ldots,K\right]\right]\\
	= &  \text{ \{ property of \ensuremath{\overrightarrow{I}  }:}\\
	& \qquad \qquad \ensuremath{\ensuremath{\overrightarrow{I_{1}+\left(I-I_{1}\right)}}=\ensuremath{\overrightarrow{I}}_{1}+\mathit{map}\left(\left(I_{1}+\right),\overrightarrow{I-I_{1}}\right)}, s.t., \ensuremath{I_{1}\leq I-1} \} \\
	& \left[\left[0\right]\right]\cup\left[\overrightarrow{\mathit{C}_{k}^{N}}\cup\mathit{map}\left(\left(\mathit{C}_{k}^{N}+\right),\overrightarrow{\mathit{C}_{k-1}^{N}}\right)\mid k\leftarrow\left[1,\ldots,K\right]\right]\\
	= & \text{  \{ $ \mathit{C}_{k}^{N+1}-1-\overleftarrow{\mathit{C}_{k-1}^{N}}=\mathit{C}_{k}^{N}+\overrightarrow{\mathit{C}_{k-1}^{N}} $ \}  } \\
	& \left[\left[0\right]\right]\cup\left[\overrightarrow{\mathit{C}_{k}^{N}}\cup\mathit{map}\left(\left(\mathit{C}_{k}^{N+1}-1\right)-,\overleftarrow{\mathit{C}_{k-1}^{N}}\right)\mid k\leftarrow\left[1,\ldots,K\right]\right]\\
	= &  \text{ \{ definition of $ \ensuremath{\mathit{reverse}} $ \} } \\
	& \left[\left[0\right]\right]\cup\left[\overrightarrow{\mathit{C}_{k}^{N}}\cup\mathit{map}\left(\left(\mathit{C}_{k}^{N+1}-1\right)-,reverse\left(\overrightarrow{\mathit{C}_{k-1}^{N}}\right)\right)\mid k\leftarrow\left[1,\ldots,K\right]\right]\\
	= & \text{ \{ definition of \ensuremath{\mathit{for}_{\text{revolInt}}} \} }\\
	& \mathit{for}_{\text{revolInt}}\left(N+1,\mathit{kcombs}_{\text{revolInt}}\left(K,N\right)\right)
\end{align*}
thus we have 
\[
\begin{aligned}\mathit{kcombs}_{\text{revolInt}} & \left(K,0\right)=\left[\left[0\right],\left[\:\right]^{K}\right]\\
	\mathit{kcombs}_{\text{revolInt}} & \left(K,N\right)=\mathit{for}_{\text{revolInt}}\left(N,\mathit{kcombs}_{\text{revolInt}}\left(K,N-1\right)\right).
\end{aligned}
\]
By simply replacing the $\mathit{for}$ in (\ref{eq: nested combs --seq})
with $\mathit{for}_{\text{revolInt}}$, and modifying the base case
to $\mathit{nestedCombs}\left(K,D,\left[\;\right]\right)=\left(\left[\left[0\right],\left[\:\right]^{D}\right],\left[\left[\left[\;\right]\right],\left[\:\right]^{K}\right]\right)$,
we obtain an integer generator for the sequential $\left(K,D\right)$-NCCG,
where the inner combinations are integers representing combinations
in revolving door order.

The generation tree of the NCCG incorporating $\mathit{for}_{\text{revolInt}}$
is shown in the middle and right panels of (\ref{fig: NCCGs-revol}).
As illustrated, the inner combinations are represented as integers
rather than explicit combinations, which not only reduces memory consumption
but also provides a natural indexing scheme for the information associated
with each combination.

\subsection{Nested generator with multiple inner combinations}

In certain applications, multiple inner combinations may need to be
used simultaneously. For example, in the decision tree problem studied
in \citet{he2025ROF,he2025ODT}, one may construct a tree using different
types of splitting rules. This can be achieved by generating outer
combinations from multiple inner combinations together.

Given a list $\mathit{Ds}=\left[D_{1},D_{2},\ldots,D_{l}\right]$
in increasing order $D_{1},D_{2},\ldots,D_{l}$, we define the generator

\begin{equation}
	\mathit{nestedCombs}^{\prime}\left(K,\mathit{Ds}\right)=\left\langle \mathit{setEmpty}\left(D_{l}\right),\mathit{kcombs}\left(K\right)\cdot!!\left(\mathit{Ds}\right)\right\rangle \cdot\mathit{kcombs}\left(D_{l}\right),\label{nested combs-list specification}
\end{equation}
where $!!\left(\mathit{Ds},\mathit{xss}\right)=\mathit{concat}\left[!\left(D,\mathit{xss}\right)\mid D\leftarrow\mathit{Ds}\right]$.
The D\&C version of (\ref{nested combs-list specification}) can be
defined by modifying a single function in (\ref{nested combs gen-abstract})
as follows

\begin{equation}
	\begin{aligned} & \bigg<\mathit{setEmpty}\left(D_{l}\right)\cdot\mathit{KcombsAlg}\left(K\right)\cdot\mathit{Ffst},\\
		& \quad\quad\mathit{\mathit{\mathit{KcombsAlg}}\left(K\right)}\cdot\left\langle \mathit{Kcombs}\left(K\right)\cdot!!\left(\mathit{Ds}\right)\cdot convol_{\text{new}}\left(\circ\right)\cdot\mathit{Ffst},\mathit{KcombsAlg}\left(K\right)\cdot\mathit{Fsnd}\right\rangle \bigg>
	\end{aligned}
	\label{eq: nested list-combs gen}
\end{equation}
where the only differences between (\ref{eq: nested list-combs gen})
and (\ref{nested combs gen-abstract}) are functions $!!\left(\mathit{Ds}\right)$
which extracts the relevant $D$-combinations for all $D\in\mathit{Ds}$,
and $convol_{\text{new}}\left(\cup,\circ,\mathit{css}_{1},\mathit{css}_{2}\right)$
which generates only \emph{new} \emph{combinations} at each recursive
step


\[
\begin{aligned}
	convol_{\text{new}}\left(\circ,\mathit{css}_{1},\mathit{css}_{2}\right)
	={}&\left[
	\mathit{concat}\left(
	\mathit{zipWith}\left(
	\circ,
	\mathit{init}\left(\mathit{tail}\left(\mathinner{css}\right)\right),
	\right.\right.\right.
	\\
	&\left.\left.\left.
	\mathit{init}\left(\mathit{tail}\left(\mathit{css}_{2}\right)\right)
	\right)\right)
	\mid\mathinner{css}\leftarrow\mathit{inits}\left(\mathit{css}_{1}\right)
	\right]
\end{aligned}
\]

In each recursive step, only the newly generated inner combinations
need to be used to construct outer combinations, as the old inner
combinations have already been used. These newly generated inner combinations
correspond to the convolution of the ``middle'' elements of $\mathinner{css}$
and $\mathit{css}_{2}$---function $\mathit{init}\left(\mathit{tail}\left(x\right)\right)$
returns the middle elements of $x$.

The proof of correctness for (\ref{eq: nested list-combs gen}) follows
directly by adapting the proof of (\ref{nested combs gen-abstract})
requiring only a modification of one function. Due to limited space,
the D\&C definition of $\mathit{nestedCombs}^{\prime}$ is provided
in Appendix \ref{subsec:Implementations-of-generators}.

\subsection{Nested permutation-combination generator}

We define the generator for generating $K$-permutations of $D$-combinations
as

\begin{equation}
	\mathit{nestedPerms}\left(K,D\right)=\left\langle \mathit{setEmpty}\left(D\right),\mathit{kperms}\left(K\right)\cdot!\left(D\right)\right\rangle \cdot\mathit{kcombs}\left(D\right),\label{nested perms  specification}
\end{equation}
The nested permutation generator (NPCG) will be almost same as the
NCCG, the proof of it will be almost same as NCCG except the cross
product $\circ$ operator is replaced to $\merge$. The D\&C generator
of (\ref{nested perms  specification}) is then defined as

\begin{equation}
	\begin{aligned}\mathit{nestedPerms} & \left(K,D,\left[\;\right]\right)=\left(\mathit{empty}\left(D\right),\left[\left[\left[\;\right]\right],\left[\:\right]^{K}\right]\right)\\
		\mathit{nestedPerms} & \left(K,D,\left[x\right]\right)=\left(\mathit{single}\left(x,D\right),\left[\left[\left[\;\right]\right],\left[\:\right]^{K}\right]\right)\\
		nestedPerms & \left(K,D,\mathit{xs}\cup\mathit{ys}\right)=\left(\mathit{setEmpty}\left(D,\mathit{pss}\right),\mathit{npss}\right),
	\end{aligned}
	\label{eq: nested perms--D=000026C}
\end{equation}
where $\mathit{pss}=\mathit{kpermsAlg}\left(D,\mathit{pss}_{1},\mathit{pss}_{2}\right)$,
such that $\mathit{nestedPerms}\left(K,D,\mathit{xs}\right)=\left(\mathit{pss}_{1},\mathit{npss}_{1}\right)$
and $\mathit{netedPerms}\left(K,D,\mathit{ys}\right)=\left(\mathit{css}_{2},\mathit{ncss}_{2}\right)$,
and $ncss$ is defined as 

\begin{equation}
	\mathit{ncss}=
	\begin{cases}
		\left[\left[\left[\;\right]\right],\left[\:\right]^{K}\right]
		& !\left(D,\mathit{pss}\right)=\left[\;\right]
		\\
		\begin{aligned}
			&\mathit{kpermsAlg}_{3}\left(
			K,
			\mathit{kpermsAlg}_{3}\left(K,npss_{1},npss_{2}\right),
			\right.
			\\
			& \qquad  \qquad \qquad \left.
			\mathit{kperms}\left(K,!\left(D,\mathit{pss}\right)\right)
			\right)
		\end{aligned}
		& \text{otherwise}.
	\end{cases}
\end{equation}

where $\mathit{kpermsAlg}_{3}\left(K,\mathit{pss}_{1},\mathit{pss}_{2}\right)=\mathit{convol}\left(\cup,\merge,\mathit{pss}_{1},\mathit{pss}_{2}\right).$

\section{Applications\label{sec:Applications}}

\subsubsection*{$K$-medoids clustering and knapsack problem}

The knapsack problem is a canonical textbook example that is directly
related to combinations. It seeks to identify a subset of items from
a given set, where each item is associated with a value and a weight,
such that the total value of the selected subset is maximized while
the total weight does not exceed a given capacity.

As a classical example, the knapsack problem has been derived constructively
from specifications based on $K$-combination generators on numerous
occasions, including approaches using semiring algebra \citet{emoto2012filter,de1995generic,little2024polymor}
or catamorphism fusion \citet{de1995generic}.

The $K$-medoids problem is another problem that is directly related
to $K$-combination, which is a variant of the classical $K$-clustering
problem. Its objective is to select a subset of $K$ data points (medoids)
$\mathcal{U}=\left \{  u_{1},u_{2}\ldots u_{K}\right \}  $ from a dataset
$\mathcal{D}$ such that, where each medoids $u_{k}$ is associated
with a unique cluster $C_{k}$ that subsumes all data points closest
to it. The $K$-medoids problem aims to find an optimal $K$-combination
of data points $\mathcal{U}$ that minimizes the sum of squared distances
between each data point and its closest medoids. Consequently, a natural
specification of this problem is to exhaustively enumerate all possible
$K$-combinations of data points and then select the optimal one.

\subsubsection*{Cell enumeration}

\begin{figure}
	\begin{centering}
		\includegraphics[scale=0.15]{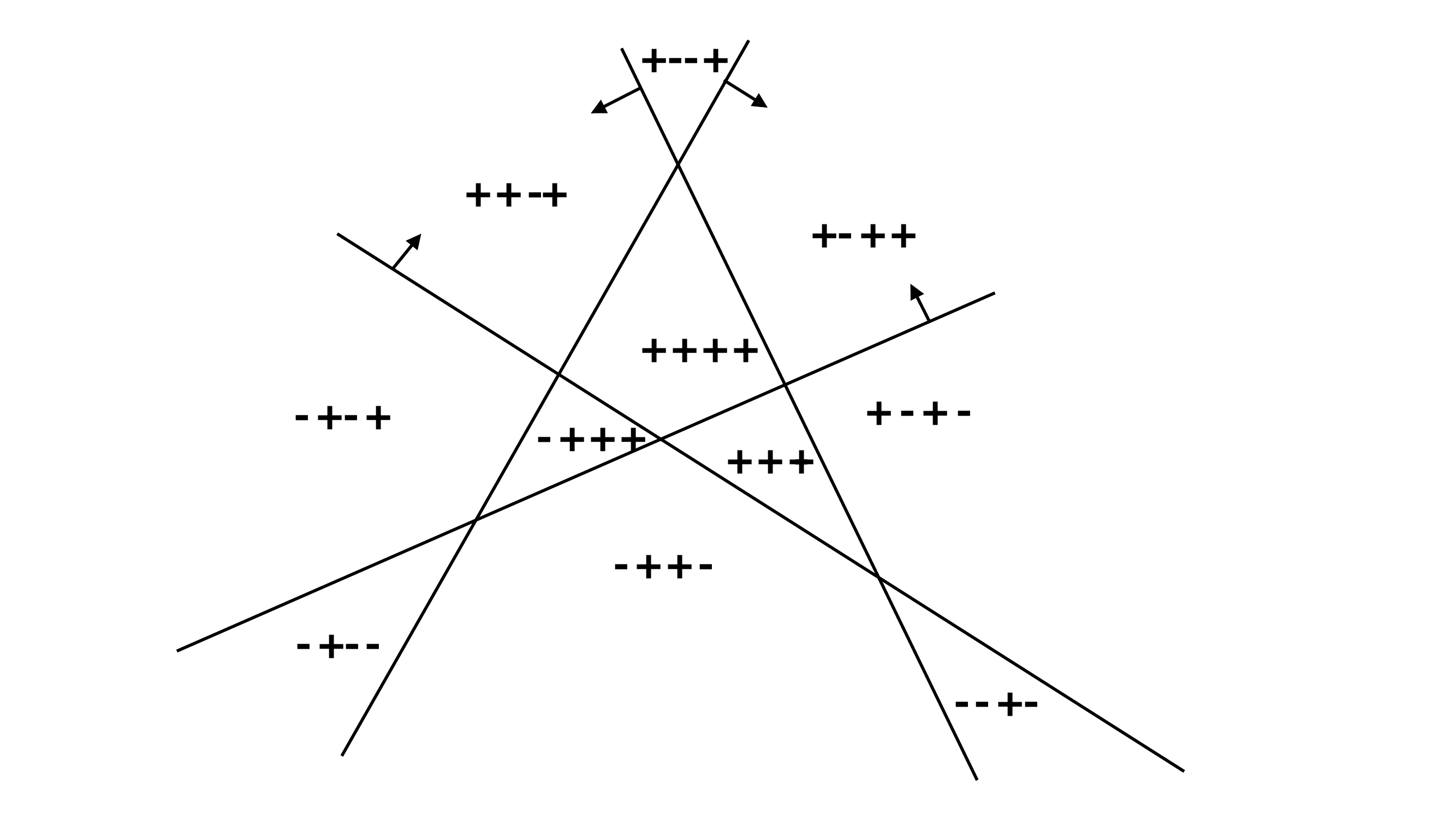}\includegraphics[scale=0.15]{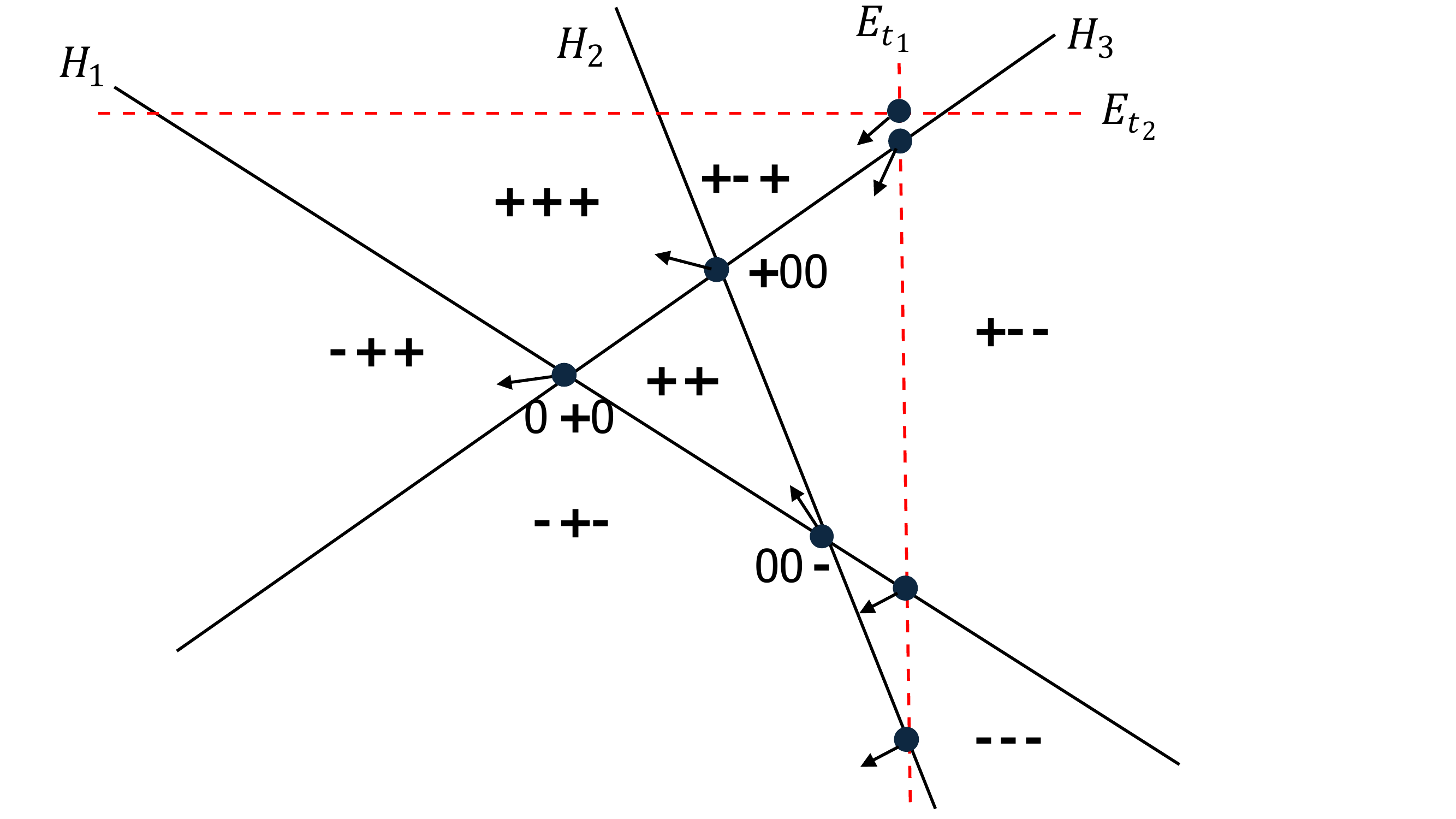}
		\par\end{centering}
	\caption{The left panel describes a hyperplane arrangement in $\mathbb{R}^{2}$
		consists of four hyperplanes with 11 cells. The right panel describes
		\citet{gerstner2006algorithms}'s approach for enumerating cells of
		an arrangement. \label{fig: arrangment and cell enumeration}}
\end{figure}

A dissection of $\mathbb{R}^{D}$ by a finite number of hyperplanes
is called a hyperplane arrangement, and \textbf{cells} are, informally,
the \emph{connected} regions into which space is divided by a collection
of hyperplanes. For instance, Figure \ref{fig: arrangment and cell enumeration}
illustrates an arrangement formed by four hyperplanes, which partitions
the space into 11 regions (\emph{cells}). The labels associated with
each region are referred to as \emph{sign vectors}. The objective
of the cell enumeration problem is to exhaustively enumerate all possible
sign vectors corresponding to the cells of the arrangement.

The classical approach to this problem is the reverse search algorithm
of \citet{avis1996reverse}, which is inherently an procedure that
starts at a random cell and explores neighboring cells by repeatedly
solving linear programs.

At first glance, this problem appears unrelated to combination generation.
Nevertheless, \citet{gerstner2006algorithms} showed that one can
construct a bounding box such that all vertices (i.e., intersection
points of hyperplanes) of the arrangement lie within it, see right
panel of Figure \ref{fig: arrangment and cell enumeration}. The box
is defined by $D$ axis-parallel hyperplanes in $\mathbb{R}^{D}$,
when combined with the \emph{original arrangement}, it yields a \emph{new
	arrangement.}

There is a one-to-one correspondence between the vertexes of the new
arrangement and the cells of the original arrangement. This forms
again a combination enumeration problem, since the vertexes of an
arrangement in $\mathbb{R}^{D}$ can be obtained by exhaustively enumerating
all \emph{$D$-combination of hyperplanes}.

As a result, by combining the construction of \citet{gerstner2006algorithms}
with the combination generators developed in this paper, it is possible
to obtain an algorithm that is potentially more efficient than the
reverse search algorithm of \citet{avis1996reverse}, since intersections
of hyperplanes can be computed via matrix inversion, which is typically
more efficient than repeatedly solving linear programs.

\subsubsection*{linear classification problem}

\begin{figure}
	\centering{}\includegraphics[viewport=100bp 200bp 1220bp 620bp,clip,scale=0.3]{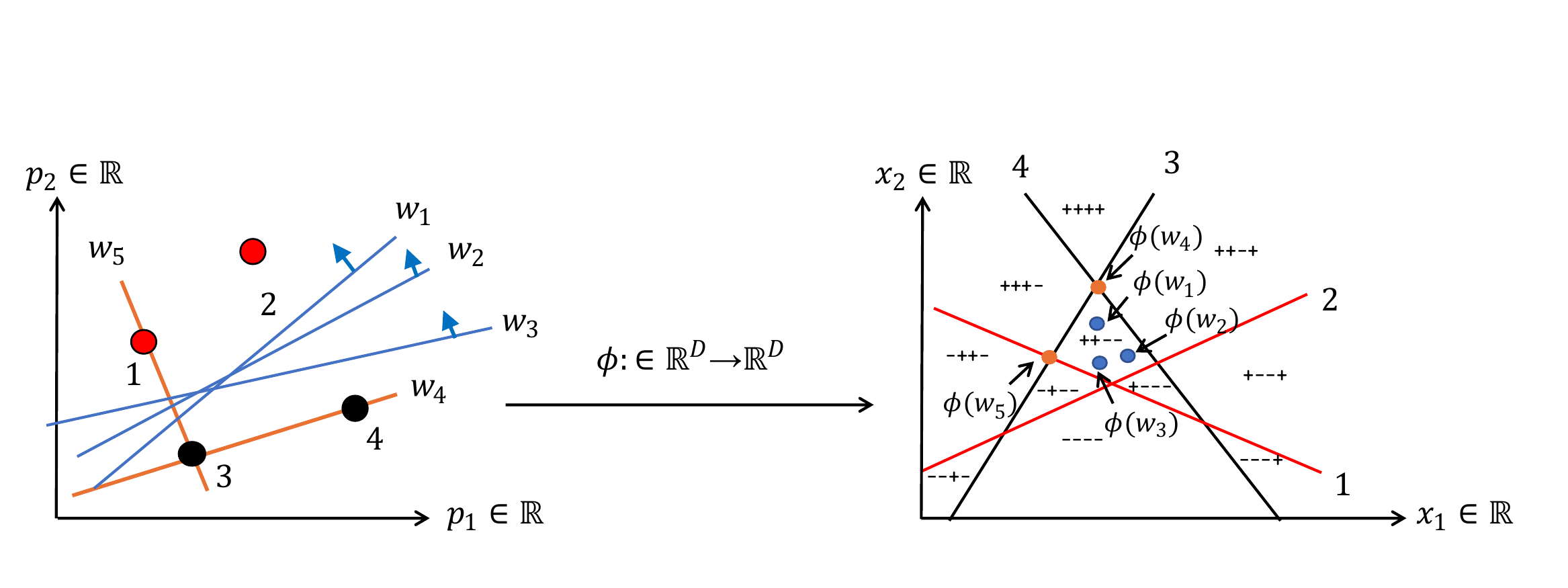}\caption{A point configuration $\mathcal{D}$ (left-panel) and its dual arrangement
		$\phi\left(\mathcal{D}\right)=\mathcal{H}_{\mathcal{D}}$ (right-panel).
		The yellow hyperplanes $w_{4}$, $w_{5}$ with two points lying on
		them in $\mathbb{R}^{D}$ correspond to the yellow points in the dual
		space, which are the intersection of corresponding dual hyperplanes
		$\phi\left(w_{4}\right)$, $\phi\left(w_{5}\right)$. For (blue) hyperplanes
		$w_{1}$, $w_{2}$, $w_{3}$ with the same prediction labels (+, +,
		-, -), their corresponding dual points $\phi\left(w_{1}\right)$,
		$\phi\left(w_{2}\right)$, $\phi\left(w_{3}\right)$. lie in the same
		cell of dual arrangement $\mathcal{H}_{\mathcal{D}}$. \label{fig: dual arrangement}}
\end{figure}

The linear classification problem seeks to identify a hyperplane that
separates a set of data points into two classes. Given that each data
point is associated with a label, the objective is to find a hyperplane
that minimizes the number of misclassifications.

\citet{he2023efficient} showed that a dataset can be transformed
into a hyperplane arrangement via a dual transformation (see Figure
\ref{fig: dual arrangement}), in which \emph{data points are mapped
	to hyperplanes} in the dual space, while \emph{hyperplanes are mapped
	to points}. In this dual representation, the linear classification
task amounts to identifying a cell whose sign vector is optimal with
respect to the true labels. This reformulation reduces the original
linear classification problem to a cell enumeration problem.

\citet{he2023efficient} further demonstrated that, when optimizing
the number of misclassifications, it suffices to enumerate the vertices
of the original arrangement, rather than those of the augmented arrangement
obtained by introducing an additional bounding box. Enumerating the
vertices of the arrangement in the dual space corresponds precisely
to enumerating all $D$-combinations of the original data points.

\subsubsection*{2-layer ReLU/Maxout neural network}

\begin{figure}
	\centering{}\includegraphics[scale=0.2]{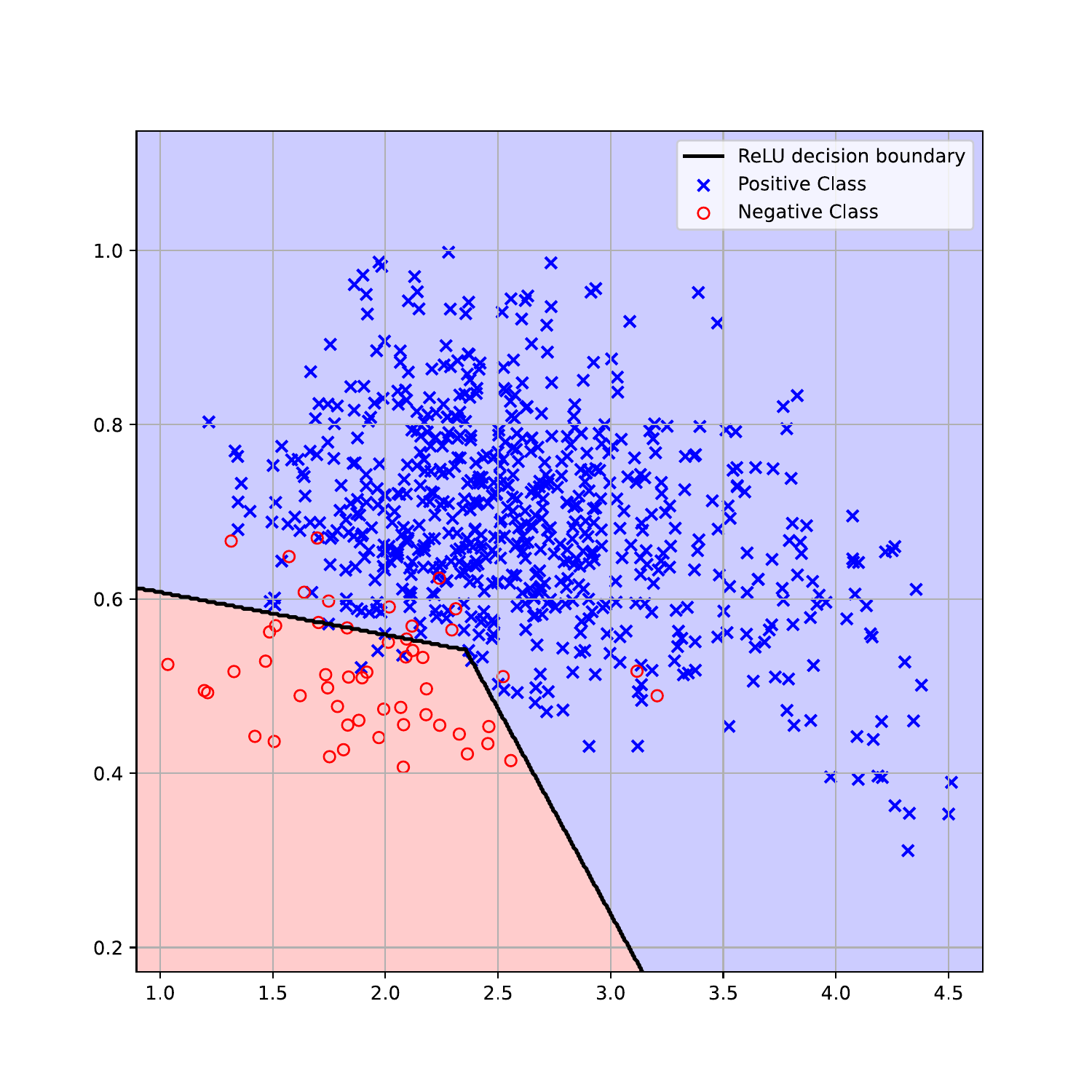}\caption{A rank-2 maxout network with one neuron on real-world datasets. The
		decision boundary is a 2-combination of hyperplanes.\label{fig:A-rank-2-maxout}}
\end{figure}

Neural networks have revolutionized nearly every scientific field
involving data analysis, and ReLU/maxout activation functions are
among the most widely used for constructing such networks. For networks
employing ReLU or maxout activations, \citet{he2025deepice} showed
that a two-layer neural network can be characterized as a \emph{combination
	of hyperplanes}. See Figure \ref{fig:A-rank-2-maxout} for illustration,
where a rank-2 maxout network is given, the decision boundaries consists
of $2$-combination of hyperplanes.

As discussed above, when optimizing the number of misclassifications,
a hyperplane itself can be characterized as a combination of data
points. Consequently, the optimal two-layer ReLU/maxout network naturally
gives rise to a problem of nested combinations. By exhaustively enumerating
all possible nested combinations, one can obtain the optimal two-layer
ReLU/maxout neural network. This addresses a long-standing problem
in machine learning, for which no publicly available solution existed
prior to the work of \citet{he2025deepice} which is based on the
enumeration of nested combinations.

\subsubsection*{Polynomial hypersurface decision tree}

\begin{figure}
	\begin{centering}
		\includegraphics[viewport=0bp 200bp 1280bp 620bp,clip,scale=0.2]{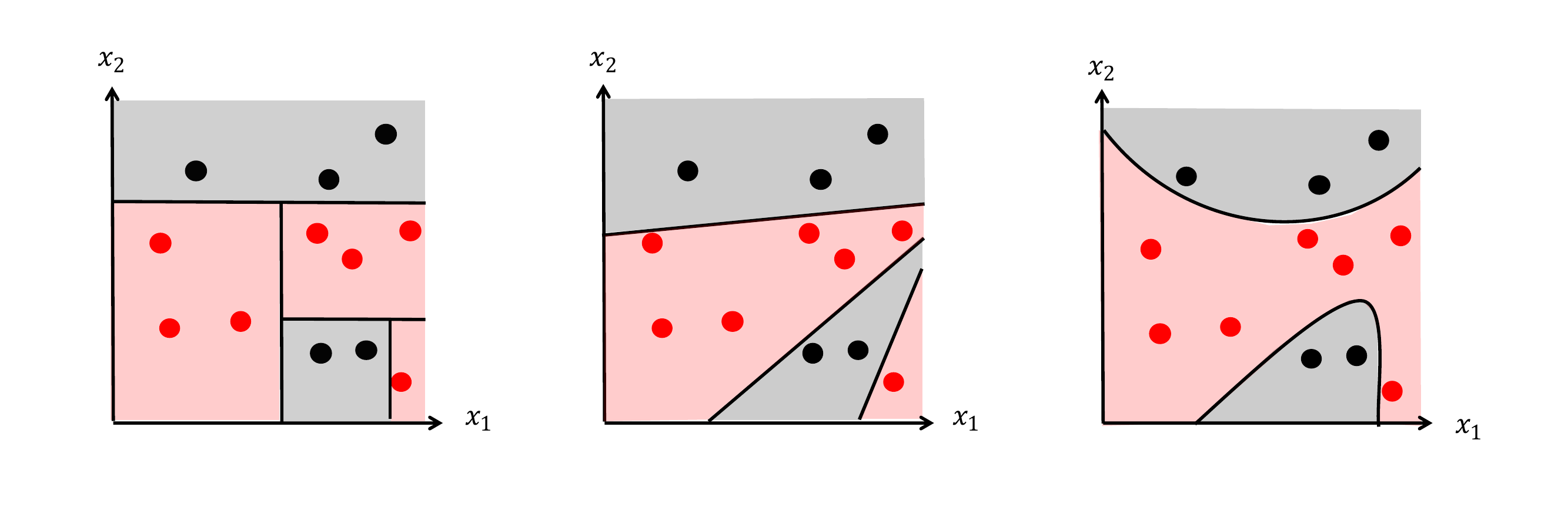}
		\par\end{centering}
	\caption{An axis-parallel decision tree model (left), a hyperplanes (oblique)
		decision tree model (middle), and a hypersurface (defined by degree-$2$
		polynomials) decision tree model (right).\label{fig: DT problems}}
\end{figure}

A decision tree is a supervised machine learning model that makes
predictions by recursively subdividing the feature space in a tree-structured
manner. It can be viewed as a flowchart or a sequence of binary (``yes''
or ``no'') questions that guide the input toward a final decision.
Geometrically, each decision rule at an internal node corresponds
to a split of the feature space into two regions based on the value
of a particular feature.

The left panel of Figure \ref{fig: DT problems} illustrates an axis-parallel
decision tree, in which the decision rules are defined by axis-parallel
hyperplanes. The middle and right panels present examples of more
expressive decision rules based on general hyperplanes and quadratic
(degree-two) hypersurfaces, respectively. These examples illustrate
that employing more complex decision rules can lead to substantially
simpler trees without sacrificing classification accuracy.

\citet{he2025ODT} showed that, when a decision tree satisfies a collection
of axioms---referred to as the proper decision tree axioms---a binary
decision tree can be transformed into a \emph{permutation of its decision
	rules }(nodes).

As a result, if a (axis-parallel) hyperplane decision tree satisfies
these axioms---which they do, as shown by \citet{he2025FoODTI}---then
the problem of constructing an optimal hyperplane decision tree reduces
to the problem of enumerating permutations of hyperplanes. Since hyperplanes
themselves can be characterized as combinations, this yields a problem
of permutations of combinations, which can be naturally formulated
using nested permutation generators.

Moreover, these results can be generalized to polynomial hypersurfaces.
\citet{he2023efficient} showed that classification by a polynomial
hypersurface is equivalent to classification by a hyperplane embedded
in a higher-dimensional feature space. Consequently, the polynomial
hypersurface classification problem also constitutes an instance of
the same combinatorial formulation.

\section{Experimental results\label{sec:Experimental-results}}

In this section, we evaluate the computational performance of our
combination generator. Our goal is to demonstrate its advantages from
the following perspectives: (a) the proposed generator is more efficient
than both list-based generators and classical lexicographic generation
algorithms \citep{kreher1999combinatorial}; (b) the proposed generator
is better suited for vectorization and parallelization, enabling more
efficient execution in downstream optimization tasks such as the $K$-medoids
problem; and (c) a simple brute-force algorithm built upon our generator
outperforms state-of-the-art mixed-integer programming (MIP) solvers.

We implement our generator in Python using the PyTorch library to
facilitate efficient vectorized computation. All experiments are conducted
on a machine equipped with an Intel Core i9 CPU, with 24 cores, 2.4-6
GHz, 32 GB RAM and GeForce RTX 4060 Ti GPU.

\subsection*{Comparison between different combinatorial generators}

\begin{figure}
	\begin{centering}
		\includegraphics[scale=0.15]{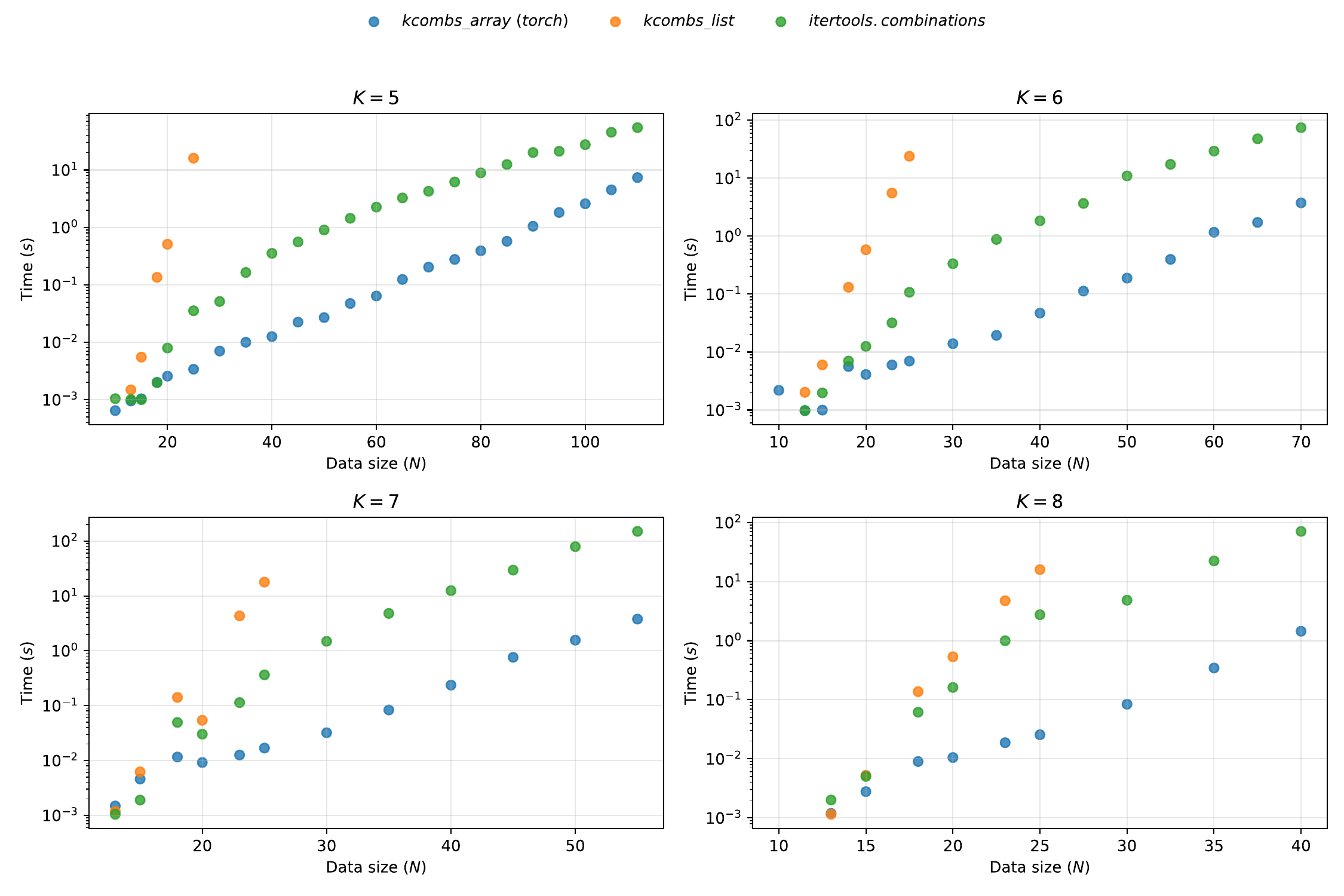}\includegraphics[scale=0.15]{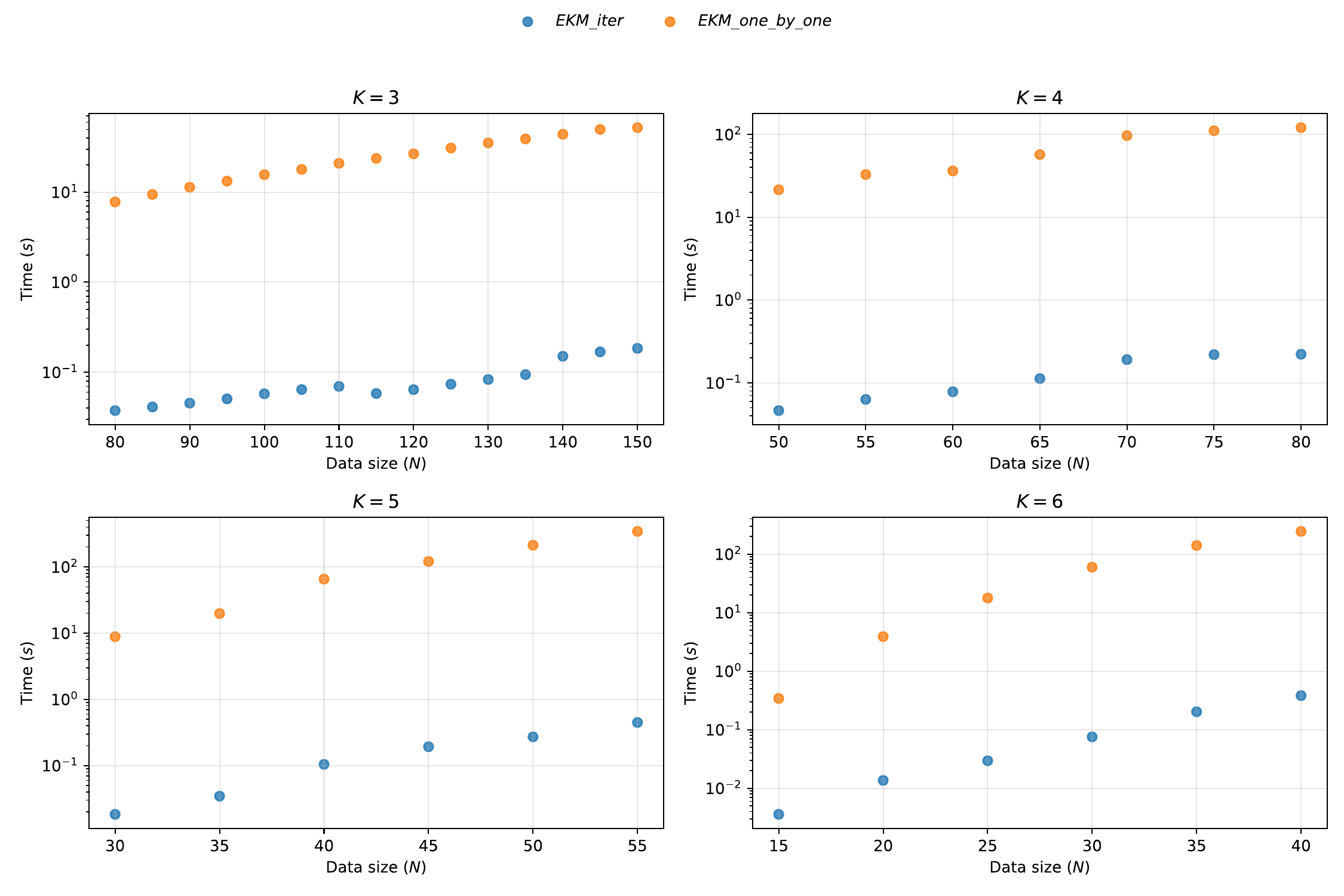}
		\par\end{centering}
	\caption{Running time comparison of the widely used Python $\mathtt{itertools}$
		library for combination generation ($\mathit{itertools.combinations}$), the classical list-based method ($\mathit{kcombs\_list}$),
		and our proposed generator ($\mathit{kcombs\_array}$) for the task of exhaustive combination
		generation (left-panel) and solving $K$-medoids problem (right-panel) . Results are reported
		on a log-linear scale. The left panels, arranged from left to right
		and top to bottom for $K=5$ to $K=8$ shows the reults of exhuastive
		generation tasks (the list-based method incurs substantial overhead
		in Python; consequently, the experiments are truncated once the memory
		limit of our machine is reached). The right panels report the performance
		on the $K$-medoids problem for $K=3$ to $K=6$. The results demonstrate
		that our Python implementation achieves up to 60-fold and 1000-fold
		speedups over the C-based $\mathtt{itertools}$ library on the CPU
		for exhaustive generation and $K$-medoids optimization, respectively.
		EKM\_iter and EKM\_one\_by\_one denotes the algorithm built upon our
		generator and $\mathtt{itertools}$ library for solving the $K$-medoids
		problem.\label{fig:Running-time-comparison}}
\end{figure}

\paragraph{Comparison on pure generation abilities}

To provide a more detailed evaluation, it is important to examine
the wall-clock runtime of different generators in practical implementations.
In this section, we compare our approach against the classical combination
generator in Python, namely the $\mathtt{itertools}$ package, which
is implemented in C and based on the lexicographic generation algorithm
\citep{kreher1999combinatorial} as well as the classical list-based
generator commonly used in the literature. We show that although these
methods share the same asymptotic complexity, their cache performance
differs substantially in practical implementations.

As shown in Figure \ref{fig:Running-time-comparison}, left-panel, our algorithm
consistently outperforms both the list-based approach and the C implementation
of the lexicographic generation algorithm. In particular, our method
achieves speedups of approximately 40-{}-60× over $\mathtt{itertools}$
and improvements of two to three orders of magnitude over the list-based
approach.

\paragraph{Comparison on the $K$-medoids problem}

To further demonstrate that the proposed array-based generator is
more suitable for vectorization and parallelization, and therefore
more amenable to efficient implementation in downstream applications,
we compare its performance against the generator provided by the Python
$\mathtt{itertools}$ library for solving the K-medoids problem (Figure \ref{fig:Running-time-comparison}, right-panel). We
omit comparisons with list-based methods due to their evident inefficiency.

Our generator achieves speedups of two to three orders of magnitude
over $\mathtt{itertools}$ when solving practical optimization problems.
These results indicate that our approach not only exhibits substantially
better cache performance, but also supports more effective vectorization.
Consequently, downstream optimization tasks can benefit from end-to-end
vectorized implementations, leading to significantly improved overall
efficiency.

\subsection*{Comparison with the state-of-art MIP solver for the $K$-medoids
	problem}

We test the performance of our EKM (short for ``exact K-meddoids'') algorithm, build on the exhaustive generator $\mathit{kcombs}$, against the approximate
algorithms partition around medoids (PAM), Faster-PAM and Clustering
Large Applications based on RANdomized Search (CLARANS)\footnote{We set the maximum number of neighbors examined as 4, and the number
	of iteration as 5.} on 18 datasets from the UCI Machine Learning Repository, two datasets
from\textcolor{blue}{{} }\citet{ren2022global} (UK, HCV), and two open-source
datasets (PR2392, HEMI) from \citep{wang2022predicting,padberg1991branch,ren2022global}.
We show that the simple brute-force algorithm is more efficient than
state-of-art MIP algorithm proposed by \citet{ren2022global}. 

The results are shown in Table \ref{tab:Empirical-comparison-on}, our EKM algorithm not only consistently outperforms the state-of-art MIP algorithm, but also outperforms approximate method (CLARANS) on many data sets.

Furthermore, we have some interesting observations regard the MIP algorithm for the $K$-medoids problem. No other algorithms can achieve better objective function values (see
Table \ref{tab:Empirical-comparison-on}), except in cases where \citet{ren2022global}'s
BnB algorithm returned incorrect solutions, which were clearly invalid
as they were several orders of magnitude lower than our exact solutions.

Our experiments included real-world datasets with a maximum size of
$N=5,000$. To the best of our knowledge, the largest dataset for
which an exact solution has been previously obtained is $N=150$,
as documented by \citet{ceselli2005branch} with $K=3$. Existing
literature on the $K$-medoids problem has only reported exact solutions
on very small datasets, primarily due to the use of BnB algorithms.
Given their unpredictable run-time and worst-case exponential time
complexity, most reported usage of BnB algorithms impose a hard computational
time limit to avoid memory overflow or intractable run times.

In summary, \citet{ren2022global}'s algorithm returned only \textbf{approximate}
solutions (with an optimality gap greater than zero), whereas our
brute-force algorithm consistently produced \textbf{provably} \textbf{exact} solutions
in \textbf{significantly} \textbf{less} \textbf{time}. Furthermore,
for challenging datasets, such as WDG, \citet{ren2022global}'s algorithm
produced a solution with an optimality gap of \textbf{800\%} even
after running for three hours! Moreover, in \emph{nearly} \emph{all}
datasets tested in our experiments but not included in \citet{ren2022global}
(e.g. IC, Yearst, WDG, wine, LD, VC, UKM and LM), their algorithm
produces obvious errorness solutions where upper bounds that were
\textbf{lower} \textbf{than} our \textbf{exact} solutions, which is
fundamentally incorrect as an upper bound cannot be lower than the
exact solution.

We believe the reason that \citet{ren2022global} claim their algorithm
can handle datasets with over a million instances is that they test
on datasets that are inherently easy to classify---so that even approximate
algorithms can obtain exact solutions. As we have demonstrated, almost
all the datasets they use can be solved exactly using PAM or Faster-PAM,
which achieve exact solutions with significantly fewer resources.
In contrast, \citet{ren2022global}'s algorithm requires an excessive
amount of computational power (6,000 CPU cores) compared to approximate
algorithms.

Moreover, we observed that \citet{ren2022global}'s algorithm exhibits
\textbf{exponential} complexity even when $K$ is fixed. This is evident
from the \textbf{non-polynomial} \textbf{growth} in running time for
experiments on datasets such as UK, BM, and Seeds. Although these
datasets have nearly identical sizes, their running time differs significantly
in \citet{ren2022global}'s algorithm. 

\begin{figure}[h]
	\begin{centering}
		\includegraphics[scale=0.25]{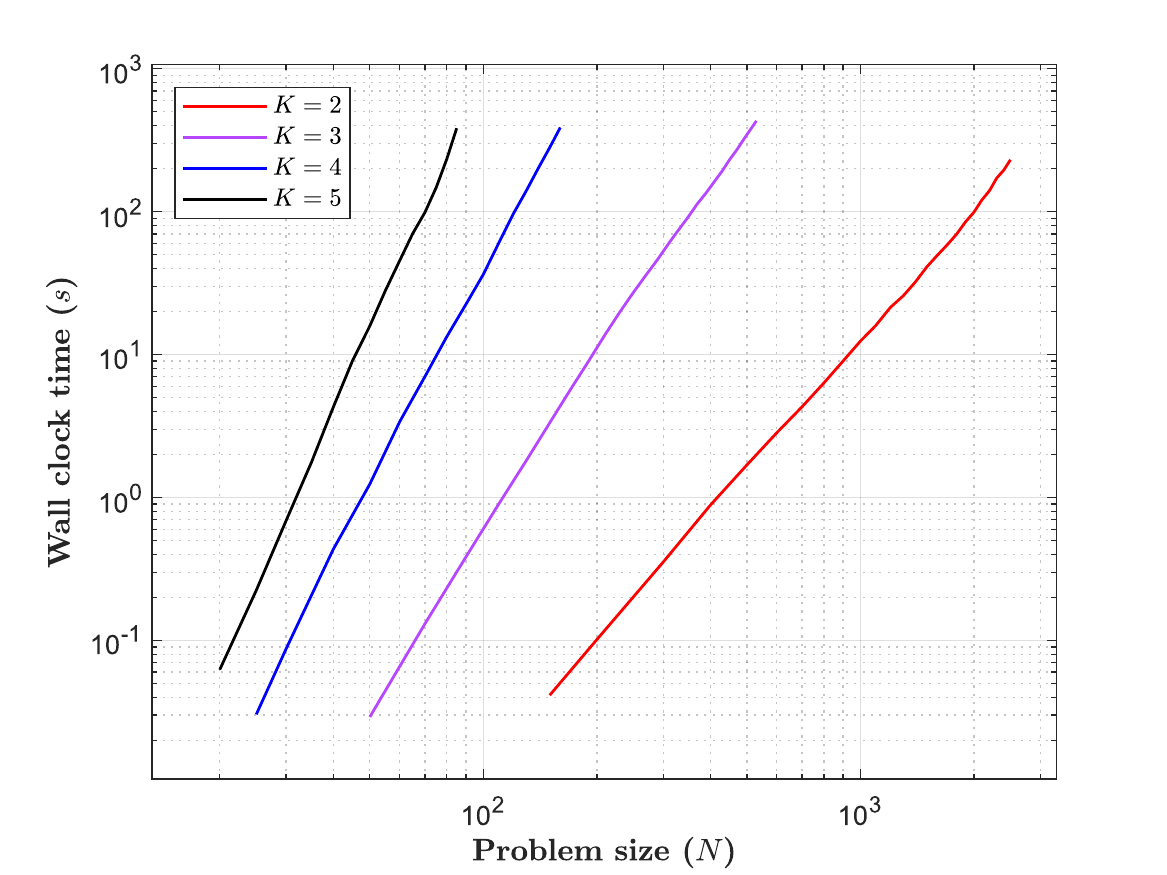}
		\includegraphics[scale=0.25]{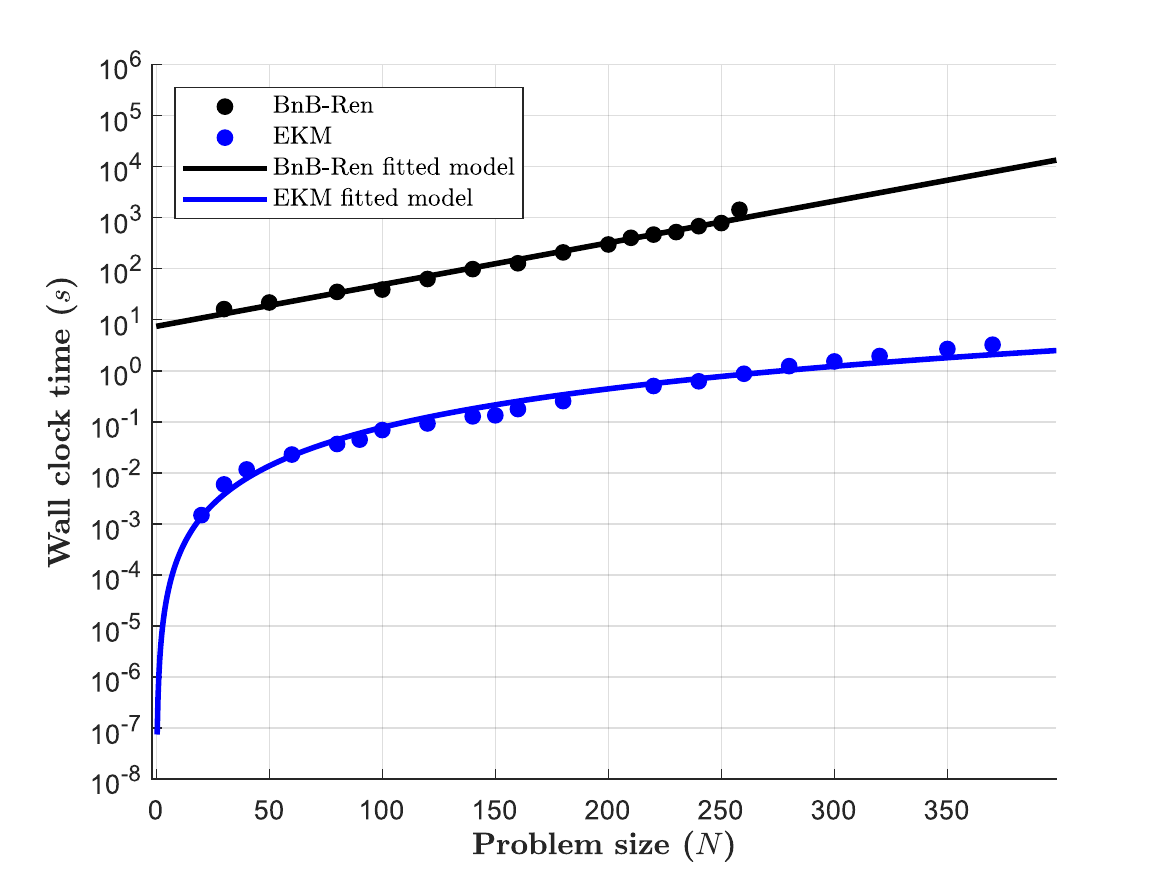}
		\par\end{centering}
	\caption{Log-log wall-clock run time (seconds) for our algorithm (EKM) tested
		on synthetic datasets (left panel). The run-time curves from left
		to right (corresponding to $K=2,3,4,5$ respectively), have slopes
		3.005, 4.006, 5.018, and 5.995, an excellent match to the predicted
		worst-case run-time complexity of $O\left(N^{3}\right)$, $O\left(N^{4}\right)$,
		$O\left(N^{5}\right)$, and $O\left(N^{6}\right)$ respectively. Log-linear
		wall-clock run-time (seconds) comparing EKM algorithm against \citet{ren2022global}'s
		algorithm by sampling UK dataset with $K=3$ (right panel). On this
		log-linear scale, exponential run-time appears as a linear function
		of problem size $N$, whereas polynomial run-time is a logarithmic
		function of $N$. \label{fig:log-log-wall-clock-run}}
\end{figure}

To verify this observation, we conduct simple experiments by sampling
the UK datasets with $K=3$. As predicted, \citet{ren2022global}'s
algorithm exhibits worst-case exponential time complexity even for
a fixed $K=3$ (Figure \ref{fig:log-log-wall-clock-run}, right panel),
whereas the simple brute-force algorithm builds on our combination
generator runs in polynomial time in the worst case.

\begin{table}
	\caption{Empirical comparison of the EKM algorithm,
		against widely-used approximate algorithms (PAM, Fast-PAM, and CLARANS)
		and the state-of-art exact BnB algorithm developed by \citet{ren2022global},
		for $K=3$, in terms of sum-of-squared errors ($E$), smaller is better.
		For \citet{ren2022global}'s aglorithm, we include both the \emph{upper}
		\emph{bound} and the \emph{optimal} \emph{gap.} The best-performing
		algorithm is highlighted in \textbf{bold}, while incorrect solutions are marked
		in \textcolor{red}{red}. Although some of the upper bounds returned by BnB algorithm are exact, these values are not marked in bold, as the optimal gap has not yet converged to zero. Wall clock time in brackets (seconds), we set a time limit $1.08\times10^{4}$ seconds for BnB algorithm.  \label{tab:Empirical-comparison-on}}
	\vskip 0.15in
	\begin{center}
		\scalebox{0.57}{
			\begin{small}
				\begin{sc}
					\begin{tabular}{>{\raggedright}p{0.1\textwidth}
							>{\centering}p{0.05\textwidth}
							>{\centering}p{0.01\textwidth}
							>{\raggedleft}p{0.15\textwidth} 
							>{\raggedleft}p{0.15\textwidth}
							>{\raggedleft}p{0.15\textwidth}
							>{\raggedleft}p{0.15\textwidth}
							>{\raggedleft}p{0.15\textwidth}}
						\toprule
						UCI dataset & $N$ & $D$ & EKM (ours) & Ren's BnB & PAM & Faster-PAM & CLARANS\tabularnewline
						\midrule
						
						LM & 338 & 3 & \textbf{$\boldsymbol{3.96\times10^{1}}$}
						
						($8.20\times10^{-1}$) & \textcolor{red}{$1.21\times10^{1}$}
						
						($2.31\times10^{1}$)
						
						$0$ & $3.99\times10^{1}$
						
						($4.02\times10^{-3}$) & $4.07\times10^{1}$
						
						($3.01\times10^{-3}$) & $5.33\times10^{1}$
						
						($6.14$)\tabularnewline
						\hline 
						UKM & 403 & 5 & \textbf{$\boldsymbol{8.36\times10^{1}}$}
						
						($1.37$) & \textcolor{red}{$5.51\times10^{1}$}
						
						($1.62\times10^{3}$)
						
						$\leq0.1\%$ & $8.44\times10^{1}$
						
						($8.57\times10^{-3}$) & $8.40\times10^{1}$
						
						($3.21\times10^{-3}$) & $1.16\times10^{2}$
						
						($4.98\times10^{1}$)\tabularnewline
						\hline 
						LD & 345 & 5 & $\boldsymbol{3.31\times10^{5}}$
						
						($8.3\times10^{-1}$) & \textcolor{red}{$1.21\times10^{1}$}
						
						($2.47\times10^{1}$)
						
						$\leq0.1\%$ & $3.56\times10^{5}$
						
						($4.11\times10^{-3}$) & \textbf{$\boldsymbol{3.31\times10^{5}}$}
						
						($3.87\times10^{-3}$) & $4.68\times10^{5}$
						
						($3.40$)\tabularnewline
						\hline 
						Energy & 768 & 8 & \textbf{$\boldsymbol{\boldsymbol{2.20}\times10^{6}}$}
						
						($1.37\times10^{1}$) & $2.20\times10^{6}$
						
						($1.68\times10^{1}$)
						
						$\leq0.1\%$ & $2.28\times10^{6}$
						
						($6.95\times10^{-3}$) & $2.28\times10^{6}$
						
						($3.94\times10^{-3}$) & $2.97\times10^{6}$
						
						($2.71$)\tabularnewline
						\hline 
						VC & 310 & 6 & \textbf{$\boldsymbol{3.13\times10^{5}}$}
						
						($6.82\times10^{-1}$) & \textcolor{red}{$1.50\times10^{5}$}
						
						($3.83\times10^{2}$)
						
						$\leq0.1\%$ & \textbf{$\boldsymbol{3.13\times10^{5}}$}
						
						($3.15\times10^{-3}$) & $3.58\times10^{5}$
						
						($5.36\times10^{-3}$) & $5.27\times10^{5}$
						
						($2.58$)\tabularnewline
						\hline 
						Wine & 178 & 13 & \textbf{$\boldsymbol{2.39\times10^{6}}$}
						
						($2.22\times10^{-1}$) & \textcolor{red}{$1.16\times10^{4}$}
						
						($5.17\times10^{1}$)
						
						$\leq0.1\%$ & \textbf{$\boldsymbol{2.39\times10^{6}}$}
						
						($1.06\times10^{-3}$) & $2.63\times10^{6}$
						
						($2.34\times10^{-3}$) & $6.86\times10^{6}$
						
						($5.56\times10^{-1}$)\tabularnewline
						\hline 
						Yeast & 1484 & 8 & \textbf{$\boldsymbol{8.37\times10^{1}}$}
						
						($1.74\times10^{2}$) & \textcolor{red}{$6.57\times10^{1}$}
						
						($1.08\times10^{4}$) $\leq39.19\%$ & $8.42\times10^{1}$
						
						($9.54\times10^{-2}$) & \textbf{$8.42\times10^{1}$}
						
						($6.08\times10^{-2}$) & \textbf{$1.05\times10^{2}$}
						
						($1.73\times10^{2}$)\tabularnewline
						\hline 
						IC & 3150 & 13 & \textbf{$\boldsymbol{6.9063\times10^{9}}$}
						
						($4.53\times10^{3}$) & \textcolor{red}{$6.18\times10^{9}$}
						
						($6.37\times10^{3}$)
						
						0 & \textbf{$6.9105\times10^{9}$}
						
						($8.68\times10^{-1}$) & \textbf{$\boldsymbol{6.9063\times10^{9}}$}
						
						($1.91\times10^{-1}$) & \textbf{$1.44\times10^{10}$}
						
						($2.70\times10^{1}$)\tabularnewline
						\hline 
						WDG & 5000 & 21 & \textbf{$\boldsymbol{1.67\times10^{5}}$}
						
						($5.23\times10^{4}$) & \textcolor{red}{$1.60\times10^{5}$}
						
						($1.08\times10^{4}$) $\leq785.05\%$ & \textbf{$\boldsymbol{1.67\times10^{5}}$}
						
						($1.34$) & \textbf{$\boldsymbol{1.67\times10^{5}}$}
						
						($1.97\times10^{-1}$) & \textbf{$2.77\times10^{5}$}
						
						($5.32\times10^{3}$)\tabularnewline
						\hline 
						IRIS & 150 & 4 & $\boldsymbol{8.40\times10^{1}}$
						
						($1.57\times10^{-1}$) & $8.46\times10^{1}$
						
						($2.51\times10^{1}$)
						
						$\leq27.1\%$ & $8.45\times10^{1}$
						
						($2.51\times10^{-3}$) & $8.45\times10^{1}$
						
						($1.03\times10^{-3}$) & $1.57\times10^{2}$
						
						($2.32\times10^{-1}$)\tabularnewline
						\hline 
						SEEDS & 210 & 7 & \textbf{$\boldsymbol{5.98\times10^{2}}$}
						
						($2.85\times10^{-1}$) & $5.98\times10^{2}$
						
						($2.42\times10^{1}$)
						
						$\leq0.1\%$ & \textbf{$\boldsymbol{5.98\times10^{2}}$}
						
						($1.14\times10^{-3}$) & \textbf{$\boldsymbol{5.98\times10^{2}}$}
						
						($3.59\times10^{-3}$) & $1.12\times10^{3}$
						
						($7.82\times10^{-1}$)\tabularnewline
						\hline 
						GLASS & 214 & 9 & \textbf{$\boldsymbol{6.29\times10^{2}}$}
						
						($2.90\times10^{-1}$) & $6.29\times10^{2}$
						
						($3.13\times10^{1}$)
						
						$\leq0.1\%$ & \textbf{$\boldsymbol{6.29\times10^{2}}$}
						
						($1.01\times10^{-3}$) & \textbf{$\boldsymbol{6.29\times10^{2}}$}
						
						($1.62\times10^{-3}$) & $1.04\times10^{3}$
						
						($2.27$)\tabularnewline
						\hline 
						BM & 249 & 6 & \textbf{$\boldsymbol{8.63\times10^{5}}$}
						
						($3.96\times10^{-1}$) & $8.63\times10^{5}$
						
						($1.19\times10^{2}$)
						
						$\leq0.1\%$ & $8.76\times10^{5}$
						
						($4.12\times10^{-3}$) & \textbf{$\boldsymbol{8.63\times10^{5}}$}
						
						($1.61\times10^{-3}$) & $1.33\times10^{6}$
						
						($1.02\times10^{1}$)\tabularnewline
						\hline 
						HF & 299 & 12 & \textbf{$\boldsymbol{7.83\times10^{11}}$}
						
						($6.26\times10^{-1}$) & \textbf{$\boldsymbol{7.83\times10^{11}}$}
						
						($5.20\times10^{1}$)
						
						$0\%$ & \textbf{$\boldsymbol{7.83\times10^{11}}$}
						
						($1.00\times10^{-3}$) & \textbf{$\boldsymbol{7.83\times10^{11}}$}
						
						($4.87\times10^{-3}$) & \textbf{$1.88\times10^{12}$}
						
						($6.55\times10^{-1}$)\tabularnewline
						\hline 
						WHO & 440 & 7 & $\boldsymbol{8.33\times10^{10}}$
						
						($1.98$) & $8.33\times10^{10}$
						
						($3.71\times10^{2}$)
						
						$\leq0.1\%$ & $\boldsymbol{8.33\times10^{10}}$
						
						($5.62\times10^{-3}$) & $\boldsymbol{8.33\times10^{10}}$
						
						($2.81\times10^{-3}$) & $1.21\times10^{11}$
						
						($8.12$)\tabularnewline
						\hline 
						UK & 258 & 5 & \textbf{$\boldsymbol{5.08\times10^{1}}$}
						
						($3.78\times10^{-1}$) & $5.08\times10^{1}$
						
						($1.43\times10^{3}$)
						
						$\leq0.1\%$ & $\boldsymbol{5.08\times10^{1}}$
						
						($2.47\times10^{-3}$) & $\boldsymbol{5.08\times10^{1}}$
						
						($2.47\times10^{-3}$) & $6.89\times10^{1}$
						
						($2.67\times10^{1}$)\tabularnewline
						\hline 
						HCV & 572 & 12 & $\boldsymbol{2.75\times10^{6}}$
						
						($5.48$) & $2.75\times10^{6}$
						
						($8.59\times10^{1}$)
						
						$\leq0.1\%$ & $\boldsymbol{2.75\times10^{6}}$
						
						($5.79\times10^{-3}$) & $\boldsymbol{2.75\times10^{6}}$
						
						($2.21\times10^{-3}$) & $4.75\times10^{6}$
						
						($6.89\times10^{1}$)\tabularnewline
						\hline 
						ABS & 740 & 19 & \textbf{$\boldsymbol{2.32\times10^{6}}$}
						
						($1.17\times10^{1}$) & $2.32\times10^{6}$
						
						($6.23\times10^{2}$)
						
						$\leq0.1\%$ & \textbf{$\boldsymbol{2.32\times10^{6}}$}
						
						($2.11\times10^{-2}$) & \textbf{$2.38\times10^{6}$}
						
						($5.00\times10^{-3}$) & $2.96\times10^{6}$
						
						($7.80\times10^{1}$)\tabularnewline
						\hline 
						TR & 980 & 10 & \textbf{$\boldsymbol{1.13\times10^{3}}$}
						
						($2.53\times10^{1}$) & $1.14\times10^{3}$
						
						($1.08\times10^{4}$)
						
						$\leq89\%$ & \textbf{$\boldsymbol{1.13\times10^{3}}$}
						
						($5.14\times10^{-2}$) & \textbf{$\boldsymbol{1.13\times10^{3}}$}
						
						($1.14\times10^{-2}$) & \textbf{$1.38\times10^{3}$}
						
						($2.59\times10^{2}$)\tabularnewline
						\hline 
						SGC & 1000 & 21 & $\boldsymbol{1.28\times10^{9}}$
						
						($3.87\times10^{1}$) & $1.28\times10^{9}$
						
						($1.75\times10^{2}$)
						
						$\leq0.1\%$ & $\boldsymbol{1.28\times10^{9}}$
						
						($1.71\times10^{-1}$) & $\boldsymbol{1.28\times10^{9}}$
						
						($4.22\times10^{-2}$) & $2.52\times10^{9}$
						
						($2.24$)\tabularnewline
						\hline 
						HEMI & 1995 & 7 & $\boldsymbol{9.91\times10^{6}}$
						
						($9.00\times10^{2}$) & $9.91\times10^{6}$
						
						($3.92\times10^{2}$)
						
						$\leq0.1\%$ & $\boldsymbol{9.91\times10^{6}}$
						
						($3.64\times10^{-1}$) & $\boldsymbol{9.91\times10^{6}}$
						
						($6.99\times10^{-2}$) & $1.66\times10^{7}$
						
						($9.53$)\tabularnewline
						\hline 
						PR2392 & 2392 & 2 & \textbf{$\boldsymbol{2.13\times10^{10}}$}
						
						($1.29\times10^{3}$) & $2.13\times10^{10}$
						
						($1.54\times10^{3}$)$\leq0.1\%$ & \textbf{$\boldsymbol{2.13\times10^{10}}$}
						
						($3.66\times10^{-1}$) & \textbf{$\boldsymbol{2.13\times10^{10}}$}
						
						($8.38\times10^{-2}$) & \textbf{$3.47\times10^{10}$}
						
						($1.15\times10^{2}$)\tabularnewline
						
						\bottomrule
					\end{tabular}
				\end{sc}
		\end{small}}
	\end{center}
\end{table}

\section{Conclusion\label{sec: Conclusion}}

We have proposed several efficient D\&C generators for producing $K$-combinations,
permutations, and nested combinations. Because these generators are
specifically designed using matrix operations and join-list homomorphisms,
it is straightforward to see why they can be easily parallelized and
executed on modern hardware, such as GPUs and easy to do memory management.

The nested combination generators, such as $\mathit{nestedCombs}$
and $\mathit{nestedPerms}$ constitute the first carefully examined
treatments beyond trivial composition, highlighting the considerable
potential of functional programming approaches in the study of combinatorial
problems. We further introduce a novel application of or integrating
revolving door ordering into nested generators, yielding several benefits
in the design and implementation of such generators. It is particularly
intriguing to consider how more sophisticated combinatorial objects,
such as binary trees, might be nested within this framework. It would
also be interesting to explore how alternative Gray code orderings
can be incorporated into the design of nested generators, such as
the Trotter--Johnson ordering, which has previously been studied
in a functional setting by \citet{bird2010pearls}.

Another promising direction is the systematic and compositional construction
of complex combinatorial generators from simpler ones. In the context
of analytic combinatorics, \citet{flajolet2009analytic} introduced
a symbolic method for deriving generating functions (GFs), which is
used to obtain counting formulas for complex combinatorial structures.
This approach constructs the GF of a complex combinatorial structure
by combining the GFs of basic structures via a small set of primitive
operations, such as \emph{product}, \emph{disjoint union} (coprodut),
\emph{Cartesian product}, \emph{set restriction}, and \emph{multiset}
formation.

These constructions correspond closely to several well-established
results in the constructive algorithmics community. For example, filter
fusion allows us to construct the combinatorial objects with set restriction,
while product fusion (the banana-split law) of \citet{bird1996algebra}
provides paradiam for constructing the results of a pair of generators
using a single recursion. Also, \citet{he2025ROF} discusses extensions
of Cartesian product fusion. These fusion technique has already process
several fundamental building blocks analogue to \citet{flajolet2009analytic}
's construction. Developing a systematic framework for constructing
combinatorial generators along these lines would be highly compelling
and could have significant impact across many areas of computer science
and discrete mathematics.

\bibliographystyle{abbrvnat}
\bibliography{Bibliography}

\appendix

\section{Haskell implementations}

\subsection{Standard functions in Haskell\label{subsec:Standard-functions-in Haskell}}

Below are the definition of standard Haskell functions used in this
paper, and their running examples.

\begin{lstlisting}[linerange={4-4},breaklines={true},breakatwhitespace={true}]

\end{lstlisting}The
\lstinline!map :: (a -> b) -> [a] -> [b]! function applies a function
$\texttt{f}$ to each element of a list $\texttt{xs}$, producing
a new list with the results. For instance, $\texttt{map (*2) [1, 2, 3] = [2, 4, 6]}$.

\begin{lstlisting}[linerange={6-7},breaklines={true},breakatwhitespace={true}]
concat (xs:xss) = xs ++ concat xss
\end{lstlisting}The
\lstinline!concat ::[[a]] -> [a]! function flattens a list of lists
into a single list by concatenating all sublists, for instance $\texttt{concat [[1, 2], [3], [4, 5]] = [1, 2, 3, 4, 5]}$.

\begin{lstlisting}[linerange={8-10},breaklines={true},breakatwhitespace={true}]
inits [] = [[]]
inits (a:x) = [[]] ++ map (a:) (inits x)
\end{lstlisting}
The \lstinline!inits ::[a] -> [[a]]! function returns all initial
segments of a list, from the empty list to the entire list, in order
of increasing length. $\texttt{inits [1,2,3] = [[],[1],[1,2],[1,2,3]]}$.

\begin{lstlisting}[linerange={12-13},breaklines={true},breakatwhitespace={true}]
tail (_:xs) = xs
tail [] = error "empty list"
\end{lstlisting}The
\lstinline!tail ::[a] -> [a]! function returns all elements of a
list except the first, or throws an error for an empty list. $\texttt{tail [1,2,3] = [2,3]}$.
Note that the definition of $\mathit{inits}$ in the main text corresponds
to \lstinline!tail . inits! in Haskell.

\begin{lstlisting}[linerange={17-17},breaklines={true},breakatwhitespace={true}]
reverse xs = foldl (\acc x -> x : acc) [] xs
\end{lstlisting}The
\lstinline!reverse ::[a] -> [a]! function reverses the order of elements
in a list.\\
 $\texttt{reverse [1,2,3] = [3,2,1]}$.

\begin{lstlisting}[linerange={20-22},breaklines={true},breakatwhitespace={true}]
zipWith _ [] _ = []
zipWith _ _ [] = []
zipWith f (x:xs) (y:ys) = f x y : zipWith f xs ys
\end{lstlisting}The
\lstinline!zipWith :: a -> b -> c -> [a] -> [b] -> [c]! function
combines two lists element-wise using a given function, stopping when
either list is empty.\\
 $\texttt{zipWith (+) [1, 2, 3] [4, 5, 6] = [5, 7, 9]}$.

\begin{lstlisting}[linerange={27-29},breaklines={true},breakatwhitespace={true}]
revInits = revInitsFrom [] where
revInitsFrom ys [] = [ys]
revInitsFrom ys (x:xs) = ys : revInitsFrom (x:ys) xs
\end{lstlisting}The
\lstinline!revInits ::[a] -> [[a]]! function returns all initial
segments of a list in reverse order, starting from the empty list
and building up by prepending elements.\\
 $\texttt{revInits [1, 2, 3] = [[],[1],[2,1],[3,2,1]]}$.

\begin{lstlisting}[linerange={32-38},breaklines={true},breakatwhitespace={true}]
setEmpty _ [] = []  
setEmpty 0 (_:xs) = [] : xs 
setEmpty d (x:xs) = x : setEmpty (d-1) xs
\end{lstlisting}The
\lstinline!setEmpty :: Int -> [[a]] -> [[a]]! function replaces the
$\texttt{d}$th element with an empty list.\\
We have $\texttt{setEmpty 2 [[1],[2],[3],[4]] = [[1],[2],[],[4]]]}$.

\subsection{Implementations of generators\label{subsec:Implementations-of-generators}}

Below are the Haskell implementations of derived generators in this
paper, and their running examples.

\subsubsection{$K$-combination/permutation generators}

\paragraph{Divide and conquer $K$-combination generator}

\begin{lstlisting}[linerange={90-96},breaklines={true},breakatwhitespace={true}]
kcombs_DC k [] = empty k
kcombs_DC k [a] = single k a
kcombs_DC k xs = convol crj (kcombs_DC k ys) (kcombs_DC k zs)
 where
  (ys,zs) = (take m xs, drop m xs)
  m = (length xs) `div` 2
\end{lstlisting}

where \lstinline!empty k = [[]]: replicate k []!, \lstinline!crj xs ys = crp (++) xs ys!
and the convolution operator is defined as

\begin{lstlisting}[linerange={37-37},breaklines={true},breakatwhitespace={true}]
convol f xss yss = [concat (zipWith f css yss)  |css <- tail (revInits xss)]
\end{lstlisting}Running
$\texttt{kcombs\_DC 2 [1,2,3]}$ gives $\texttt{[[[]],[[1],[2],[3]],[[1,2],[1,3],[2,3]]]}$.

\paragraph{Divide and conquer $K$-permutation generator}

\begin{lstlisting}[linerange={105-112},breaklines={true},breakatwhitespace={true}]
kperms_DC k [] = [[]]: replicate k [] 
kperms_DC k [a] = [[]]:[[a]]: replicate (k-1) [] 
kperms_DC k xs = convol crm (kperms_DC k ys) (kperms_DC k zs)
 where
  crm x y = concat [interleave (a, b) | a <- x, b <- y]
  (ys,zs) = (take m xs, drop m xs)
  m = (length xs) `div` 2
\end{lstlisting}

\begin{lstlisting}[linerange={101-103},breaklines={true},breakatwhitespace={true}]
interleave ([ ], ys) =[ys]
interleave (x : xs,y : ys) = map (x:) (interleave (xs, y : ys))++ map (y:) (interleave (x : xs, ys))
\end{lstlisting}We
have\\
 $\texttt{kperms\_DC 2 [1,2,3]=[[[]],[[1],[2],[3]],[[1,2],[2,1],[1,3],[3,1],[2,3],[3,2]]]}$.

\paragraph{Sequential $K$-combination generator}

\begin{lstlisting}[linerange={70-73},breaklines={true},breakatwhitespace={true}]
kcombs :: Int -> [a] -> Css a
kcombs k [] = empty k
kcombs k (x:xs) = for x (kcombs k xs)
 where for a css = [[]]:[map (a:) (css!!(i-1)) ++ (css!!i) | i <- [1..k]]
\end{lstlisting}We
have $\texttt{kcombs 2 [1,2,3]= [[[]],[[1],[2],[3]],[[1,2],[1,3],[2,3]]]}$.

\paragraph{Sequential $K$-combination generator in revolving door ordering}

\begin{lstlisting}[linerange={76-79},breaklines={true},breakatwhitespace={true}]
kcombs_revol :: Int -> [a] -> Css a
kcombs_revol k [] = empty k
kcombs_revol k (x:xs) = for_revol x (kcombs_revol k xs)
 where for_revol a css  = [[]]:[(css!!i) ++ (map (++[a]) (reverse (css!!(i-1)))) | i <- [1..k]]
\end{lstlisting}We
have $\texttt{kcombs\_revol 2 [3,2,1]= [[[]],[[1],[2],[3]],[[1,2],[2,3],[1,3]]]}$.

\paragraph{Integer $K$-combination generator in revolving door ordering}

\begin{lstlisting}[linerange={81-86},breaklines={true},breakatwhitespace={true}]
kcombs_revol_int :: Int -> Int -> [[Int]]
kcombs_revol_int 0 k = [0]:replicate k []
kcombs_revol_int n k = for_rev_int n (kcombs_revol_int (n-1) k)
 where
  for_rev_int n css  = [0]:[(css!!i) ++ (map ((nk_max n i)-) (reverse (css!!(i-1)))) | i <- [1..k]]
  nk_max i j = (i `choose` j) -1
\end{lstlisting}We
have $\texttt{kcombs\_revol\_int 3 2= [[0],[0,1,2],[0,1,2]]}$.

\subsubsection{Nested generators}

\paragraph{Divide and conquer nested $\text{\ensuremath{\left(K,D\right)}}$-combination-combination
	generator}

\begin{lstlisting}[linerange={124-136},breaklines={true},breakatwhitespace={true}]
nestedcombs_DC k d [] = (empty d, empty k)
nestedcombs_DC k d [x] = (single d x, empty k)
nestedcombs_DC k d xs = ((setEmpty d css), ncss)
 where
  ncss 
   | null (css!!d) = empty k
   | otherwise = cvcrj (cvcrj ncss1 ncss2) (kcombs k (css!!d))
  css = cvcrj css1 css2
  (css1,ncss1) = nestedcombs_DC k d ys
  (css2,ncss2) = nestedcombs_DC k d zs
  (ys,zs) = (take m xs, drop m xs)
  m = (length xs) `div` 2
\end{lstlisting}where
$\texttt{cvcrj = convol crj}$. Running $\texttt{nestedcombs\_DC 2 2 [1,2,3]}$
we have\\
$\texttt{([[[]],[[1],[2],[3]],[]],[[[]],[[[2,3]],[[1,2]],[[1,3]]],[[[2,3],[1,2]],...)}$.

\paragraph{Sequential nested $\text{\ensuremath{\left(K,D\right)}}$-combination-combination
	generator}

\begin{lstlisting}[linerange={139-147},breaklines={true},breakatwhitespace={true}]
nestedcombs k d [] = (empty d, empty k)
nestedcombs k d (x:xs) = ((setEmpty d css), ncss)
 where
  ncss 
   | null (css!!d) = empty k
   | otherwise = cvcrj (kcombs k (css!!d)) ncss'
  css = cvcrj (single d x) css'
  (css',ncss') = nestedcombs k d xs
\end{lstlisting}
Running $\texttt{nestedcombs 2 2 [1,2,3]}$ we have\\
$\texttt{([[[]],[[1],[2],[3]],[]],[[[]],[[[1,2]],[[1,3]],[[2,3]]],[[[1,2],[1,3]],...)}$.

\paragraph{Sequential nested $\text{\ensuremath{\left(K,D\right)}}$-combination-combination
	generator with inner combinations in revolving door ordering}

\begin{lstlisting}[linerange={162-170},breaklines={true},breakatwhitespace={true}]
nestedcombs_revol k d (x:xs) = ((setEmpty d css), ncss)
 where
  ncss 
   | null (css!!d) = empty k
   | otherwise = cvcrj (kcombs k (css!!d)) ncss'
  css = [[]]:[(css'!!i) ++ (map (++ [x]) (reverse (css'!!(i-1)))) | i <- [1..d]]
  (css',ncss') = nestedcombs_revol k d xs

\end{lstlisting}
Running $\texttt{nestedcombs\_revol 2 2 [1,2,3]}$ we have\\
$\texttt{([[[]],[[1],[2],[3]],[]],[[[]],[[[2,3]],[[1,3]],[[1,2]]],[[[2,3],[1,3]],...)}$.

\paragraph{Integer nested $\text{\ensuremath{\left(K,D\right)}}$-combination-combination
	generator}

\begin{lstlisting}[linerange={173-182},breaklines={true},breakatwhitespace={true}]
nestedcombs_revol_int k d n = ((setEmpty d css), ncss)
 where
  ncss 
   | null (css!!d) = empty k
   | otherwise = cvcrj (kcombs k (css!!d)) ncss'
  css = [0]:[(css'!!i) ++ (map ((nk_max n i)-) (reverse (css'!!(i-1)))) | i <- [1..d]]
  nk_max i j = (i `choose` (j)) -1
  (css',ncss') = nestedcombs_revol_int k d (n-1)

\end{lstlisting}
Running $\texttt{nestedcombs\_revol\_int 2 2 3}$ we have\\
$\texttt{([[0],[0,1,2],[]],[[[]],[[1],[2],[0]],[[1,2],[1,0],[2,0]]])}$.

\paragraph{Divide and conquer nested $\text{\ensuremath{\left(K,D\right)}}$-permutation-combination
	generator}

\begin{lstlisting}[linerange={189-201},breaklines={true},breakatwhitespace={true}]
nestedperms_DC k d [] = (empty d, empty k)
nestedperms_DC k d [x] = (single d x, empty k)
nestedperms_DC k d xs = ((setEmpty d pss), npss)
 where
  npss 
   | null (pss!!d) = empty k
   | otherwise = cvcrm (cvcrm npss1 npss2) (kperms_DC k (pss!!d))
  pss = cvcrm pss1 pss2
  (pss1,npss1) = nestedperms_DC k d ys
  (pss2,npss2) = nestedperms_DC k d zs
  (ys,zs) = (take m xs, drop m xs)
  m = (length xs) `div` 2
\end{lstlisting}where
$\texttt{cvcrj = convol crm}$. Running $\texttt{nestedperms 2 2 [1,2]}$
we have\\
$\texttt{([[[]],[[1],[2]],[]],[[[]],[[[1,2]],[[2,1]]],[[[1,2],[2,1]],[[2,1],[1,2]]]])}$.

\paragraph{Divide and conquer nested combination-combination generator with
	multiple inner combinations}

\begin{lstlisting}[linerange={83-97},breaklines={true},breakatwhitespace={true}]
nestedcombs_DC' :: Int -> [Int] -> [a] -> (Css a,NCss a)
nestedcombs_DC' k ds [] = (empty (maximum ds), empty k)
nestedcombs_DC' k ds [x] = (single (maximum ds) x, empty k)
nestedcombs_DC' k ds xs = (setEmpty (maximum ds) css, ncss)
 where
  ncss 
   | null (css!!(minimum ds)) = empty k
   | otherwise = cvcrj (cvcrj ncss1 ncss2) (kcombs k cs')
  cs' = concat [css_new!!d | d <- ds]
  css_new = convol_new crj css1 css2
  css = cvcrj css1 css2
  (css1,ncss1) = nestedcombs_DC' k ds ys
  (css2,ncss2) = nestedcombs_DC' k ds zs
  (ys,zs) = (take m xs, drop m xs)
  m = (length xs) `div` 2
\end{lstlisting}

where $\texttt{convol\_new}$ is defined as

\begin{lstlisting}[linerange={36-37},breaklines={true},breakatwhitespace={true}]
convol_new f xss yss = [new css yss | css <- tail (revInits xss)]
 where new ass bss = concat (zipWith f (init' (tail ass)) (init' (tail bss)))
\end{lstlisting}
where $\texttt{init}$ is redefined as

\begin{lstlisting}[linerange={41-43},breaklines={true},breakatwhitespace={true}]
init' [] = [[]]
init' [_] = []
init' (x:xs) = x : init' xs
\end{lstlisting}to
handle the empty list in $\texttt{css}$. Running $\texttt{ nestedcombs\_DC' 2 [2,3] [1,2,3]}$
we have $\texttt{([[[]],[[1],[2],[3]],[[1,2],[1,3],[2,3]],[]],}$ \\
	$\texttt{[[[]],[[[2,3]],[[1,2]],[[1,3]],[[1,2,3]]]...)}$.

\section{Proofs}

\subsection{Notations and laws used in the proof}

\paragraph{Definitions}

This section considers only the join-list functor $\mathbf{F}_{\mathbf{A}}=\mathbf{1}+\mathbf{A}+\mathbf{id}\times\mathbf{id}$
where boldface is used to denote functors. Since this section is concerned
solely with reasoning about the correctness of the recursive case---namely,
the join part ($\mathbf{id}\times\mathbf{id}$) of the join-list functor---we
restrict our attention accordingly. To avoid unnecessary abstraction,
we treat the morphism $\mathbf{F}_{\mathbf{A}}\left(f\right)$ on
the join component as a function on pairs, and denote it by $Ff$
obtained by prefixing the function $f$ with $F$ . For example, we
previously defined the functions $\mathit{Ffst}$ and $\mathit{Fsnd}$
function as following

\paragraph{
	\begin{align*}
		\mathit{Ffst}\left(\left(a,b\right),\left(c,d\right)\right) & =\left(a,c\right)\\
		\mathit{Fsnd}\left(\left(a,b\right),\left(c,d\right)\right) & =\left(b,d\right)
	\end{align*}
}

Similarly, we define the corresponding morphisms for $\mathit{SE}$,
$!$ as following

\paragraph{
	\begin{align*}
		\mathit{FSE}\left(d,a,b\right) & =\left(\mathit{SE}\left(d,a\right),\mathit{SE}\left(d,b\right)\right)\\
		\mathit{F!}\left(D,a,b\right) & =\left(!\left(D,a\right),!\left(D,b\right)\right)
	\end{align*}
}

Lastly, we adopt the similar definitions for $\mathit{FKcombs}$ and
$\mathit{FKcombsAlg}$.

\paragraph{Laws}

We adapt several laws presented in \citet{bird1996algebra}

\begin{align}
	\text{ Product fusion: } & \left\langle f,g\right\rangle \cdot h=\left\langle f\cdot h,g\cdot h\right\rangle \label{eq: product fusion}\\
	\text{\ensuremath{\times} absorption law:} & \ensuremath{\left(f\times g\right)\cdot\left\langle p,q\right\rangle =\left\langle f\cdot p,g\cdot q\right\rangle }\label{eq: times absorption}\\
	\text{\ensuremath{\times} preserve composition:} & \ensuremath{\left(f\times g\right)\cdot p\times q=f\cdot p\times g\cdot q}\label{eq: times preserves composition}
\end{align}
where $\left\langle f,g\right\rangle \left(a\right)=\left(f\left(a\right),g\left(a\right)\right)$
and $f\times g=\left\langle f\cdot\mathit{fst},g\cdot\mathit{snd}\right\rangle $.
Also, we can define $\mathit{Ffst}$ and $\mathit{Fsnd}$ using $\times$

\begin{align}
	\mathit{fst}\times\mathit{fst} & =\left\langle \mathit{fst}\cdot\mathit{fst},\mathit{fst}\cdot\mathit{snd}\right\rangle =\mathit{Ffst}\label{eq: Ffst definition}\\
	\mathit{snd}\times\mathit{snd} & =\left\langle \mathit{snd}\cdot\mathit{fst},\mathit{snd}\cdot\mathit{snd}\right\rangle =\mathit{Fsnd}\label{eq: Fsnd definition}
\end{align}
Composing $\mathit{Ffst}$ with $\left\langle f,g\right\rangle $
and $f\times g$, it is straightforward to derive the following useful
rules
\begin{align}
	\mathit{Ffst}\cdot\left\langle \left\langle f_{1},g_{1}\right\rangle ,\left\langle f_{2},g_{2}\right\rangle \right\rangle  & =\left\langle f_{1},f_{2}\right\rangle \label{eq: Ffst<>,<>>}\\
	\mathit{Fsnd}\cdot\left\langle \left\langle f_{1},g_{1}\right\rangle ,\left\langle f_{2},g_{2}\right\rangle \right\rangle  & =\left\langle g_{1},g_{2}\right\rangle \label{eq:Fsnd<<>,<>>}
\end{align}
and similarly 

\begin{equation}
	\begin{aligned}
		\mathit{Ffst}\cdot\left\langle f_{1},g_{1}\right\rangle \times\left\langle f_{2},g_{2}\right\rangle 
		&= \mathit{Ffst}\cdot\left\langle \left\langle f_{1}\cdot\mathit{fst},g_{1}\cdot\mathit{fst}\right\rangle ,\left\langle f_{2}\cdot\mathit{snd},g_{2}\cdot\mathit{snd}\right\rangle \right\rangle \\
		&= f_{1}\times f_{2}
	\end{aligned}
	\label{eq: Ffst <> x <>}
\end{equation}

\begin{equation}
	\begin{aligned}
		\mathit{Fsnd}\cdot\left\langle f_{1},g_{1}\right\rangle \times\left\langle f_{2},g_{2}\right\rangle 
		&= \mathit{Fsnd}\cdot\left\langle \left\langle f_{1}\cdot\mathit{fst},g_{1}\cdot\mathit{fst}\right\rangle ,\left\langle f_{2}\cdot\mathit{snd},g_{2}\cdot\mathit{snd}\right\rangle \right\rangle \\
		&= g_{1}\times g_{2}
	\end{aligned}
	\label{eq: Fsnd <> x <>}
\end{equation}

\begin{equation}
	\begin{aligned}
		\mathit{Ffst}\cdot\left(f_{1}\times g_{1}\right)\times\left(f_{2}\times g_{2}\right)
		&= \mathit{Ffst}\cdot\left\langle \left\langle f_{1}\cdot\mathit{fst},g_{1}\cdot\mathit{snd}\right\rangle \cdot\mathit{fst},\left\langle f_{2}\cdot\mathit{fst},g_{2}\cdot\mathit{snd}\right\rangle \cdot\mathit{snd}\right\rangle \\
		&= \left(f_{1}\cdot\mathit{fst}\right)\times\left(f_{2}\cdot\mathit{fst}\right)
	\end{aligned}
	\label{eq: Ffst x x x}
\end{equation}

\begin{equation}
	\begin{aligned}
		\mathit{Fsnd}\cdot\left(f_{1}\times g_{1}\right)\times\left(f_{2}\times g_{2}\right)
		&= \mathit{Fsnd}\cdot\left\langle \left\langle f_{1}\cdot\mathit{fst},g_{1}\cdot\mathit{snd}\right\rangle \cdot\mathit{fst},\left\langle f_{2}\cdot\mathit{fst},g_{2}\cdot\mathit{snd}\right\rangle \cdot\mathit{snd}\right\rangle \\
		&= \left(f_{1}\cdot\mathit{snd}\right)\times\left(f_{2}\cdot\mathit{snd}\right)
	\end{aligned}
	\label{eq: Fsnd x x x}
\end{equation}

\subsection{Proof of the nested combination generator\label{sec:Proofs}}

Given $\mathit{nestedCombsAlg}\left(k,d\right)$ defined as

\begin{equation}
	\begin{aligned} & \bigg<\mathit{setEmpty}\left(D\right)\cdot\mathit{KcombsAlg}\left(D\right)\cdot\mathit{Ffst}\\
		& \qquad\mathit{\mathit{\mathit{KcombsAlg}}\left(K\right)}\cdot\left\langle \mathit{Kcombs}\left(K\right)\cdot!\left(D\right)\cdot\mathit{KcombsAlg}\left(D\right)\cdot\mathit{Ffst},\mathit{KcombsAlg}\left(K\right)\cdot\mathit{Fsnd}\right\rangle \bigg>
	\end{aligned}
\end{equation}

We need to verify the following fusion condition

\begin{equation}
	f\cdot\mathit{KcombsAlg}\left(D\right)=\mathit{nestedCombsAlg}\left(K,D\right)\cdot f\times f,
\end{equation}
where $f=\left\langle \mathit{setEmpty}\left(D\right),\mathit{Kcombs}\left(K\right)\cdot!\left(D\right)\right\rangle $.
In other words, we need to prove that the following diagram commutes

\[
\xymatrix{\mathit{Css}\ar[d]_{f} &  &  &  & \left(\mathit{Css},\mathit{Css}\right)\ar[d]^{f\times f}\ar[llll]_{\mathit{\mathit{kcombsAlg}\left(D\right)}}\\
	\left(\mathit{Css},\mathit{NCss}\right) &  &  &  & \left(\left(\mathit{Css},\mathit{NCss}\right),\left(\mathit{Css},\mathit{NCss}\right)\right)\ar[llll]^{\mathit{nestedCombsAlg}\left(K,D\right)}
}
\]

However, proving that the above diagram commutes is challenging. Instead,
we expand the diagram by presenting all intermediate stage explicitly

\subsubsection*{
	\[
	\protect\xymatrix@C=1.2em@R=2.2em{\mathit{Css}\ar[d]^{\left\langle \mathit{SE}\left(D\right),!\left(D\right)\right\rangle } &  &  &  &  &  &  & \left(\mathit{Css},\mathit{Css}\right)\ar[d]_{\left\langle \mathit{SE}\left(D\right),!\left(D\right)\right\rangle \times\left\langle \mathit{SE}\left(D\right),!\left(D\right)\right\rangle }\ar[lllllll]^{\mathit{KCsA}\left(D\right)}\\
		\left(\mathit{Css},\mathit{Cs}\right)\ar[d]^{\mathit{id}\times\mathit{KCs}\left(K\right)} &  & \left(\mathit{Css},\left(\mathit{Cs},\mathit{Cs}\right)\right)\ar[ll]^{\mathit{SE}\left(D\right)\times\cup} &  &  &  &  & \left(\left(\mathit{Css},\mathit{Cs}\right),\left(\mathit{Css},\mathit{Cs}\right)\right)\ar[d]_{\left(\mathit{id}\times\mathit{KCs}\left(K\right)\right)\times\left(\mathit{id}\times\mathit{KCs}\left(K\right)\right)}\ar[lllll]^{\texttt{\texttt{\ensuremath{\left\langle \mathit{KCsA}\left(D\right)\cdot\mathit{Ffst},\left\langle !\left(D\right)\cdot\mathit{KCsA}\left(D\right)\cdot\mathit{Ffst},\cup\cdot\mathit{Fsnd}\right\rangle \right\rangle }}}}\\
		\left(\mathit{Css},\mathit{NCss}\right) &  & \left(\mathit{Css},\left(\mathit{NCss},\mathit{NCss}\right)\right)\ar[ll]^{\mathit{SE}\left(D\right)\times\mathit{KCsA}\left(K\right)} &  &  &  &  & \left(\left(\mathit{Css},\mathit{NCss}\right),\left(\mathit{Css},\mathit{NCss}\right)\right)\ar[lllll]^{\texttt{\ensuremath{\left\langle \mathit{KCsA}\left(D\right)\cdot\mathit{Ffst},\left\langle \mathit{KCs}\left(K\right)\cdot!\left(D\right)\cdot\mathit{KCsA}\left(D\right)\cdot\mathit{Ffst},\mathit{KCsA}\left(K\right)\cdot\mathit{Fsnd}\right\rangle \right\rangle }}}
	}
	\]
}

where $\cup\left(a,b\right)=a\cup b$, and $\mathit{\mathit{SE}}$,
$\mathit{KCs}$ and $\mathit{KCsA}$ are short for $\mathit{setEmpty}$,
$\mathit{Kcombs}$ and $\mathit{KcombsAlg}$.

To prove the fusion condition, we first need to verify the two paths
between $\left(\mathit{Css},\mathit{Css}\right)$ and $\left(\mathit{Css},\mathit{Cs}\right)$.
In other words, we need to prove
\begin{equation}
	\begin{aligned} & \left\langle \mathit{SE}\left(D\right),!\left(D\right)\right\rangle \cdot\mathit{KCsA}\left(D\right)=  SE\left(D\right)\times \\ 
		& \qquad \left(\cup\cdot\texttt{\texttt{\ensuremath{\left\langle \mathit{KCsA}\left(D\right)\cdot\mathit{Ffst},\left\langle !\left(D\right)\cdot\mathit{KCsA}\left(D\right)\cdot\mathit{Ffst},\cup\cdot\mathit{Fsnd}\right\rangle \right\rangle }}}\right)\cdot\left(\left\langle \mathit{SE}\left(D\right),!\left(D\right)\right\rangle \times\left\langle \mathit{SE}\left(D\right),!\left(D\right)\right\rangle \right)
	\end{aligned}
\end{equation}

This can be proved by following equational reasoning

\begin{align*}
	& \mathit{SE}\left(D\right)\times\cup\cdot\\
	&\qquad \texttt{\texttt{\ensuremath{\left\langle \mathit{KCsA}\left(D\right)\cdot\mathit{Ffst},\left\langle !\left(D\right)\cdot\mathit{KCsA}\left(D\right)\cdot\mathit{Ffst},\cup\cdot\mathit{Fsnd}\right\rangle \right\rangle }}}\cdot\left(\left\langle SE\left(D\right),!\left(D\right)\right\rangle \times\left\langle \mathit{SE}\left(D\right),!\left(D\right)\right\rangle \right)\\
	= &  \text{ \{  \ensuremath{\times} absorption law }(\ref{eq: times absorption}) \} \\
	& \texttt{\texttt{\ensuremath{\left\langle \mathit{SE}\left(D\right)\cdot\mathit{KCsA}\left(D\right)\cdot\mathit{Ffst},\cup\cdot\left\langle !\left(D\right)\cdot\mathit{KCsA}\left(D\right)\cdot\mathit{Ffst},\cup\cdot\mathit{Fsnd}\right\rangle \right\rangle }}}\cdot\left(\left\langle SE\left(D\right),!\left(D\right)\right\rangle \times\left\langle \mathit{SE}\left(D\right),!\left(D\right)\right\rangle \right)\\
	= &  \text{ \{   Product fusion (\ref{eq: product fusion})} \} \\
	& \bigg<\mathit{SE}\left(D\right)\cdot\mathit{KCsA}\left(D\right)\cdot\mathit{Ffst}\cdot\left(\left\langle SE\left(D\right),!\left(D\right)\right\rangle \times\left\langle \mathit{SE}\left(D\right),!\left(D\right)\right\rangle \right),\\
	& \qquad\quad\texttt{\texttt{\ensuremath{\cup\cdot\left\langle !\left(D\right)\cdot\mathit{KCsA}\left(D\right)\cdot\mathit{Ffst},\cup\cdot\mathit{Fsnd}\right\rangle \cdot\left(\left\langle SE\left(D\right),!\left(D\right)\right\rangle \times\left\langle \mathit{SE}\left(D\right),!\left(D\right)\right\rangle \right)\bigg>}}}\\
	= &  \text{ \{   (\ref{eq: Ffst <> x <>}) and (\ref{eq: Fsnd <> x <>}) \} }\\
	& \bigg<\mathit{SE}\left(D\right)\cdot\mathit{KCsA}\left(D\right)\cdot SE\left(D\right)\times SE\left(D\right),\\
	& \qquad\quad\texttt{\texttt{\ensuremath{\cup\cdot\left\langle !\left(D\right)\cdot\mathit{KCsA}\left(D\right)\cdot SE\left(D\right)\times SE\left(D\right),\cup\cdot!\left(D\right)\times!\left(D\right)\right\rangle }}}\bigg>
\\
	= &  \text{ \{  apply \ensuremath{SE\left(D\right)} twice is same as once} \} \\
	& \bigg<\mathit{SE}\left(D\right)\cdot\mathit{KCsA}\left(D\right),\\
	& \qquad\quad\texttt{\texttt{\ensuremath{\cup\cdot\left\langle !\left(D\right)\cdot\mathit{KCsA}\left(D\right)\cdot SE\left(D\right)\times SE\left(D\right),\cup\cdot!\left(D\right)\times!\left(D\right)\right\rangle }}}\bigg>\\
	= &  \text{ \{  definition of combination:}\\
	& \qquad \qquad !\ensuremath{\left(D\right)\cdot}\ensuremath{\mathit{KCsA}\left(D\right)} = \ensuremath{\cup\cdot\left\langle !\left(D\right)\cdot\mathit{KCsA}\left(D\right)\cdot SE\left(D\right)\times SE\left(D\right),\cup\cdot!\left(D\right)\times!\left(D\right)\right\rangle } \} \\
	& \left\langle \mathit{SE}\left(D\right)\cdot\mathit{KCsA}\left(D\right),!\left(D\right)\cdot\mathit{KCsA}\left(D\right)\right\rangle \\
	= &  \text{ \{  Product fusion }(\ref{eq: product fusion}) \} \\
	& \left\langle \mathit{SE}\left(D\right),!\left(D\right)\right\rangle \cdot\mathit{KCsA}\left(D\right)
\end{align*}
Note that, $!\left(D\right)\cdot\mathit{KCsA}\left(D\right)=\cup\cdot\left\langle !\left(D\right)\cdot\mathit{KCsA}\left(D\right)\cdot SE\left(D\right)\times SE\left(D\right),\cup\cdot!\left(D\right)\times!\left(D\right)\right\rangle $
holds by the definition of combination. In particular, $\cup\cdot!\left(D\right)\times!\left(D\right)$
corresponds to the size $D$ combinations generated in the final recursive
step. If these ``old combinations'' are removed using $SE\left(D\right)\times SE\left(D\right)$,
and new combinations are subsequently generated via $\mathit{KCsA}\left(D\right)$.
Thus the ``newly generated $D$-combinations'' ($!\left(D\right)\cdot\mathit{KCsA}\left(D\right)\cdot SE\left(D\right)\times SE\left(D\right)$)
together with the retained “old combinations” constitute the complete
set of $D$ combinations.

Next, we prove the two paths between $\left(\left(\mathit{Css},\mathit{Cs}\right),\left(\mathit{Css},\mathit{Cs}\right)\right)$
and $\left(\mathit{Css},\mathit{NCss}\right)$ are equivalent.
\begin{align*}
	& \mathit{SE}\left(D\right)\times\left(\mathit{KCsA}\left(K\right)\right)\cdot\left\langle \mathit{KCsA}\left(D\right)\cdot\mathit{Ffst},\left\langle \mathit{KCs}\left(K\right)\cdot\left(D\right)\cdot\mathit{KCsA}\left(D\right)\cdot\mathit{Ffst},\mathit{KCsA}\left(K\right)\cdot\mathit{Fsnd}\right\rangle \right\rangle \cdot\\
	& \qquad\qquad\left(\mathit{id}\times\mathit{KCs}\left(K\right)\right)\times\left(\mathit{id}\times\mathit{KCs}\left(K\right)\right)\\
	= &  \text{ \{  Product fusion }(\ref{eq: product fusion}) \} \\
	& \mathit{SE}\left(D\right)\times\left(\mathit{KCsA}\left(K\right)\right)\cdot\bigg<\mathit{KCsA}\left(D\right)\cdot\mathit{Ffst}\cdot\left(\mathit{id}\times\mathit{KCs}\left(K\right)\right)\times\left(\mathit{id}\times\mathit{KCs}\left(K\right)\right),\\
	& \qquad\qquad\left\langle \mathit{KCs}\left(K\right)\cdot\left(D\right)\cdot\mathit{KCsA}\left(D\right)\cdot\mathit{Ffst}\cdot,\mathit{KCsA}\left(K\right)\cdot\mathit{Fsnd}\right\rangle \cdot\left(\mathit{id}\times\mathit{KCs}\left(K\right)\right)\times\left(\mathit{id}\times\mathit{KCs}\left(K\right)\right)\bigg>\\
	= &  \text{ \{  unfold the definition of \ensuremath{\left(\mathit{id}\times\mathit{KCs}\left(K\right)\right)\times\left(\mathit{id}\times\mathit{KCs}\left(K\right)\right)} , (\ref{eq: Ffst x x x}) and (\ref{eq: Fsnd x x x}) \} }\\
	& \mathit{SE}\left(D\right)\times\left(\mathit{KCsA}\left(K\right)\right)\cdot\bigg<\mathit{KCsA}\left(D\right)\cdot\left(\mathit{id}\cdot\mathit{fst}\right)\times\left(\mathit{id}\cdot\mathit{fst}\right),\\
	& \qquad\qquad 
	\big<\mathit{KCs}\left(K\right)\cdot\left(D\right)\cdot\mathit{KCsA}\left(D\right)\cdot\left(\mathit{id}\cdot\mathit{fst}\right)\times\left(\mathit{id}\cdot\mathit{fst}\right),\\
	& \qquad \qquad \mathit{KCsA}\left(K\right)\cdot\left(\mathit{KCs}\left(K\right)\cdot\mathit{snd}\right)\times\left(\mathit{KCs}\left(K\right)\cdot\mathit{snd}\right)\big> \bigg>\\
	= &  \text{ \{  \ensuremath{\times} absorption law (\ref{eq: times absorption}) and definition of \ensuremath{\mathit{Ffst}}} \} \\
	& \bigg<\mathit{SE}\left(D\right)\cdot\mathit{KCsA}\left(D\right)\cdot\ensuremath{\mathit{Ffst}},\\
	& \qquad\qquad\mathit{KCsA}\left(K\right)\cdot\left\langle \mathit{KCs}\left(K\right)\cdot\left(D\right)\cdot\mathit{KCsA}\left(D\right)\cdot\ensuremath{\mathit{Ffst}}\cdot,\mathit{KCsA}\left(K\right)\cdot\left(\mathit{KCs}\left(K\right)\cdot\mathit{snd}\right)\times\left(\mathit{KCs}\left(K\right)\cdot\mathit{snd}\right)\right\rangle \bigg>\\
	= &  \text{ \{  \ensuremath{\times} preserves composition (backwards) (\ref{eq: times preserves composition})} \} \\
	& \bigg<\mathit{SE}\left(D\right)\cdot\mathit{KCsA}\left(D\right)\cdot\ensuremath{\mathit{Ffst}},\\
	& \qquad\qquad\mathit{KCsA}\left(K\right)\cdot\left\langle \mathit{KCs}\left(K\right)\cdot\left(D\right)\cdot\mathit{KCsA}\left(D\right)\cdot\ensuremath{\mathit{Ffst}}\cdot,\mathit{KCsA}\left(K\right)\cdot\left(\mathit{KCs}\left(K\right)\times\mathit{KCs}\left(K\right)\right)\cdot\left(\mathit{snd}\times\mathit{snd}\right)\right\rangle \bigg>\\
	= &  \text{ \{  definition of \ensuremath{\mathit{Fsnd}} (\ref{eq: Fsnd definition})} \} \\
	& \texttt{\ensuremath{\bigg<}\ensuremath{\mathit{SE}\left(D\right)\cdot}\ensuremath{\mathit{KCsA}\left(D\right)\cdot}\ensuremath{\mathit{Ffst}},}\\
	& \qquad\qquad\mathit{KCsA}\left(K\right)\cdot\left\langle \mathit{KCs}\left(K\right)\cdot!\left(D\right)\cdot\mathit{KCsA}\left(D\right)\cdot\mathit{Ffst},\mathit{KCsA}\left(K\right)\cdot\mathit{KCs}\left(K\right)\times\mathit{KCs}\left(K\right)\cdot\mathit{Fsnd}\right\rangle \bigg>\\
	= &  \text{ \{  definition of \ensuremath{\mathit{KCs}}: \ensuremath{\mathit{KCs}\left(K\right)\cdot\cup\ }= \ensuremath{\mathit{KCsA}\left(K\right)\cdot}\ensuremath{\mathit{KCs}\left(K\right)\times}\ensuremath{\mathit{KCs}\left(K\right)}} \} \\
	& \texttt{\ensuremath{\bigg<}\ensuremath{\mathit{SE}\left(D\right)\cdot}\ensuremath{\mathit{KCsA}\left(D\right)\cdot}\ensuremath{\mathit{Ffst}},}\\
	& \qquad\qquad \mathit{KCsA}\left(K\right)\cdot\left\langle \mathit{KCs}\left(K\right)\cdot!\left(D\right)\cdot\mathit{KCsA}\left(D\right)\cdot\mathit{Ffst},\mathit{KCs}\left(K\right)\cdot\cup\cdot\mathit{Fsnd}\right\rangle \bigg>\\
	= &  \text{ \{  \ensuremath{\times} absorption law (backwards) (\ref{eq: times absorption})} \} \\
	& \texttt{\ensuremath{\bigg<}\ensuremath{\mathit{SE}\left(D\right)\cdot}\ensuremath{\mathit{KCsA}\left(D\right)\cdot}\ensuremath{\mathit{Ffst}},}\\
	& \qquad\qquad\mathit{KCsA}\left(K\right)\cdot\mathit{KCs}\left(K\right)\times\mathit{KCs}\left(K\right)\cdot\left\langle !\left(D\right)\cdot\mathit{KCsA}\left(D\right)\cdot\mathit{Ffst},\cup\cdot\mathit{Fsnd}\right\rangle \bigg>\allowdisplaybreaks[1]\\
	= &  \text{ \{  definition of \ensuremath{\mathit{KCs}} (backwards)} \} \\
	& \texttt{\ensuremath{\bigg<}\ensuremath{\mathit{SE}\left(D\right)\cdot}\ensuremath{\mathit{KCsA}\left(D\right)\cdot}\ensuremath{\mathit{Ffst}},}\\
	& \qquad\qquad\left(\mathit{KCs}\left(K\right)\cdot\cup\right)\cdot\left\langle !\left(D\right)\cdot\mathit{KCsA}\left(D\right)\cdot\mathit{Ffst},\cup\cdot\mathit{Fsnd}\right\rangle \bigg>\\
	= &  \text{ \{  \ensuremath{\times} absorption law (backwards) (\ref{eq: times absorption})} \} \\
	& \mathit{SE}\left(D\right)\times\left(\mathit{KCs}\left(K\right)\cdot\cup\right)\cdot\texttt{\ensuremath{\left\langle \mathit{KCsA}\left(D\right)\cdot\mathit{Ffst},\left\langle !\left(D\right)\cdot\mathit{KCsA}\left(D\right)\cdot\mathit{Ffst},\cup\cdot\mathit{Fsnd}\right\rangle \right\rangle }}\\
	= &  \text{ \{  \ensuremath{\times}preserves composition (\ref{eq: times preserves composition})} \} \\
	& \mathit{id}\times\mathit{KCs}\left(K\right)\cdot\mathit{SE}\left(D\right)\times\cup\cdot\texttt{\ensuremath{\left\langle \mathit{KCsA}\left(D\right)\cdot\mathit{Ffst},\left\langle !\left(D\right)\cdot\mathit{KCsA}\left(D\right)\cdot\mathit{Ffst},\cup\cdot\mathit{Fsnd}\right\rangle \right\rangle }}
\end{align*}

\subsection{Proof of Theorem 1\label{Proof-of-Theorem}}
\begin{theorem*}
	\emph{Equation (\ref{eq: seq-kcombs revol}) generates $\mathit{cs}_{k}^{N}$
		in revolving door order for all $1\le k\leq K$.}
\end{theorem*}
\begin{proof}
	It is straightforward to verify that (\ref{eq: seq-kcombs revol})
	generates all valid combinations, as it merely reorders the list $\mathit{cs}_{k}^{n}$
	and $\mathit{map}\left(\cup\left[x\right]:,\mathit{cs}_{k-1}^{n}\right)$,
	the $\mathit{reverse}$ function only affects the ordering of elements,
	not the combinations themselves.
	
	To prove that (\ref{eq: seq-kcombs revol}) satisfies the revolving
	door property, we proceed by induction. This property requires that
	every pair of adjacent combinations in $\mathit{cs}_{k}$, including
	the first and last, differ by exactly two elements.
	
	By induction, $\mathit{cs}_{k}^{N-1}$ and $\mathit{cs}_{k-1}^{N-1}$
	satisfy the revolving door property for all $k\in\left[1,\ldots,K\right]$.
	In order to prove $\mathit{cs}_{k}^{N}$ satisfy the revolving door
	property, we aim to show
	
	(1) The first element of $\mathit{cs}_{k}^{N-1}$ and the last element
	of $\mathit{reverse}\left(\mathit{map}\left(\cup\left[x\right]:,\mathit{cs}_{k-1}^{N-1}\right)\right)$
	differ by exactly two elements.
	
	(2) The last element of $\mathit{cs}_{k}^{N-1}$ and the first element
	of $\mathit{reverse}\left(\mathit{map}\left(\cup\left[x\right]:,\mathit{cs}_{k-1}^{N-1}\right)\right)$
	also differ by exactly two elements.
	
	It is easy to verify that $\mathit{cs}_{k}^{k}=\left[\left[1,\ldots,k\right]\right]$
	since there is only one $k$-combination from a size $k$ set and
	we recursively apply $\cup\left[k\right]$ to $\left[1,\ldots,k-1\right]$
	in (\ref{eq: seq-kcombs revol}). Let $\mathit{fst}\left(xs\right)$
	and $\mathit{last}\left(xs\right)$ denote the first and last elements
	of a list $xs$, respectively. Then:
	
	\begin{align*}
		& \quad\;\mathit{map}\left(\mathit{fst},\mathit{kcombs}_{\text{revol}}\left(K,\left[N,N-1,\ldots,1\right]\right)\right)\\
		& =\mathit{map}\left(\mathit{fst},\left[\left[\left[\:\right]\right]\right]\cup\left[\mathit{cs}_{k}^{N-1}\cup\mathit{reverse}\left(\mathit{map}\left(\cup\left[N\right],\mathit{cs}_{k-1}^{N-1}\right)\right)\mid k\leftarrow\left[1,\ldots,K\right]\right]\right)\\
		& =\left[\:\right]\cup\left[\mathit{fst}\left(\mathit{cs}_{k}^{N-1}\cup\mathit{reverse}\left(\mathit{map}\left(\cup\left[N\right],\mathit{cs}_{k-1}^{N-1}\right)\right)\right)\mid k\leftarrow\left[1,\ldots,K\right]\right]\\
		& =\left[\:\right]\cup\left[\mathit{fst}\left(\mathit{cs}_{k}^{N-1}\right)\mid k\leftarrow\left[1,\ldots,K\right]\right]\\
		& =\left[\:\right]\cup\left[\mathit{fst}\left(\mathit{cs}_{k}^{N-2}\cup\mathit{reverse}\left(\mathit{map}\left(\cup\left[N-1\right],\mathit{cs}_{k-1}^{N-2}\right)\right)\right)\mid k\leftarrow\left[1,\ldots,K\right]\right]\\
		& =\left[\:\right]\cup\left[\mathit{fst}\left(\mathit{cs}_{k}^{N-2}\right)\mid k\leftarrow\left[1,\ldots,K\right]\right]\\
		& \vdots\\
		& =\left[\:\right]\cup\left[\mathit{fst}\left(\mathit{cs}_{k}^{k}\right)\mid k\leftarrow\left[1,\ldots,K\right]\right]\\
		& =\left[\:\right]\cup\left[\left[1\ldots k\right]\mid k\leftarrow\left[1,\ldots,K\right]\right]
	\end{align*}
	
	For the last element, we have:
	\begin{equation}
		\begin{aligned} & \mathit{map}\left(\mathit{last},\mathit{kcombs}_{\text{revol}}\left(K,\left[N,N-1,\ldots,1\right]\right)\right)\\
			= &  \text{ \{  definition of \ensuremath{\mathit{\mathit{kcombs}}}} \} \\
			& \mathit{map}\left(\mathit{last},\left[\left[\left[\:\right]\right]\right]\cup\left[\mathit{cs}_{k}^{N-1}\cup\mathit{reverse}\left(\mathit{map}\left(\cup\left[N\right],\mathit{cs}_{k-1}^{N-1}\right)\right)\mid k\leftarrow\left[1,\ldots,K\right]\right]\right)\\
			= &  \text{ \{  definition of \ensuremath{\mathit{map}}} \} \\
			& \left[\left[\:\right]\right]\cup\left[\mathit{last}\left(\mathit{cs}_{k}^{N-1}\cup\mathit{reverse}\left(\mathit{map}\left(\cup\left[N\right],\mathit{cs}_{k-1}^{N-1}\right)\right)\right)\mid k\leftarrow\left[1,\ldots,K\right]\right]\\
			= &  \text{ \{  definition of \ensuremath{\mathit{map}}} \} \\
			& \left[\left[\:\right]\right]\cup\left[\mathit{cs}_{k}^{N-1}\cup\mathit{last}\left(\mathit{reverse}\left(\mathit{map}\left(\cup\left[N\right],\mathit{cs}_{k-1}^{N-1}\right)\right)\right)\mid k\leftarrow\left[1,\ldots,K\right]\right]\\
			= &  \text{ \{  definition of \ensuremath{\mathit{reverse}}} \} \\
			& \left[\left[\:\right]\right]\cup\left[\mathit{fst}\left(\mathit{map}\left(\cup\left[N\right],\mathit{cs}_{k-1}^{N-1}\right)\right)\mid k\leftarrow\left[1,\ldots,K\right]\right]\\
			= &  \text{ \{  definition of \ensuremath{\mathit{map}}} \} \\
			& \left[\left[\:\right]\right]\cup\left[\mathit{fst}\left(\mathit{cs}_{k-1}^{N-1}\right)\cup\left[N\right]\mid k\leftarrow\left[1,\ldots,K\right]\right]\vdots\\
			= &  \text{ \{ $\mathit{fst}\left(\mathit{cs}_{k}^{N}\right)=\left[1\ldots k\right]$  \} } \\
			& \left[\left[\:\right]\right]\cup\left[\left[1\ldots k-1\right]\cup\left[N\right]\mid k\leftarrow\left[1,\ldots,K\right]\right]
		\end{aligned}
		\label{eq: Last}
	\end{equation}
	Thus, for every $k\in\left[1,\ldots,K\right]$, $\mathit{fst}\left(\mathit{cs}_{k}^{n}\right)$
	and $\mathit{last}\left(\mathit{cs}_{k}^{n}\right)$ differ in exactly
	two elements: $k-1$ and $N$. Similarly, from (\ref{eq: Last}),
	we have $\mathit{last}\left(\mathit{cs}_{k}^{N}\right)=\left[1\ldots k-1\right]\cup\left[N\right]$,
	it follows immediately that $\mathit{last}\left(\mathit{cs}_{k}^{N-1}\right)=\left[1\ldots k-1\right]\cup\left[N-1\right]$
	and 
	\begin{align*}
		\mathit{fst}\left(\mathit{reverse}\left(\mathit{map}\left(\cup\left[N\right],\mathit{cs}_{k-1}^{N-1}\right)\right)\right)= & \mathit{last}\left(\mathit{cs}_{k-1}^{N-1}\right)\cup\left[N\right]\\
		= &  \{ \mathit{last}\left(\mathit{cs}_{k}^{N}\right)=\left[1\ldots k-1\right]\cup\left[N\right] \} \\
		& \left[1\ldots k-2\right]\cup\left[N-1\right]\cup\left[N\right]
	\end{align*}
	Hence, $\mathit{last}\left(\mathit{cs}_{k}^{N-1}\right)$ and $\mathit{fst}\left(\mathit{reverse}\left(\mathit{map}\left(\cup\left[N\right],\mathit{cs}_{k-1}^{N-1}\right)\right)\right)$
	differ in exactly two elements: $N$ and $k-1$.
\end{proof}

\subsection{Proof of fusion theorem\label{subsec:Proof-of-fusion-condition}}
\begin{theorem}
	Fusion\emph{ }Theorem\emph{. Let $f$ be a function and let $h$ is
		a recursive function defined by the algebras $\mathit{alg}$, in the
		form
		\begin{equation}
			\begin{aligned}h & \left(\left[\;\right]\right)=alg_{1}\left(\left[\;\right]\right)\\
				h & \left(\left[a\right]\right)=alg_{2}\left(\left[a\right]\right)\\
				h & \left(x\cup y\right)=alg_{3}\left(h\left(x\right),h\left(y\right)\right).
			\end{aligned}
		\end{equation}
		Let $\mathit{hom}\left(alg_{1},alg_{2},alg_{3}\right)$ be the unique
		solution in $h$ to the equation \ref{eq: join-list fusion condition},
		and $\mathit{hom}$ and $\mathit{hom}^{\prime}$ are short for $\mathit{hom}\left(alg_{1},alg_{2},alg_{3}\right)$
		and $\mathit{hom}\left(alg_{1}^{\prime},alg_{2}^{\prime},alg_{3}^{\prime}\right)$.
		The fusion theorem states that $f\cdot\mathit{hom}=\mathit{hom}^{\prime}$
		if }$f\left(alg_{1}\left(\left[\;\right]\right)\right)=alg_{1}^{\prime}\left(\left[\;\right]\right)$
	\emph{and} $f\left(alg_{2}\left(\left[a\right]\right)\right)=alg_{2}^{\prime}\left(\left[a\right]\right)$\emph{
		and the recursive pattern satisfies}
	
	\emph{
		\begin{equation}
			f\left(alg_{3}\left(\mathit{hom}\left(x\right),\mathit{hom}\left(y\right)\right)\right)=alg_{3}^{\prime}\left(f\left(\mathit{hom}\left(x\right)\right),f\left(\mathit{hom}\left(y\right)\right)\right),\label{eq: join-list fusion condition-1}
		\end{equation}
		(\ref{eq: join-list fusion condition-1}) is known as the }fusion
	condition\emph{.}
\end{theorem}
\begin{proof}
	The recursive case can be proved by following equational reasoning
	\begin{align*}
		& f\left(alg_{3}\left(\mathit{hom}\left(x\right),\mathit{hom}\left(y\right)\right)\right)\\
		= &  \text{ \{  fusion condition} \} \\
		& \mathit{alg}_{3}^{\prime}\left(f\left(\mathit{hom}\left(x\right)\right),f\left(\mathit{hom}\left(y\right)\right)\right)\\
		= &  \text{ \{  definition of \ensuremath{\mathit{hom}^{\prime}}} \} \\
		& \mathit{alg}_{3}^{\prime}\left(\mathit{hom}^{\prime}\left(x\right),\mathit{hom}^{\prime}\left(y\right)\right)\\
		= &  \text{ \{  definition of \ensuremath{\mathit{hom}^{\prime}}} \} \\
		& \mathit{hom}^{\prime}\left(x\cup y\right)
	\end{align*}
\end{proof}

\end{document}